\documentclass{jfm}
   
\usepackage{graphicx}
\usepackage{natbib}
\usepackage{mathtools}
\usepackage{amssymb}
\usepackage{amsbsy}
\usepackage{overpic}
\usepackage{dsfont}
\usepackage{epsfig}
\usepackage{color}
\usepackage[normalem]{ulem}

\usepackage{tikz}
\usetikzlibrary{arrows}
\usetikzlibrary{shapes.geometric}
\usetikzlibrary{shapes.misc}
\usetikzlibrary{decorations.pathreplacing,decorations.markings}
\tikzset{cross/.style={cross out, draw=black, minimum size=2*(#1-\pgflinewidth), inner sep=0pt, outer sep=0pt},
cross/.default={3pt}}

\ifCUPmtlplainloaded \else
  \checkfont{eurm10}
  \iffontfound
    \IfFileExists{upmath.sty}
      {\typeout{^^JFound AMS Euler Roman fonts on the system,
                   using the 'upmath' package.^^J}%
       \usepackage{upmath}}
      {\typeout{^^JFound AMS Euler Roman fonts on the system, but you
                   dont seem to have the}%
       \typeout{'upmath' package installed. JFM.cls can take advantage
                 of these fonts,^^Jif you use 'upmath' package.^^J}%
      }
  \else
  \fi
\fi

\ifCUPmtlplainloaded \else
  \checkfont{msam10}
  \iffontfound
    \IfFileExists{amssymb.sty}
      {\typeout{^^JFound AMS Symbol fonts on the system, using the
                'amssymb' package.^^J}%
       \usepackage{amssymb}%
         \let\leq=\leqslant
         
      }{}
  \fi
\fi

\ifCUPmtlplainloaded \else
  \IfFileExists{amsbsy.sty}
    {\typeout{^^JFound the 'amsbsy' package on the system, using it.^^J}%
     \usepackage{amsbsy}}
    {\providecommand\boldsymbol[1]{\mbox{\boldmath $##1$}}}
\fi

\usepackage{bbold}

\usepackage{my_macros}

\graphicspath{{./}{figures/}}

\title{On the boundary layer structure near a highly permeable porous interface}

\author[M. P. Dalwadi, S. J. Chapman, S. L. Waters, and J. M. Oliver]%
{M\ls O\ls H\ls I\ls T\ns P.\ns D\ls A\ls L\ls W\ls A\ls D\ls I$^{1,2}$%
  \thanks{Email address for correspondence: mohit.dalwadi@nottingham.ac.uk}, \ns 
  S.\ns J\ls O\ls N\ls A\ls T\ls H\ls A\ls N\ns C\ls H\ls A\ls P\ls M\ls A\ls N$^{2}$,
  S\ls A\ls R\ls A\ls H\ns L.\ns W\ls A\ls T\ls E\ls R\ls S$^{2}$
  \and J\ls A\ls M\ls E\ls S\ns M.\ns O\ls L\ls I\ls V\ls E\ls R$^{2}$}

\affiliation{$^1$ Synthetic Biology Research Centre, The University of Nottingham, University Park, Nottingham, NG7 2RD, UK
\\$^2$ OCIAM, Mathematical Institute, Woodstock Road, Oxford, OX2 6GG, UK\\[\affilskip]}
 
\date{?; revised ?; accepted ?. - To be entered by editorial office}
\begin{document}

\maketitle

\begin{abstract}
The method of matched asymptotic expansions is used to study the canonical problem of steady laminar flow through a narrow two-dimensional channel blocked by a tight-fitting finite-length highly permeable porous obstacle. We investigate the behaviour of the local flow close to the interface between the single-phase and porous regions (governed by the incompressible Navier--Stokes and Darcy flow equations, respectively). We solve for the flow in these inner regions in the limits of low and high Reynolds number, facilitating an understanding of the nature of the transition from Poiseuille to plug to Poiseuille flow in each of these limits. Significant analytical progress is made in the high-Reynolds-number limit, and we explore in detail the rich boundary layer structure that occurs. We derive general results for the interfacial stress and for the conditions that couple the flow in the outer regions away from the interface. We consider the three-dimensional generalization to unsteady laminar flow through and around a tight-fitting highly permeable cylindrical porous obstacle within a Hele-Shaw cell. For the high-Reynolds-number limit, we give the coupling conditions and interfacial stress in terms of the outer flow variables, allowing information from a nonlinear three-dimensional problem to be obtained by solving a linear two-dimensional problem. Finally, we illustrate the utility of our analysis by considering the specific example of time-dependent forced far-field flow in a Hele-Shaw cell containing a porous cylinder with a circular cross-section. We determine the internal stress within the porous obstacle, which is key for tissue engineering applications, and the interfacial stress on the boundary of the porous obstacle, which has applications to biofilm erosion. In the high-Reynolds-number limit, we demonstrate that the fluid inertia can result in the cylinder experiencing a time-independent net force, even when the far-field forcing is periodic with zero mean.
\end{abstract}

\begin{keywords}
%Authors should not enter keywords on the manuscript, as these must be chosen by the author during the online submission process and will then be added during the typesetting process (see http://journals.cambridge.org/data/\linebreak[3]relatedlink/jfm-\linebreak[3]keywords.pdf for the full list)
\end{keywords}
 
\section{Introduction}
\label{sec: intro}

Flow through a narrow channel containing a porous blockage is a canonical problem with numerous practical applications. For example, this flow configuration is used: to purify colloids in separation science, where it is also known as dead-end filtration~\citep{van2003review,mccarthy1998effect,bessiere2005low}; to deliver nutrient to cells growing in a porous tissue construct in tissue engineering~\citep{haj1990cellular,o2009multiphase,jaasma2008design}; and to erode porous biofilms that have grown within a pipe, where the erosion is dependent on the interfacial shear stress~\citep{picioreanu2001two,duddu2009two,telgmann2004influence}. In many of these problems, the flow near the interface between the fluid and the porous blockage is of experimental or industrial interest. For example, the stress acting on the interface may be important because it is often coupled to some interesting physical phenomenon, such as erosion, mechanotransduction, or movement of the entire porous blockage. However, in many models the behaviour of the flow near the interface is ignored, leading to incomplete information about the flow. As the lubrication equations used to approximate the flow away from the interface are reduced in order, approximated interfacial boundary conditions must be applied to couple the flows away from the interface~\citep{Cummings&Waters,waters2006tissue,cummings2009tracking,o2009multiphase,o2013interplay}. A detailed understanding of the flow behaviour close to the interface will enable the correct coupling conditions to be applied and (dynamic) interfacial effects to be accurately included. It is, therefore, of fundamental interest to understand the flow near the interface of a porous blockage.

\begin{figure}
\centering
    	\includegraphics[width=\textwidth]{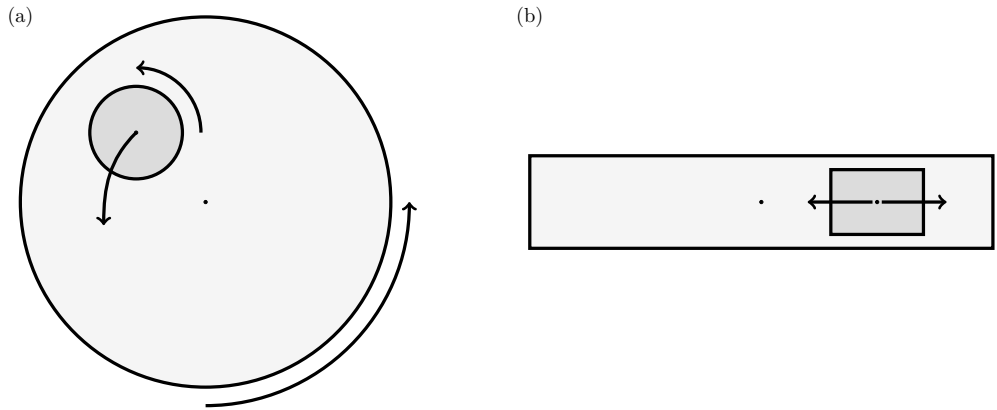}
\caption{Schematic of a high aspect ratio vessel bioreactor, containing a porous scaffold (darker region). The bioreactor has a similar shape to a petri dish turned on its side. (a) Face view. (b) Side view showing the gaps between the scaffold and the bioreactor. Cells are seeded within the scaffold, and the entire bioreactor is filled with a nutrient-rich fluid and rotated around its axis, causing the scaffold to move.}
\label{fig: HARV_schematic}
\end{figure}

Our main motivation arises from a tissue engineering problem. One approach to \emph{in vitro} tissue engineering is to seed cells onto a porous biomaterial scaffold, which is then cultured within a bioreactor. The combination of cells and scaffold is referred to as a tissue construct and one particular aim of tissue engineering is to make porous tissue constructs as permeable as possible whilst maintaining their structural integrity. This is to enhance nutrient delivery to the cells residing in the interior of the construct via advection. Moreover, many cells are mechanosensitive, and their proliferation rate depends on the stress that they experience. In a high aspect ratio vessel bioreactor, the porous tissue construct is placed within a bioreactor shaped like a petri dish turned on its side, and the entire set up is saturated with a nutrient-rich fluid, and rotated around the bioreactor axis (figure \ref{fig: HARV_schematic}). The construct moves according to the force it experiences and, in the short term, the construct undergoes periodic motion. However, over a long time, the tissue construct drifts from its periodic orbit, an effect that can be attributed to weak inertia. To predict the construct trajectory and, ultimately, the nutrient transport in such a bioreactor, we must determine the forces acting on the construct. Thus, determining the flow through the porous construct is of particular importance. To gain insights into this problem, we consider the flow past a tissue construct held at a fixed position, with a particular view to investigating the effect of weak inertia. Such an analysis can then be used to determine the construct trajectory when it is free to move. Another application of interest is the interfacial erosion of a porous biofilm, which is proportional to the square root of the fluid shear stress evaluated at the interface \citep{duddu2009two}. Thus, it is of interest to determine both the internal stress within the porous obstacle for tissue engineering applications and the interfacial stress acting on the obstacle boundary for biofilm erosion problems.

The main aim of this paper is to investigate steady laminar flow through a narrow two-dimensional channel blocked by a tightly fitting finite-length porous obstacle, and then to generalise this analysis to investigate a time-dependent three-dimensional flow past a porous cylinder with an arbitrary smooth cross-section within a Hele-Shaw cell, as illustrated in figure~\ref{fig: schematic set up model}. In particular, we are interested in the behaviour of the flow near the interface between the single-phase and porous regions, governed by the Navier--Stokes and Darcy equations, respectively. Transition from plug to Poiseuille flow in a two-dimensional channel is a classic problem, with the asymptotic structure examined by~\citet{van1970entry} and~\citet{wilson1971entry}, and numerical solutions given by, for example, \citet{brandt1966magnetohydrodynamic} and \citet{bodoia1961finite}. Although~\citet{van1970entry} mentions that a porous mesh would have to be used experimentally to induce uniform flow at the entry to a channel, the nature of the coupled flow between single-phase and porous regions remains an open question -- one which we will answer in this paper.

\citet{thompson1968secondary} used the method of matched asymptotic expansions to analyse the three-dimensional flow near a solid circular cylinder within a Hele-Shaw cell, exploiting the small aspect ratio of the cell. Due to the no-flux condition on the solid cylinder, the normal flow near the interface is small. We are motivated by the tissue engineering application in which there is flow within a highly permeable porous obstruction of arbitrary smooth cross-section. The normal flow, in general, is then significant near the interface and the flows inside and outside the obstacle are coupled through their boundary conditions at the interface. Although we are interested in a highly permeable obstacle, we do not consider the Brinkman extension to the Darcy equations in this paper because the porous construct must maintain its structural integrity. That is, we consider a porous medium whose underlying solid matrix is connected, so that Brinkman's equations do not apply \citep{auriault2009domain}.

We shall make extensive use of the method of matched asymptotic expansions by exploiting the small aspect ratio of the channel/Hele-Shaw cell -- one of our goals is to understand the flow near the interface (in an `inner' region) given a general flow away from the interface (in an `outer' region). Although the outer problems satisfy reduced equations, the inner problems are quasi-two-dimensional in each plane perpendicular to the interface. In particular, we determine the behaviour of inherently local properties, such as the stress acting on the interface. Additionally, we derive systematically the conditions that couple the outer equations for a general far-field forcing and for a general smooth cross-section of the porous obstacle. Due to their generality, the results obtained in this paper can significantly reduce the computational expense required to solve flow problems with coupled single-phase and porous regions, whilst retaining important interfacial information.

There is a large literature addressing the question of interfacial conditions on the boundary of a porous obstacle (and it remains an area of active research); see, for example, \citet{nield2006convection} for a comprehensive review. We do not address this question here, and use as our interfacial boundary conditions: continuity of flux and continuity of pressure, together with a no-tangential-slip condition. The first two are derived in \citet{levy1975boundary}, the last is a special case of the general tangential slip condition derived in \citet{carraro2015effective}.

The structure of this paper is as follows. In \S\ref{sec: 2D Flow}, we consider the steady two-dimensional case of flow through a channel containing a tightly-fitting highly permeable porous obstacle. In \S\ref{sec:modelsetup}, we formulate the mathematical problem. In \S\ref{sec: General asymptotic structure}, we describe the asymptotic structure in the small-aspect-ratio limit and present the problems to be solved to couple the outer single-phase and porous regions. In \S\ref{sec: Small Rey} and \S\ref{sec: Rey gg 1}, we investigate the asymptotic behaviour of these coupling conditions in the small- and large- Reynolds number limits, respectively. In particular, we find that the large-Reynolds-number limit induces a further boundary-layer structure, and investigate this in full. In \S\ref{sec: 3D flow} we extend our analysis to an unsteady three-dimensional flow in a Hele-Shaw cell past a tight-fitting highly permeable cylindrical porous obstacle. If the curvature of the cross-sectional boundary of the porous obstacle is not large (in a sense to be made precise later), the boundary layer structure in each plane perpendicular to the interface is equivalent to the two-dimensional case considered in \S\ref{sec: 2D Flow}, and we are able to generalize the two-dimensional results to the three-dimensional case. In \S\ref{sec: Example} we apply the general results of \S\ref{sec: 3D flow} to a time-dependent forced far-field flow in a Hele-Shaw cell containing a porous cylinder with circular cross-section. We determine the internal and interfacial stress within and on the porous construct, and show that fluid inertia can result in the generation of a net force on the cylinder, even when the far-field flow is periodic with zero mean. If the cylinder were allowed to move, this would cause a long-term drift, thus highlighting the singular perturbative nature of fluid inertia, and how it can be important in dynamical problems arising in lubrication flow.

\section{Steady two-dimensional flow}
\label{sec: 2D Flow}
\subsection{Model formulation}
\label{sec:modelsetup}

\begin{figure}
\centering
    	\includegraphics[width=\textwidth]{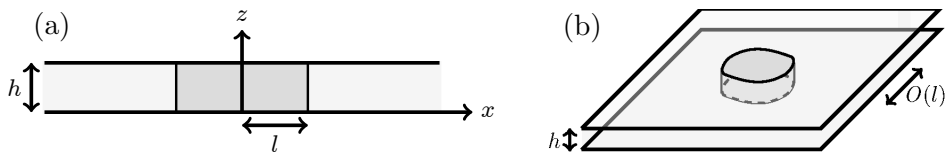}
\caption{The model geometry for flow within (a) a channel and (b) a Hele-Shaw cell. The light region is the exterior fluid and the dark region is the porous obstacle.}
\label{fig: schematic set up model}
\end{figure}

We consider the \emph{steady} two-dimensional flow of an incompressible Newtonian fluid with density $\fluiddensity$ and viscosity $\viscosity$ in an infinitely long channel of height $\hgt$. The channel contains a fully saturated porous obstacle with rectangular cross-section, height $\hgt$ and length $2 \leng$, as illustrated in figure~\ref{fig: schematic set up model}(a). The aspect ratio $\eps= \hgt/\leng$ is assumed to be small. We use Cartesian coordinates $(x,z)$ along and across the channel, as illustrated in figure~\ref{fig: schematic set up model}a, such that in the channel $z \in [0, \hgt]$ and the plug lies in $x \in [-\leng,\leng]$, $z \in [0, \hgt]$. The flow is driven by a prescribed upstream pressure gradient, such that in the far-field we have Poiseuille flow with cross-sectionally averaged velocity of magnitude $\V$.

The flow exterior to the porous plug (the light regions in figure~\ref{fig: schematic set up model}(a)), which we refer to as the `exterior flow', is governed by the incompressible steady Navier--Stokes equations with no body force. The exterior fluid velocity, pressure, and stress tensor are denoted by $\uu = u \bs{e}_x + w \bs{e}_z$, $\pres$, and $\stresstensor$, respectively, where $\bs{e}_x$ and $\bs{e}_z$ are the unit vectors in the $x$- and $z$-directions, respectively. The fluid inside the porous plug (the dark region in figure~\ref{fig: schematic set up model}a), which we refer to as the `interior flow', is governed by the incompressible Darcy equations, again with no body force. The Darcy (volume-averaged) velocity and (interstitial) pressure within the porous plug are denoted by $\QQ = U \bs{e}_x + W \bs{e}_z$ and $\Pres$, respectively. The porous plug has constant permeability $\permeability$. We do not use the porosity of the plug as a parameter as it is absorbed into the definition of the volume-averaged velocity, but note that it is assumed to be constant.
 
We nondimensionalize by setting $x = \leng x'$, $z = \eps \leng z'$, $u = \V u'$, $U = \V U'$, $w = \eps \V w'$, $W = \eps \V W'$, $\pres = (\viscosity \V/(\eps\hgt)) \pres'$, $\Pres = (\viscosity \V/\eps \hgt) \Pres'$, and $\stresstensor = (\viscosity \V/(\eps\hgt)) \stresstensor'$, where primes denote dimensionless quantities. Since all variables are dimensionless henceforth, we drop the primes without risk of confusion.

The dimensionless governing equations in the exterior region $\left|x\right|>1$, $0 < z < 1$ are the scaled Navier--Stokes equations given by
\begin{subequations}
\label{eq: 2D Navier Stokes}
\begin{align}
\label{eq: 2D Navier Stokes x-mom}
\eps \Rey \left(u u_x +  w u_z \right) &= -\pres_{x} + u_{zz} + \eps^2 u_{xx}, \\
\label{eq: 2D Navier Stokes z-mom}
\eps^3 \Rey \left(u w_x + w w_z \right) &= -\pres_{z} + \eps^2 w_{zz} + \eps^4 w_{xx}, \\
\label{eq: 2D Navier Stokes cont}
0 &= u_x + w_z,
\end{align}
\end{subequations}
where $\Rey = \fluiddensity \V \hgt /\viscosity$ is the modified Reynolds number based on the channel height. The governing equations in the porous plug $\left|x\right|<1$, $0 < z < 1$ are the scaled Darcy equations given by
\refstepcounter{equation}
\[
U  = - \Darcy \Pres_x, \quad \eps^2 W =  - \Darcy \Pres_z,\quad  0 = U_x + W_z,
\eqno{(\theequation{\mathit{a}-\mathit{c}})}\label{eq: 2D Darcy}
\]
where the dimensionless permeability $\Darcy = \permeability/\hgt^2$. For the Darcy equations to be valid, we must have $\Darcy \ll 1$. We discuss the typical sizes of the three dimensionless parameters appearing in our model, $\eps$, $\Rey$, and $\Darcy$, in \S\ref{sec: Parameter values}.

\subsubsection{Boundary conditions}
\label{sec: 2D BC}
On the channel walls we impose no-flux and no-slip conditions on the exterior flow and a no-flux condition on the Darcy flow, so that
\refstepcounter{equation}
\[
\uu = \bs{0} \quad \text{ on } z = 0, 1 \text{ for }  \left|x\right|>1; \qquad W = 0 \quad \text{on } z = 0, 1 \text{ for } \left|x\right|<1.
\eqno{(\theequation{\mathit{a},\mathit{b}})}\label{eq: no flux or slip BC}
\]
On the inflow ($x = -1$) and outflow ($x = 1$) interfaces, continuity of flux, continuity of pressure, and a no-tangential-slip condition \citep{levy1975boundary,carraro2015effective} are given by
\refstepcounter{equation}
\[
u = U, \quad \pres = \Pres, \quad w = 0 \quad \text{on } x = \pm 1 \text{ for } 0 < z < 1.
\eqno{(\theequation{\mathit{a}-\mathit{c}})}\label{eq: BC Cont of flux inflow}
\]

The flow is driven by imposing a constant, upstream pressure gradient in the far-field, with
\begin{align}
\label{eq: far-field pressure}
\dbyd{\pres}{x} \to -12 \quad \text{ as } x \ttni,
\end{align}
ensuring that the cross-sectionally averaged horizontal component of velocity has magnitude $1$ in the far-field. 

\subsection{Parameter values}
\label{sec: Parameter values}

Within a bioreactor, values of the three dimensionless parameters $\eps$, $\Rey$, and $\Darcy$ can vary greatly in magnitude (table \ref{tab: Parameter values}). We will investigate the dimensionless problem defined by~\eqref{eq: 2D Navier Stokes}--\eqref{eq: far-field pressure} in specific regions of this parameter space.

\begin{table}
  \begin{center}
\def~{\hphantom{0}}
  \begin{tabular}{l l l}
      Parameter  & Range & Reference  \\ \hline
      $\hgt$ & $2.5 \times 10^{-4}$ -- $6 \times 10^{-3}\, \mathrm{m}$ & \citet{pazzano2000comparison,pathi2005role} \\
      & & \citet{pierre2008theoretical}\\
      $\leng$ & $10^{-3}$ -- $10^{-1}\, \mathrm{m}$ & \citet{pazzano2000comparison,pathi2005role}\\
      & & \citet{pierre2008theoretical} \\
      $\permeability$ &  $10^{-12}$ -- $10^{-6} \, \mathrm{m}^2$ & \citet{vsimavcek1996permeability} \citet{sucosky2004fluid} \\
      & & \citet{nabovati2009general}\\
      $\V$ & $3 \times 10^{-9}$ -- $3 \times 10^{-2}\, \mathrm{m} \, \mathrm{s}^{-1}$ & \citet{zhao2005effects,pierre2008theoretical}  \\ 
      & & \citet{chung2007enhancement} \\
      $\fluiddensity$ & $10^{3} \, \mathrm{kg} \, \mathrm{m}^{-3}$ & \\
      $\viscosity$ & $8.9 \times 10^{-4} \, \mathrm{kg} \, \mathrm{m}^{-1} \, \mathrm{s}^{-1}$ &\\
      \hline
      $\eps $ & $2.5 \times 10^{-3}$ -- $6 \times 10^{0}$  & \\
      $\Rey $ & $10^{-6}$ -- $2 \times 10^{2}$& \\
      $\Darcy $ & $2.7 \times 10^{-8}$ --  $8.3 \times 10^{-2}$ & \\
  \hline
  \end{tabular} 
  \caption{Typical dimensional parameter values and the corresponding values of the dimensionless parameters.  The absolute upper bound of $1/12 \approx 8.3 \times 10^{-2}$ for $\Darcy$ can be obtained by considering unobstructed Poiseuille flow through a channel. We use the density and viscosity values of water at room temperature for $\fluiddensity$ and $\viscosity$, respectively.}
  \label{tab: Parameter values}
  \end{center}
\end{table}

The dimensionless problem is characterised by two lengthscales: the obstacle half-length, $1$, and the channel height, $\eps$. We proceed by considering the case in which $\eps \ll 1$, the channel height then being much smaller than the half-length of the obstacle.

The value of $\Darcy$ must be small enough for the underlying solid matrix to be connected, resulting in Darcy's equations governing the flow within the porous medium \citep{auriault2009domain}, but large enough to enhance nutrient transfer via fluid advection to the cells within the plug. There is an absolute upper bound of $\Darcy < 1/12$, and this can be seen by considering unobstructed Poiseuille flow through a channel. In reality, $\Darcy \ll 1$, and we work in this limit. We see in the three-dimensional case, considered in \S\ref{sec: 3D flow}, that $\Darcy = \order{\eps}$ corresponds to an obstacle which is impermeable to the flow at leading order. Hence, this limit reduces to the leading-order problem given in \citet{thompson1968secondary}, who considered the three-dimensional flow near a solid circular cylinder within a Hele-Shaw cell. We consider instead the limit in which $\eps \ll \Darcy \ll 1$, corresponding to the regime in which the plug is long, thin, and permeable at leading order in $\eps$. Examples of long, thin tissues include skin \citep{lei2011nasa}, urothelial tissue \citep{gabouev2003vitro}, and the cornea \citep{su2003situ}. Although we do assume that $\Darcy \ll 1$ throughout this paper, we effectively treat $\Darcy = \order{1}$ as $\eps \ttz$ in our following asymptotic analysis for mathematical expediency.

We must also define the relative scaling of $\Rey$. There is a distinguished limit when $\Rey = \order{1}$, and the investigation of this limit is the subject of the next section. Further analytic progress can be made when $\Rey$ is either small or large: we consider the asymptotic sub-limit $\Rey \ll 1$ in \S\ref{sec: Small Rey}, and the asymptotic sub-limit $1 \ll 1/\Darcy \ll \Rey \ll 1/\eps$ in \S\ref{sec: Rey gg 1}. We note that Darcy's law \eqref{eq: 2D Darcy} and the continuity of pressure condition (\ref{eq: BC Cont of flux inflow}b) would require modification if we were to generalize this work to significantly larger Reynolds numbers \citep{kaviany2012principles}.

\subsection{Asymptotic structure for $\Rey = \order{1}$}
\label{sec: General asymptotic structure}

\begin{figure}
\centering
    	\includegraphics[width=\textwidth]{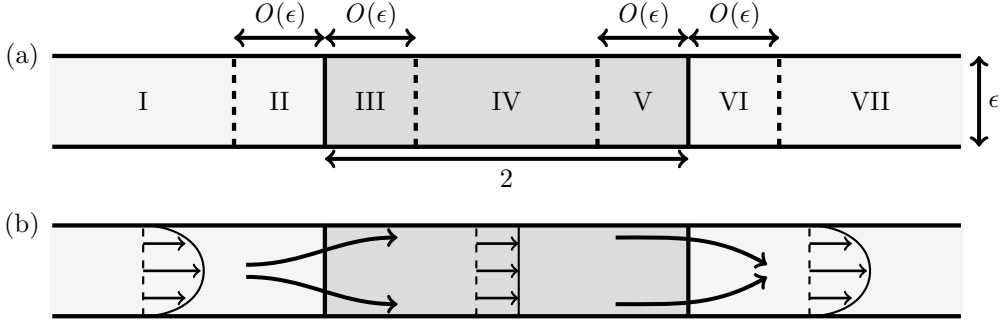}
\caption{The boundary-layer structure for $\Rey = \order{1}$. The light region is the exterior fluid and the dark region is the porous obstacle. (a) The different asymptotic regions are indicated and labelled in Roman numerals. (b) A schematic representation of the flow.}
\label{fig: schematic BL}
\end{figure}

We seek a solution using the method of matched asymptotic expansions with small parameter $\eps$. The asymptotic structure is shown in figure~\ref{fig: schematic BL}a, where each of seven asymptotic regions are labelled $\mathrm{I}$--$\mathrm{VII}$. We describe the regions as follows, noting that in each case the relevant transverse lengthscale is the channel height, $\eps$. There are three `outer' regions characterised by an $\order{1}$ axial lengthscale: two for the exterior fluid (regions $\mathrm{I}$ and $\mathrm{VII}$) in which we have Poiseuille flow at leading order, and one for the interior fluid (region $\mathrm{IV}$) in which we have plug flow at leading order. Thus, there is a transition from Poiseuille to plug to Poiseuille flow, between regions $\mathrm{I}$, $\mathrm{IV}$, and $\mathrm{VII}$, respectively, as illustrated schematically in figure~\ref{fig: schematic BL}b. The inflow transition from Poiseuille to plug flow occurs near the interface $x=-1$, where there are two `inner' regions, one on each side of the interface, each characterised by an axial lengthscale of $\order{\eps}$. The exterior inflow inner region is region $\mathrm{II}$, and the porous inflow inner region is region $\mathrm{III}$. In a similar manner, the outflow transition from plug to Poiseuille flow occurs near the interface $x=1$, where again there are two inner regions, one on each side of the interface, and each characterised by an axial lengthscale of $\order{\eps}$. The porous outflow inner region is region $\mathrm{V}$, and the exterior outflow inner region is region $\mathrm{VI}$.

\subsubsection{Outer regions}

In the outer regions regions $\mathrm{I}$, $\mathrm{IV}$, and $\mathrm{VII}$, we begin by posing asymptotic expansions of the form
\begin{align}
\label{eq: outer expansions}
(\uu,\QQ,\pres,\Pres) &= (\term{\uu}{0},\term{\QQ}{0},\term{\pres}{0},\term{\Pres}{0}) + \eps (\term{\uu}{1},\term{\QQ}{1},\term{\pres}{1},\term{\Pres}{1}) + \order{\eps^2} \quad \text{as } \eps \ttz.
\end{align}
Substituting~\eqref{eq: outer expansions} into~\eqref{eq: 2D Navier Stokes}, we find that the leading-order governing equations for the exterior flow are simply the lubrication equations
\refstepcounter{equation}
\[
0 = -\term{\pres}{0x} + \term{u}{0zz}, \quad 0 = -\term{\pres}{0z}, \quad 0 = \term{u}{0x} + \term{w}{0z} \quad \text{ for } \left|x\right|>1 \text{, } 0 < z < 1.
\eqno{(\theequation{\mathit{a},\mathit{b},\mathit{c}})}\label{eq: order 1 equations 2D}
\]
Substituting~\eqref{eq: outer expansions} into~\eqref{eq: 2D Darcy}, the leading-order governing equations for the interior flow are
\refstepcounter{equation}
\[
\term{U}{0} = - \Darcy \term{\Pres}{0x}, \quad 0 = - \Darcy \term{\Pres}{0z}, \quad 0 = \term{U}{0x} + \term{W}{0z} \quad \text{ for } \left|x\right|<1 \text{, } 0 < z < 1.
\eqno{(\theequation{\mathit{a},\mathit{b},\mathit{c}})}\label{eq: 2d PO equations leading order}
\]
At leading-order, the channel wall boundary conditions~\eqref{eq: no flux or slip BC} become\begin{subequations}
\label{eq: no flux or slip BC order 1 outer}
\begin{alignat}{3}
\label{eq: no flux or slip BC NS order 1 outer}
\uu_0 &= \bs{0} \quad &\text{on } z = 0, 1 \quad & \text{for } \left|x\right|>1, \\
\label{eq: no flux or slip BC Darcy order 1 outer}
W_0 &= 0 \quad &\text{on } z = 0, 1 \quad &\text{for } \left|x\right|<1,
\end{alignat}
\end{subequations}
while the far-field condition~\eqref{eq: far-field pressure} becomes
\begin{align}
\label{eq: far-field pressure leading order outer}
\dbyd{\pres_0}{x} \to -12 \quad \text{ as } x \ttni.
\end{align}
We are unable to impose at leading-order the interfacial boundary conditions~\eqref{eq: BC Cont of flux inflow} in the outer regions; these conditions are satisfied within the inner regions described in \S\ref{sec: Rey = order 1 inner regions}. For the outer problems, we must instead impose matching conditions (using the matching law given in~\citet{van1975perturbation}). These are determined in \S\ref{sec: Rey = order 1 inner regions}, and we state them here to close the leading-order outer problem. As the outer pressure is independent of $z$ and the fluid velocity is proportional to the pressure gradient, we find that we may impose continuity of total flux and pressure at each interface (corresponding to Neumann and Dirichlet conditions on the pressure, respectively). That is,
\refstepcounter{equation}
\[
\int_0^1 \! u_0 \, \mathrm{d}z = \int_0^1 \! U_0 \, \mathrm{d}z, \quad \pres_0 = \Pres_0 \quad \text{on } x = \pm 1.
\eqno{(\theequation{\mathit{a},\mathit{b}})}\label{eq: BC outer interfacial}
\]
We note that whilst the conditions~\eqref{eq: BC outer interfacial} close the outer problem, they give no information about, for example, the leading-order stress acting on the interfaces.

The leading-order solutions are readily obtained as follows: in \region{I},
\refstepcounter{equation}
\[
\term{\uu}{0} = 6z(1-z)\bs{e}_x, %\quad \term{w}{0} = 0,
\quad \term{\pres}{0} = -12x; % \quad \text{ for } x<-1 \text{, } 0 < z < 1;
\eqno{(\theequation{\mathit{a},\mathit{b}})}\label{eq: u_0 inflow}
\]
in \region{IV},
\refstepcounter{equation}
\[
\term{\UU}{0} = \bs{e}_x, %\quad \term{W}{0} = 0,
\quad \term{\Pres}{0} = -\dfrac{1+x}{\Darcy} + 12; % \quad \text{ for } \left|x\right|<1 \text{, } 0 < z < 1;
\eqno{(\theequation{\mathit{a},\mathit{b}})}\label{eq: Outer Darcy leading order sol}
\]
and in \region{VII},
\refstepcounter{equation}
\[
\term{\uu}{0} = 6z(1-z)\bs{e}_x, % \quad \term{w}{0} = 0,
\quad \term{\pres}{0} = -12x + \left(24 - \dfrac{2}{\Darcy}\right).% \quad \text{ for } x>1 \text{, } 0 < z < 1.
\eqno{(\theequation{\mathit{a},\mathit{b}})}\label{eq: u_0 outflow}
\]
The pressure is determined up to an arbitrary constant which (at this order) we fix by supposing $\pres_0 + 12 x \ttz$ as $x \ttni$, without loss of generality. We remark that the presence of the porous plug introduces a jump in pressure of $24 - 2/ \Darcy \sim -2/\Darcy$ (as $\Darcy$ is small). Equivalently, an $\order{1}$ pressure drop creates an $\order{\Darcy}$ flow through the porous plug.

We note that the horizontal components of velocity~(\ref{eq: u_0 inflow}a) and (\ref{eq: Outer Darcy leading order sol}a) are incompatible with the continuity of flux condition~(\ref{eq: BC Cont of flux inflow}a) on the interfacial inflow boundary for the full problem. Similarly for outflow, we find that (\ref{eq: Outer Darcy leading order sol}a) and (\ref{eq: u_0 outflow}a) are incompatible with~(\ref{eq: BC Cont of flux inflow}a). In the next section we introduce inner regions to resolve this issue, but for the remainder of this section we consider higher-order terms in the outer regions as the higher-order pressure is coupled to the leading-order flow in the inner regions.

Using the leading-order solutions~\eqref{eq: u_0 inflow} and~\eqref{eq: u_0 outflow}, the $\order{\eps}$ terms in the exterior-flow equations~\eqref{eq: 2D Navier Stokes} become
\refstepcounter{equation}
\[
0 = -\term{\pres}{1x} + \term{u}{1zz}, \quad 0 = -\term{\pres}{1z}, \quad 0 = \term{u}{1x} + \term{w}{1z} \quad \text{ for } \left|x\right|>1 \text{, } 0 < z < 1.
\eqno{(\theequation{\mathit{a}-\mathit{c}})}\label{eq: order 1 equations 2D HO}
\]
Similarly, using the leading-order solutions~\eqref{eq: Outer Darcy leading order sol}, the $\order{\eps}$ terms in the interior-flow equations~\eqref{eq: 2D Darcy} become
\refstepcounter{equation}
\[
\term{U}{1} = - \Darcy\term{\Pres}{1x}, \quad 0 = - \Darcy \term{\Pres}{1z}, \quad 0 = \term{U}{1x} + \term{W}{1z} \quad \text{ for } \left|x\right|<1 \text{, } 0 < z < 1.
\eqno{(\theequation{\mathit{a}-\mathit{c}})}\label{eq: 2d PO equations HO}
\]
The channel wall boundary conditions~\eqref{eq: no flux or slip BC} at this order are as follows
\begin{subequations}
\label{eq: no flux or slip BC order 1}
\begin{alignat}{3}
\label{eq: no flux or slip BC NS order 1}
\uu_1 &= \bs{0} \quad &\text{on } z = 0, 1 \quad & \text{for } \left|x\right|>1, \\
\label{eq: no flux or slip BC Darcy order 1}
W_1 &= 0 \quad &\text{on } z = 0, 1 \quad &\text{for } \left|x\right|<1,
\end{alignat}
\end{subequations}
while the far-field condition~\eqref{eq: far-field pressure} implies
\begin{align}
\label{eq: far-field pressure leading order outer 1O}
\dbyd{\pres_1}{x} \ttz \quad \text{ as } x \ttni.
\end{align}

The matching conditions across each interface are determined in \S\ref{sec: Rey = order 1 inner regions}, and we give them here to close the problem. For inflow, we have
\refstepcounter{equation}
\[
\int_0^1 \! u_1 \, \mathrm{d}z = \int_0^1 \! U_1 \, \mathrm{d}z, \quad \pres_1 - \Pres_1 = \presfny(\Rey,\Darcy) \quad \text{on } x = -1,
\eqno{(\theequation{\mathit{a},\mathit{b}})}\label{eq: BC outer interfacial HO inflow}
\]
and for outflow we have
\refstepcounter{equation}
\[
\int_0^1 \! u_1 \, \mathrm{d}z = \int_0^1 \! U_1 \, \mathrm{d}z, \quad \pres_1 - \Pres_1 = \presfnz(\Rey,\Darcy) \quad \text{on } x = 1,
\eqno{(\theequation{\mathit{a},\mathit{b}})}\label{eq: BC outer interfacial HO outflow}
\]
where the pressure jumps $\presfny$ and $\presfnz$ are functions of the parameters $\Rey$ and $\Darcy$. In \S\ref{sec: Rey = order 1 inner regions}, we specify the problems that need to be solved to determine these functions, while in \S\ref{sec: Small Rey} and \S\ref{sec: Rey gg 1} we determine their asymptotic behaviour for small and large $\Rey$.

Given $\Pi^{\pm}$, the solution to \eqref{eq: order 1 equations 2D HO}--\eqref{eq: BC outer interfacial HO outflow} is readily determined: in \region{I},
\refstepcounter{equation}
\[
\term{\uu}{1} = \bs{0}, \quad \term{\pres}{1} = \presfny; % \quad \text{ for } x<-1 \text{, } 0 < z < 1,
\eqno{(\theequation{\mathit{a},\mathit{b}})}\label{eq: u_1 inflow}
\]
in \region{IV},
\refstepcounter{equation}
\[
\term{\QQ}{1} = \bs{0}, \quad \term{\Pres}{1} = 0; % \quad \text{ for } \left|x\right|<1 \text{, } 0 < z < 1,
\eqno{(\theequation{\mathit{a},\mathit{b}})}\label{eq: Outer Darcy HO sol}
\]
and in \region{VII},
\refstepcounter{equation}
\[
\term{\uu}{1} = \bs{0}, \quad \term{\pres}{1} = \presfnz. % \quad \text{ for } x>1 \text{, } 0 < z < 1.
\eqno{(\theequation{\mathit{a},\mathit{b}})}\label{eq: u_1 outflow}
\]
As with the leading-order solutions~\eqref{eq: u_0 inflow}--\eqref{eq: u_0 outflow}, the pressure field is unique up to an arbitrary constant. For convenience, we choose the constant such that $\Pres_1(0,0.5) = 0$, without loss of generality. We proceed by considering the inner regions, highlighting the problems that would need to be solved numerically to determine the functions $\presfny$ and $\presfnz$.

\subsubsection{Inner regions}
\label{sec: Rey = order 1 inner regions}
 
The appropriate scalings in the inflow and outflow inner regions are $x = \mp1 + \eps X^{\mp}$, respectively, with $(w, W) = (\wi, \Wi)/\eps$. To make it clear that we are working in the inner regions we introduce overbars on the other dimensionless variables, writing $(u, U) = (\ui, \Ui)$, $(\pres, \Pres) = (\pri, \Pri)$ and $(\bs{\ui}, \Qi) = (\ui, \Ui) \bs{e}_x + (\wi, \Wi) \bs{e}_z$. Using vector operators in terms of $X^{\mp}$ and $z$ as appropriate, the exterior fluid governing equations~\eqref{eq: 2D Navier Stokes} become
\refstepcounter{equation}
\[
\Rey \left(\bs{\ui} \bcdot \nabla \right)\bs{\ui} = -\eps^{-1}\nabla \pri + \nabla^2 \bs{\ui}, \quad 0 = \nabla \bcdot \bs{\ui} \quad \text{in regions $\mathrm{II}$ and $\mathrm{VI}$},
\eqno{(\theequation{\mathit{a},\mathit{b}})}\label{eq: Inner NS}
\]
which are valid when $X^{-} < 0$ for inflow (\region{II}) and $X^{+}> 0$ for outflow (\region{VI}), with $0 < z < 1$ in both cases. The interior fluid governing equations~\eqref{eq: 2D Darcy} become
\refstepcounter{equation}
\[
\Qi = - \eps^{-1} \Darcy \nabla \Pri, \quad 0 = \nabla \bcdot \Qi  \quad \text{in regions $\mathrm{III}$ and $\mathrm{V}$},
\eqno{(\theequation{\mathit{a},\mathit{b}})}\label{eq: Inner Darcy}
\]
which are valid for $X^{-} > 0$ for inflow (\region{III}) and $X^{+} < 0$ for outflow (\region{V}), with $0 < z < 1$ in both cases.
 
Similarly to~\eqref{eq: outer expansions}, we form inner expansions in powers of $\eps$ as follows:
\begin{align}
\label{eq: inner expansions}
(\bs{\ui},\Qi,\pri,\Pri) &= (\term{\bs{\ui}}{0},\term{\Qi}{0},\term{\pri}{0},\term{\Pri}{0}) + \eps (\term{\bs{\ui}}{1},\term{\Qi}{1},\term{\pri}{1},\term{\Pri}{1}) + \order{\eps^2} \quad \text{as } \eps \ttz.
\end{align}
At leading order, it follows from~(\ref{eq: Inner NS}a) that
\begin{align}
\label{eq: leading order bulk pressure inner}
\bs{0} = -\nabla \term{\pri}{0} \quad \text{in regions $\mathrm{II}$ and $\mathrm{VI}$},
\end{align}
while~(\ref{eq: Inner Darcy}a) gives, similarly,
\begin{align}
\label{eq: leading order porous pressure inner}
\bs{0} = - \Darcy \nabla \term{\Pri}{0} \quad \text{in regions $\mathrm{III}$ and $\mathrm{V}$}.
\end{align}

The appropriate interfacial conditions on $X^{\mp}=0$ for the leading-order pressure are given by the leading-order terms in~(\ref{eq: BC Cont of flux inflow}b), which become
\begin{align}
\label{eq: leading order pressure on interface}
 \pri_0 = \Pri_0 \quad \text{on } X^{\mp} = 0.
\end{align}
Thus $\pres_0$ and $\Pres_0$ are constant and equal in regions $\mathrm{II}$ and $\mathrm{III}$ and in regions $\mathrm{V}$ and $\mathrm{VI}$, leading to the matching condition~(\ref{eq: BC outer interfacial}b). Using the leading-order outer solutions~\eqref{eq: u_0 inflow}--\eqref{eq: u_0 outflow} for the pressure, we find
\refstepcounter{equation}
\[
\term{\pri}{0} = 12 \quad \text{in \region{II}}, \qquad \term{\Pri}{0} = 12 \quad \text{in \region{III}},
\eqno{(\theequation{\mathit{a},\mathit{b}})}\label{eq: Leading order inflow pressures}
\]
for inflow, and
\refstepcounter{equation}
\[
\term{\Pri}{0} = 12 - \dfrac{2}{\Darcy} \quad \text{in \region{V}}, \qquad \term{\pri}{0} = 12 - \dfrac{2}{\Darcy} \quad \text{in \region{VI}}.
\eqno{(\theequation{\mathit{a},\mathit{b}})}\label{eq: Leading order outflow pressures}
\]
for outflow. We note that these leading-order pressures are constant on the interface, as predicted by \citet{levy1975boundary}.

\begin{figure}
\centering
\includegraphics[width=\textwidth]{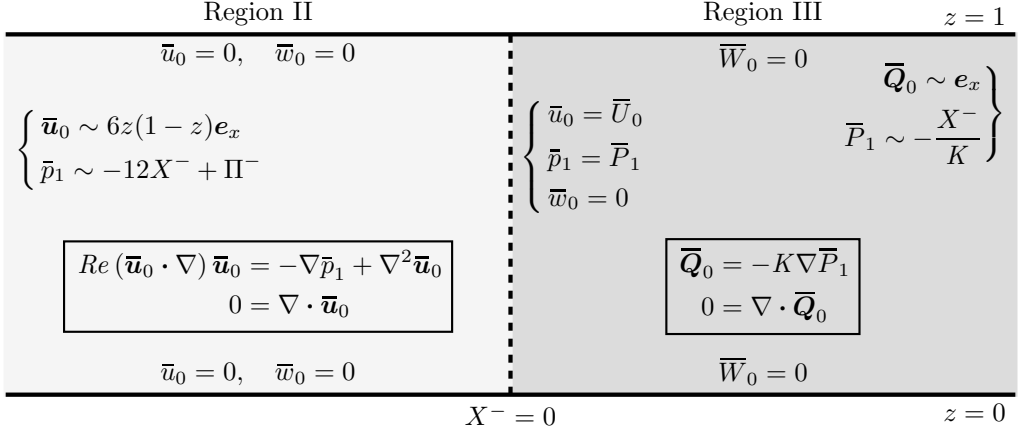}
\caption{The inner inflow problem for $\Rey = \order{1}$. The flow is from left to right.}
\label{fig: inflow leading order}
\end{figure}

\begin{figure}
\centering
\includegraphics[width=\textwidth]{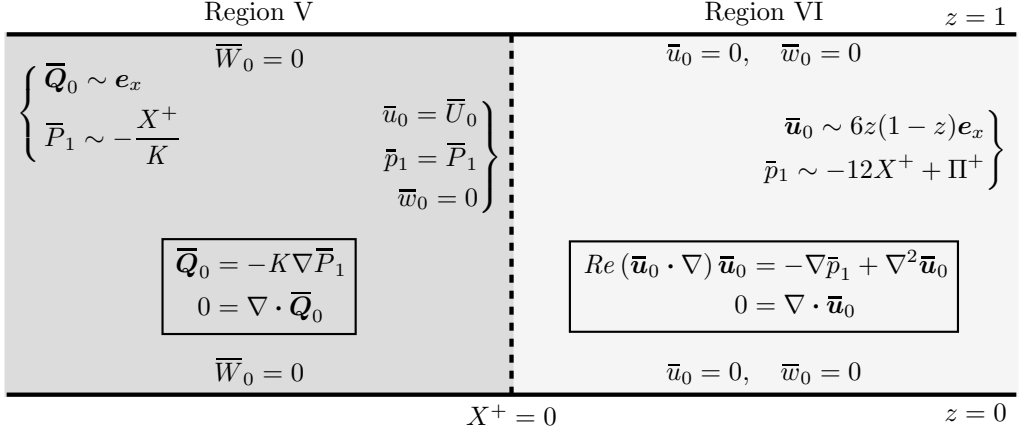}
\caption{The inner outflow problem for $\Rey = \order{1}$. The flow is from left to right.}
\label{fig: outflow leading order}
\end{figure}

As the leading-order fluid pressures are constant within each inner region, the leading-order inner flow is driven by the first correction to pressure, and these equations are given by the $\order{1}$ terms in \eqref{eq: Inner NS} and \eqref{eq: Inner Darcy}. We illustrate the leading-order flow problems in figure~\ref{fig: inflow leading order} for inflow, and in figure~\ref{fig: outflow leading order} for outflow. In both cases, the inner flow transition problem is governed by the two-dimensional steady Navier--Stokes equations coupled to the Darcy equations. To fully determine the inflow and outflow pressure jumps $\presfny$ and $\presfnz$, we would have to solve the problems indicated in figures~\ref{fig: inflow leading order} and~\ref{fig: outflow leading order} numerically for two parameters: $\Rey$ and $\Darcy$.

However, to couple the outer regions, we require another condition on the pressure. We note that we can analytically determine suitable coupling conditions for the horizontal components of velocity averaged over the channel height by appealing to global conservation of mass, rather than solving the full problem. By integrating the continuity equations across the channel height, using the conditions for the flow in the normal direction at each boundary, and matching with the outer solutions, we deduce the following coupling conditions for the outer horizontal components of velocity
\refstepcounter{equation}
\[
\int_0^1 \! u_0 \, \mathrm{d}z = \int_0^1 \! U_0 \, \mathrm{d}z, \quad \int_0^1 \! u_1 \, \mathrm{d}z = \int_0^1 \! U_1 \, \mathrm{d}z \quad \text{on } x = \pm 1.
\eqno{(\theequation{\mathit{a},\mathit{b}})}\label{eq: Cont of flux 2D macro}
\]

We proceed by determining the inner flow in the asymptotic sub-limits $\Rey \ll 1$ and $1 \ll 1/\Darcy \ll \Rey \ll 1/\eps$, starting with the former. This will allow us to calculate quantities that are unobtainable from sole consideration of the outer problem, such as the stress acting on the interface and the limiting behaviour of the functions $\presfny$, $\presfnz$.

\subsection{Small Reynolds number: $\Rey \ll 1$}
\label{sec: Small Rey}
In the sub-limit in which $\Rey \ll 1$, the leading-order problems outlined in figures~\ref{fig: inflow leading order} and~\ref{fig: outflow leading order} become Stokes flow coupled with Darcy flow. Therefore, the flow equations are now reversible, and it suffices to solely consider inflow. We therefore use $X^{-} = X$ for ease of notation. Further, we note that the inflow and outflow pressure jumps are now linked via the expression $\presfnz = -\presfny$.

We solve the resulting leading-order system numerically (for the streamfunction within the exterior flow and for the pressure within the porous flow) using a second-order accurate central finite-difference scheme. Due to the elliptic nature of the governing equations, we use an iterative method to bypass some of the cost of solving the fully coupled elliptic problems in one step. That is, we iterate between solving Laplace's equation for the pressure in \region{III}, and the biharmonic equation for the streamfunction in \region{II}, and use the solution found in the previous iteration to determine the boundary conditions for the current iteration. We truncate the infinite domain to $[-1, 0]$ for $X$ in \region{II} and $[0, 1]$ for $X$ in \region{III} (we note that an extended domain adds little accuracy to the scheme), using 300 grid points for $X$ in each region and 300 grid points for $z$. The simulation is halted once the difference between successive iterations at each grid point is less than $10^{-4}$.

\begin{figure}
\centering
\includegraphics[width=\textwidth]{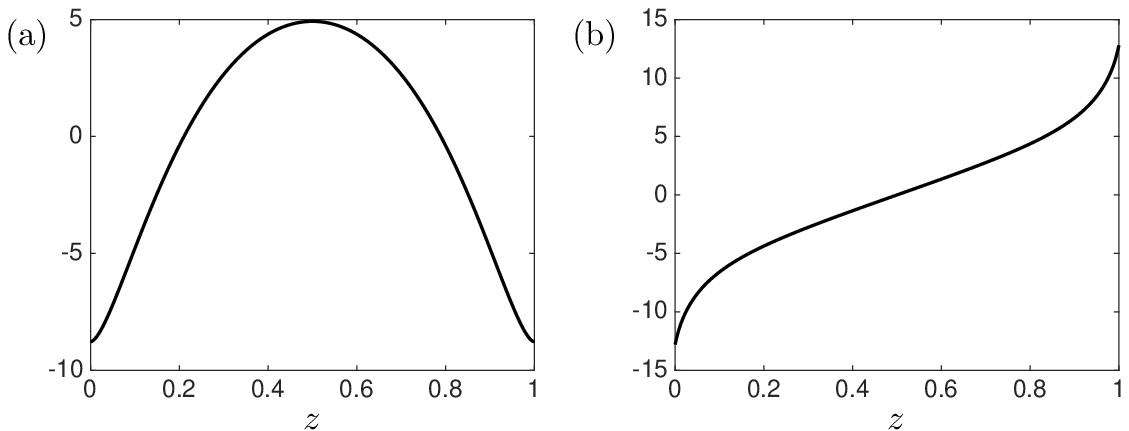}
\caption{The $\order{\eps}$ normal and shear stresses acting on the interface $X = 0$, for $\Darcy = 0.01$.  (a) $\pri_1 - 2\ui_{0X}$ (b) $-\left(\ui_{0z} + \wi_{0X}\right)$.}
\label{fig: All_low_Rey}
\end{figure}

The normal and shear components of stress acting on the interface are
\begin{subequations}
\label{eq: low rey stress}
\begin{align}
\label{eq: low rey normal stress}
\stress_{xx} &\sim -12 + \eps \left(-\pri_1 + 2 \ui_{0X} \right) , \\
\stress_{xz} &\sim \eps \left(\ui_{0z} + \wi_{0X} \right),
\end{align}
\end{subequations}
respectively, as $\eps, \Rey \ttz$. We show the $\order{\eps}$ stress results for $\Darcy = 0.01$ in figure \ref{fig: All_low_Rey}, and calculate $\presfny = 1.92$ to three significant figures. The leading-order normal stress term in \eqref{eq: low rey normal stress} takes the specific value of $-12$ because we have imposed $\Pres(0,z) = 0$ to uniquely define the pressure. In fact, it is the $\order{\eps}$ terms in \eqref{eq: low rey stress} which are more interesting, as these are the dominant terms which vary in $z$ and, unlike the leading-order pressure, cannot simply be determined from the outer solution. The boundary layer analysis is essential to obtain these results, which cannot be determined solely from the outer solutions. The reason the outer solutions cannot make good predictions of the $z$-variation of interfacial stress is because of the tendency of the exterior flow to move towards the channel edges in transition between Poiseuille and plug flow in the boundary region. The interfacial normal stress is greater/smaller than predicted by the outer flow towards the centre/edges of the channel, whilst the magnitude of interfacial shear stress is greater than predicted by the outer solution.

\subsection{Large Reynolds number: $1 \ll 1/ \Darcy \ll \Rey \ll 1/\eps$}
\label{sec: Rey gg 1}
 
We now consider the sub-limit in which $1 \ll 1/\Darcy \ll \Rey \ll 1/\eps$. We do not consider further the sub-limits in which $\Rey \Darcy = \order{1}$ or $1 \ll \Rey \ll 1/ \Darcy \ll 1/ \eps$. This is because we are interested in the tissue engineering problem with porous plugs whose values of $\Darcy$ are not too small. Moreover, the above sub-limits yield coupled nonlinear problems which are not amenable to analytic study. We emphasize that the limit we consider is when the external fluid has a large Reynolds number, but the pore Reynolds number is small, so that the Darcy equations and continuity of pressure conditions are valid.

In this section we perform an asymptotic analysis using the small parameter $\epsb = 1/\Rey \ll 1$ for ease of notation. Therefore, when we refer to, for example, `leading-order' in this section, it is meant in reference to $\epsb$. In the limit $\epsb \ttz$, there is a significant change in boundary layer structure within the inner regions $\mathrm{II}$ and $\mathrm{VI}$ compared to the $\order{1}$ Reynolds number case. The limit as $\epsb \ttz$ is singular, and we expect viscous effects to become important in regions near each boundary. Furthermore, as fluid momentum dominates within regions $\mathrm{II}$ and $\mathrm{VI}$, the flow direction is important. Indeed, we obtain a different boundary layer structure for inflow and outflow, and therefore split the analysis into these two cases, starting with inflow.

In the limit of small $\epsb$, the first-correction terms $\uu_1$, $\QQ_1$, $\pres_1$, and $\Pres_1$ from the asymptotic series in the outer regions~\eqref{eq: outer expansions} are now scaled with $\epsb^{-1}$ to be consistent with the new inner scalings, as follows:
\begin{align}
\label{eq: asymptotic series for large Rey}
(\uu_1, \QQ_1, \pres_1, \Pres_1) = \epsb^{-1}(\uu_{10}, \QQ_{10}, \pres_{10}, \Pres_{10}) + (\uu_{11}, \QQ_{11}, \pres_{11}, \Pres_{11}) + \order{\epsb}.
\end{align}
The scalings for the pressure jumps $\presfny$ and $\presfnz$ will turn out to be
\refstepcounter{equation}
\[
\presfny = \presfny_1 + \order{\epsb}, \quad
\presfnz = \epsb^{-1} \presfnz_0 + \order{1} \quad \text{as } \epsb \ttz,
\eqno{(\theequation{\mathit{a},\mathit{b}})}\label{eq: asymptotic series for pres jump}
\]
our objective being to determine $\presfny_1$ and $\presfnz_0$.

Though we have not yet fully described the inner asymptotic structure, we give the form of the asymptotic expansions within each region in Table~\ref{tab: Asymptotic expansions} for ease of reference.

\begin{table}
  \begin{center}
\def~{\hphantom{0}}
  \begin{tabular}{l l l}
      Region  & Asymptotic expansions & Relative scalings  \\ \hline  
 $\mathrm{I}$ & $\left(\uu,\pres\right) = \left(\uu_0, \pres_0 \right) + \eps \left(\uu_1, \pres_1 \right) + \order{\eps^2}$ & \\ 
              & $(\bs{u}_1,\pres_1) = \epsb^{-1}\left(\uu_{10}, \pres_{10} \right) + \left(\uu_{11}, \pres_{11} \right) + \order{\epsb}$ &  \\ \hline
$\mathrm{II}$ & $\left(\bs{\ui},\pri \right) = \left(\bs{\ui}_0,\pri_0 \right) + \eps \left(\bs{\ui}_1,\pri_1 \right) + \order{\eps^{2}}$ & $u = \ui$, $w = \eps^{-1} \wi$ \\
               & $\bs{\ui}_0 = 6z(1-z) \bs{e}_x + \epsb \bs{\ui}_{01} + \order{\epsb^{4/3}}$ & $x = \eps X$, $\pres = \pri$ \\
             & $\pri_1 = \pri_{11} + \order{\epsb^{1/3}}$ & \\ \hline
$\mathrm{IIa}$ & $\left(\bs{\uii},\prii \right) = \left(\bs{\uii}_0,\prii_0 \right) + \eps \left(\bs{\uii}_1,\prii_1 \right) + \order{\eps^{2}}$ & $u = \uii$, $w = \eps^{-1} \epsb \wii$ \\
               & $\uii_{0} = 6z(1-z) + \epsb \ui_{01}(0,z) + \order{\epsb^{4/3}}$ & $x = \eps \epsb \Ri$,  $\pres = \prii$ \\
               & $\wii_{0} = \wii_{01} + \order{\epsb^{1/3}}$ &  \\
          & $\prii_1 = \pri_{11}(0,z) + \order{\epsb^{1/3}}$ & \\ \hline
$\mathrm{III}$ & $\left(\Qi,\Pri \right) = \left(\Qi_0,\Pri_0 \right) + \eps \left(\Qi_1,\Pri_1 \right) + \order{\eps^{2}}$ & $U = \Ui$, $W = \eps^{-1} \Wi$ \\
               & $\Qi_0 = \Qi_{00} + \epsb \Qi_{01} + \order{\epsb^{4/3}}$ & $x = \eps X$, $\Pres = \Pri$\\
           & $\Pri_1 = \Pri_{11} + \order{\epsb^{1/3}}$ & \\ \hline
$\mathrm{IV}$ & $\left(\QQ,\Pres\right) = \left(\QQ_0, \Pres_0 \right) + \eps \left(\QQ_1, \Pres_1 \right) + \order{\eps^2}$ &  \\ 
           & $(\QQ_1,\Pres_1) = \epsb^{-1}\left(\QQ_{10}, \Pres_{10} \right) + \left(\QQ_{11}, 		\Pres_{11} \right) + \order{\epsb}$ & \\ \hline
$\mathrm{V}$  & $\left(\Qi,\Pri \right) = \left(\Qi_0,\Pri_0 \right) + \eps \left(\Qi_1,\Pri_1 \right) + \order{\eps^{2}}$ & $U = \Ui$, $W = \eps^{-1} \Wi$ \\
               & $\Qi_0 \sim \bs{e}_x + \epsb^{1/2} \Qi_{0a}$ & $x = \eps X$, $\Pres = \Pri$ \\
          & $\Pri_1 \sim -X^{-}/ \Darcy + \epsb^{1/2} \Pri_{1b}$ & \\ \hline
$\mathrm{VI}$  & $\left(\bs{\ui},\pri \right) = \left(\bs{\ui}_0,\pri_0 \right) + \eps \left(\bs{\ui}_1,\pri_1 \right) + \order{\eps^{2}}$ & $u = \ui$, $w = \eps^{-1} \wi$\\
               & $\bs{\ui}_0 \sim \bs{e}_x + \epsb^{1/2} \bs{\ui}_{0a}$ & $x = \eps X$, $\pres = \pri$ \\
           & $\pri_1 \sim \epsb^{-1/2} \pri_{1a}$ & \\ \hline 
       $\mathrm{VIa}$  & $\left(\bs{\uiii},\priii \right) = \left(\bs{\uiii}_0,\priii_0 \right) + \eps \left(\bs{\uiii}_1,\priii_1 \right) + \order{\eps^{2}}$ & $u = \uiii$, $w = \eps^{-1} \epsb \wiii$\\
               & $ \uiii_0 = \uiii_{00} + \order{\epsb^{1/2}}$ & $x = \eps \epsb^{-1} \Rii$, $\pres = \priii$ \\
               & $ \wiii_0 = \wiii_{01} + \order{\epsb^{1/2}}$ & \\
             & $\priii_1 = \epsb^{-1}\priii_{10} + \order{\epsb^{-1/2}}$ & \\ \hline
          $\mathrm{VII}$ & $\left(\uu,\pres\right) = \left(\uu_0, \pres_0 \right) + \eps \left(\uu_1, \pres_1 \right) + \order{\eps^2}$ & \\ 
               & $(\bs{u}_1,\pres_1) = \epsb^{-1}\left(\uu_{10}, \pres_{10} \right) + \order{\epsb^{-1/2}}$ &  \\ \hline
  \end{tabular} 
  \caption{Asymptotic expansions within boundary layers in the large-Reynolds-number limit.}
  \label{tab: Asymptotic expansions}
  \end{center}
\end{table}

\subsubsection{Inflow}
\label{sec: Inflow}

\begin{figure}
\centering
\includegraphics[width=0.7\textwidth]{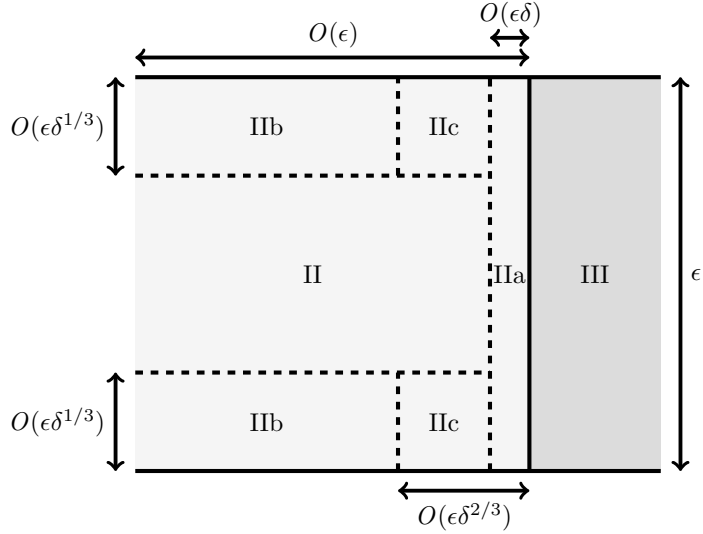}
\caption{Schematic diagram of the inner regions $\mathrm{IIa}$-$\mathrm{c}$ within the inflow boundary \region{II}. The light region is the exterior fluid and the dark region is the porous obstacle. The size of the boundary layers has been exaggerated for illustrative purposes}
\label{fig: inflow inner BL}
\end{figure}

The inflow inner problem (within regions $\mathrm{II}$ and $\mathrm{III}$) is shown in figure~\ref{fig: inflow leading order}. The limit as $\epsb \ttz$ yields leading-order inviscid flow in \region{II} away from the boundaries, with five additional boundary layers being required to satisfy all of the boundary conditions. Two of these viscous boundary layers are on the channel walls, one is on the porous interface, whilst the remaining boundary layers occur in the corners between the channel wall and porous interface. The boundary layer structure for the inflow problem is shown in figure~\ref{fig: inflow inner BL}.
 
We start by considering the coupled regions $\mathrm{II}$ and $\mathrm{III}$, and show that the transition to Poiseuille flow occurs solely within \region{III} at leading order in $\epsb$.

\subsubsection{Regions $\mathrm{II}$ and $\mathrm{III}$}
\label{sec: Regions II and III}

Within regions $\mathrm{II}$ and $\mathrm{III}$, we make the following asymptotic expansions
\begin{subequations}
\label{eq: High Rey II and III asymptotic expansions}
\begin{align}
\term{\bs{\ui}}{0} &= 6z(1-z) \bs{e}_x + \epsb \term{\bs{\ui}}{01} + \epsb^{4/3} \term{\bs{\ui}}{02} + \order{\epsb^{5/3}}, \\
\term{\Qi}{0} &= \term{\Qi}{00} + \epsb \term{\Qi}{01} + \epsb^{4/3} \term{\Qi}{02} + \order{\epsb^{5/3}}, \\
\label{eq: regions II and III pressure ae}
(\term{\pri}{1}, \term{\Pri}{1}) &= (\term{\pri}{11},\term{\Pri}{11}) + \epsb^{1/3} (\term{\pri}{12},\term{\Pri}{12})+ \order{\epsb^{2/3}}
\end{align}
\end{subequations}
as $\epsb \ttz$, where the non-integer powers of $\epsb$ come from corrections to the flow that arise due to the presence of a boundary layer near the channel walls (regions $\mathrm{IIb}$ in figure~\ref{fig: inflow inner BL}). These terms are used in Appendix~\ref{sec: Inflow app}, but are not required further in this section.

Substituting the asymptotic expansions~\eqref{eq: High Rey II and III asymptotic expansions} into the governing equations presented in figure~\ref{fig: inflow leading order}, the $\order{1/\epsb}$ terms in \region{II} are trivially satisfied, but so far $\Qi_{00}$ is unknown. Proceeding with the expansions, we find at $\order{1}$ that $\bs{\ui}_{01} = (\ui_{01},\wi_{01})$ is governed by
\refstepcounter{equation}
\[
\term{\ui}{00} \term{\ui}{01X} + \term{\ui}{00z}\term{\wi}{01} = -\term{\pri}{11X} -12, \quad \term{\ui}{00} \term{\wi}{01X} = - \term{\pri}{11z}, \quad \nabla \bcdot \term{\bs{\ui}}{01} = 0 \quad \text{in \region{II}},
\eqno{(\theequation{\mathit{a}-\mathit{c}})}\label{eq: High Rey Inner inflow bulk 2O}
\]
where $\term{\ui}{00} = 6z(1-z)$, while $\term{\Qi}{00} = (\Ui_{00}, \Wi_{00})$ is governed by
\refstepcounter{equation}
\[
\term{\Qi}{00} = - \Darcy \nabla\term{\Pri}{11}, \quad 0 = \nabla \bcdot \term{\Qi}{00} \quad \text{in \region{III}}.
\eqno{(\theequation{\mathit{a},\mathit{b}})}\label{eq: High Rey Inner inflow porous 2O}
\]
The boundary conditions we impose at this order in \region{II} are again obtained via matching. On the channel walls (via matching with \region{IIb} in Appendix~\ref{sec: Inflow app}), we have no flux, so that
\begin{align}
\label{eq: no flux High Rey inner bulk 2O}
\wi_{01} = 0 \quad \text{on } z = 0, 1 \text{ for } X^{-} < 0,
\end{align}
while in the far-field we match with \region{I}, to obtain
\refstepcounter{equation}
\[
\bs{\ui}_{01} \to \bs{0}, \quad \pri_{11} \sim -12X^{-} + \presfny_{1} \quad \text{as } X^{-} \ttni \text{ for } 0 < z < 1.
\eqno{(\theequation{\mathit{a},\mathit{b}})}\label{eq: matching High Rey inner 2O}
\]
Similarly, the boundary conditions in \region{III} are
\begin{align}
\label{eq: no flux High Rey inner porous 2O}
\Wi_{00} = 0 \quad \text{on } z = 0, 1 \text{ for } X^{-} > 0,
\end{align}
while matching with \region{IV} gives
\begin{align}
\label{eq: matching High Rey porous inner 2O}
\Pri_{11} \sim -\dfrac{X^{-}}{\Darcy} \quad \text{as } X^{-} \tti \text{ for } 0 < z < 1.
\end{align}
We note that the far-field matching condition $\Qi_{00} \sim \bs{e}_x$ as $X^{-} \tti$ follows consistently from~\eqref{eq: matching High Rey porous inner 2O} using~(\ref{eq: High Rey Inner inflow porous 2O}a).

The interfacial conditions at $X^{-} = 0$ for \eqref{eq: High Rey Inner inflow bulk 2O}--\eqref{eq: High Rey Inner inflow porous 2O} are formally derived from consideration of the inner-inner boundary layer (\region{IIa}) in \S\ref{sec: Smaller boundary layer}, within which we are able to satisfy the no-tangential-slip condition, but additionally find that the leading-order flow and the first correction to the pressure $\pri_{11}$ are unchanged. From \S\ref{sec: Smaller boundary layer}, the correct conditions to couple regions $\mathrm{II}$ and $\mathrm{III}$ are continuity of flux and pressure, with
\refstepcounter{equation}
\[
6z(1-z) = \Ui_{00}, \quad \pri_{11} = \Pri_{11} \qquad \text{on } X^{-} = 0 \text{ for } 0 < z < 1.
\eqno{(\theequation{\mathit{a},\mathit{b}})}\label{eq: cont of pres interface high Rey 2O}
\]

The system to solve in regions $\mathrm{II}$ and $\mathrm{III}$ (up to the $\order{\epsb}$ terms) is given by~\eqref{eq: High Rey Inner inflow bulk 2O}--\eqref{eq: cont of pres interface high Rey 2O}. We see that regions $\mathrm{II}$ and $\mathrm{III}$ have decoupled: the boundary condition (\ref{eq: cont of pres interface high Rey 2O}a) allows us to solve for $\Qi_{00}$ and $\Pri_{11}$ using~\eqref{eq: High Rey Inner inflow porous 2O} with \eqref{eq: no flux High Rey inner porous 2O}--\eqref{eq: matching High Rey porous inner 2O}, and then we can use the boundary condition (\ref{eq: cont of pres interface high Rey 2O}b) to determine $\bs{\ui_{01}}$ and $\pri_{11}$ using \eqref{eq: High Rey Inner inflow bulk 2O} with \eqref{eq: no flux High Rey inner bulk 2O}--\eqref{eq: matching High Rey inner 2O}. The governing equations~\eqref{eq: High Rey Inner inflow porous 2O}, with boundary conditions \eqref{eq: no flux High Rey inner porous 2O}--\eqref{eq: matching High Rey porous inner 2O} and~(\ref{eq: cont of pres interface high Rey 2O}a), are solved by
\begin{align}
\label{eq: Solution to Pri_11}
-\Darcy\term{\Pri}{11} = X^{-} + \sum_{k=1}^{\infty} A_k \exp(-2 k \pi X^{-}) \cos(2 k \pi z), \quad A_k = \dfrac{3}{\left(\pi k\right)^3}.
\end{align}
Since $\bs{\ui}_{00}$ is independent of $X^{-}$ within \region{II}, the leading-order transition from Poiseuille to plug flow occurs entirely within the porous medium in \region{III}, and is governed by~\eqref{eq: Solution to Pri_11}.

We now complete the solution in \region{II} by determining $\ui_{01}$ and $\pri_{11}$. As well as allowing us to determine $\presfny_1$, knowledge of $\pri_{11}$ gives the normal stress acting on the interface up to $\order{\eps}$, the first order at which the stress varies in the $z$-direction. A rearrangement of the exterior flow equations~\eqref{eq: High Rey Inner inflow bulk 2O} allows us to decouple the \region{II} system (where $-\infty < X^{-} < 0$ and $0 < z < 1$) into
\begin{subequations}
\label{eq: region II first correction gov eq}
\begin{alignat}{3}
\label{eq: wi_01 governing equation}
z(1-z)\nabla^2 \term{\wi}{01} +  2\term{\wi}{01} &= 0, \\
\label{eq: pri_11 governing equation}
\nabla^2 \term{\pri}{11} &= -2\term{\ui}{00z} \term{\wi}{01X^{-}}.
\end{alignat}
\end{subequations}

We solve~\eqref{eq: wi_01 governing equation} for $\wi_{01}$ first, using boundary conditions~\eqref{eq: no flux High Rey inner bulk 2O} on $z = 0$, $1$ and (\ref{eq: matching High Rey inner 2O}a) as $X^{-} \ttni$. We derive a consistent boundary condition for $\wi_{01}$ on $X^{-}=0$ using the boundary condition~(\ref{eq: cont of pres interface high Rey 2O}b) for the pressure on $X^{-} = 0$, in combination with the governing equation~(\ref{eq: High Rey Inner inflow bulk 2O}b), giving
\begin{align}
\label{eq: BC wi at R=0}
\wi_{01X^{-}} = -\dfrac{1}{\Darcy \ui_{00}} \sum_{k=1}^{\infty}2 k \pi A_k \sin(2 k \pi z) \quad \text{on } X^{-} = 0 \text{ for } 0 < z < 1.
\end{align}
The boundary condition~\eqref{eq: BC wi at R=0} ensures that $\wi_{01}$ does not vanish on $X^{-} = 0$, inducing a slip velocity on this interface within \region{II}. This issue is taken care of in an inner-inner boundary layer closer to the interface (\region{IIa} in figure~\ref{fig: inflow inner BL}) in \S\ref{sec: Smaller boundary layer}. The linearity of the governing equation~\eqref{eq: wi_01 governing equation}, with boundary conditions~\eqref{eq: no flux High Rey inner bulk 2O}, (\ref{eq: matching High Rey inner 2O}a), and \eqref{eq: BC wi at R=0}, also allows us to deduce that $\term{\wi}{01}$ scales with $\Darcy^{-1}$. Further, from the continuity equation~(\ref{eq: High Rey Inner inflow bulk 2O}c) it follows that $\ui_{01}$ also scales with $\Darcy^{-1}$.

\begin{figure}
\centering
\includegraphics[width=\textwidth]{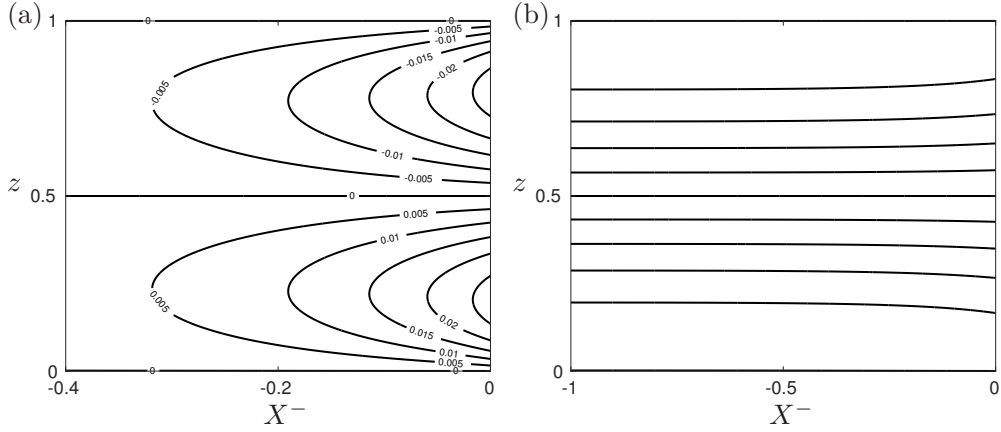}
\caption{The streamlines for inflow near the interface. (a) The streamlines for the scaled first-correction velocity $\Darcy \bs{\ui}_{01}$, which is independent of $\Darcy$. The fluid moves from the centre of the channel to the channel walls. (b) The streamlines for the full velocity up to first-correction $\bs{\ui}_{00} + \epsb \bs{\ui}_{01}$ (with a large value of $\epsb/\Darcy = 1$ taken to emphasise the effect of the correction). The fluid moves from left to right.}
\label{fig: inflow SL}
\end{figure}

We solve for $\wi_{01}$ using a standard central finite-difference scheme. By numerically integrating $\term{\wi}{01}$ with respect to $X^{-}$, we are able to plot the streamlines in figure~\ref{fig: inflow SL}. In figure~\ref{fig: inflow SL}a, we show the streamlines for the first-correction to the velocity, and in figure~\ref{fig: inflow SL}b, we show the streamlines for the full velocity up to this first-correction. We see that the fluid moves from the centre of the channel towards the channel walls, as expected. Therefore, adjustment from Poiseuille to plug flow occurs in \region{II}, but at higher order. This movement induces a horizontal slip velocity on the channel walls, which is resolved in Appendix~\ref{sec: Inflow app} by introducing a boundary layer near the channel walls in which $z = \order{\epsb^{1/3}}$ on the bottom wall, with a similar layer near the top wall. These regions are labelled as $\mathrm{IIb}$ in figure \ref{fig: inflow inner BL}. In Appendix~\ref{sec: Inflow app}, we show that the flow near $z = 0$ has the form
\begin{align}
\label{eq: Solution of flow near z = 0 full text}
\ui_0 \sim 6z(1-z) + \epsb \displaystyle\sum_{k=1}^{\infty} C_k \exp(\eigvalII X^{-}) g'_k(0) \dfrac{\int_0^{z/\epsb^{1/3}} \Ai(a_k v) \, \mathrm{d} v}{\int_0^\infty \Ai(a_k s) \, \mathrm{d} s} \quad \text{as } z \to 0^{+}, 
\end{align} 
where $C_k$, $\eigvalII$, and $g_k$ are described in~\eqref{eq: Region II velocities at epsa}--\eqref{eq: SL eig val problem}, $\Ai$ is the Airy function of the first kind, and $a_k$ is defined in~(\ref{eq: Inflow channel SF BL near wall}b). From~\eqref{eq: Solution of flow near z = 0 full text}, we see that the leading-order solution is preserved near the channel wall. In fact, the slip is resolved within higher-order terms in the boundary layer, and does not affect the leading-order solution in \region{II}.

%/home/dalwadi/Desktop/MATLAB code for project/Rotating Darcy/Flow in channel/Inflow_SL(500,500,1,-1)

\begin{figure}
\centering
\includegraphics[width=0.7\textwidth]{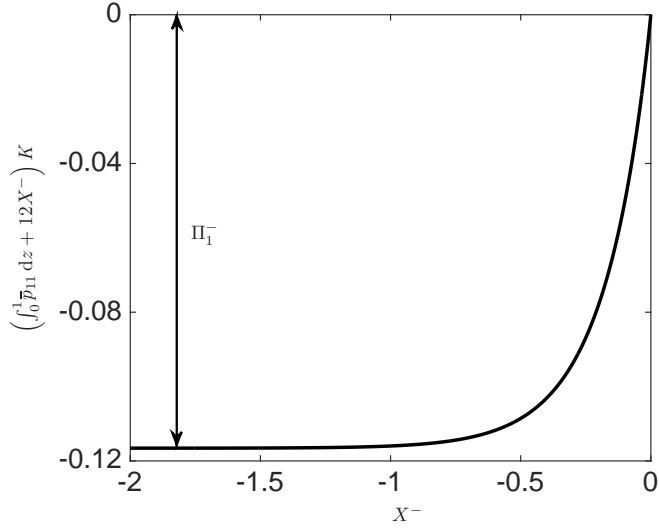}
\caption{The function $\left(\int_0^1 \! \pri_{11}\,\mathrm{d}z + 12 X^{-}\right) \Darcy$ for varying $X^{-}$ in \region{II}, which allows us to determine the outer pressure drop for inflow.}
\label{fig: pressure drop inflow}
\end{figure}

The problem for $\pri_{11}$ is given by~\eqref{eq: pri_11 governing equation} with boundary conditions (\ref{eq: matching High Rey inner 2O}b) and (\ref{eq: cont of pres interface high Rey 2O}b). We obtain consistent boundary conditions for $\pri_{11}$ on the channel walls by substituting the leading-order flow solution into the governing equation~(\ref{eq: High Rey Inner inflow bulk 2O}b) to obtain
\begin{align}
\label{eq: pri_11z on channel walls}
\term{\pri}{11z} = 0 \quad \text{on } z = 0, 1 \text{ for } X^{-} < 0.
\end{align}
We solve the full system for $\pri_{11}$, given by \eqref{eq: pri_11 governing equation} with boundary conditions~(\ref{eq: matching High Rey inner 2O}b), (\ref{eq: cont of pres interface high Rey 2O}b), and~\eqref{eq: pri_11z on channel walls} numerically using a central finite-difference scheme. Using a similar scaling argument as for $\wi_{01}$, we find that $\presfny_1$ scales with $1/\Darcy$. We are able to calculate $\presfny_1$ by plotting $\left(\pri_{11} + 12X^{-}\right) \Darcy$, and using the matching condition~(\ref{eq: matching High Rey inner 2O}b). As $\presfny_1$ is independent of $z$, and to more easily show the pressure drop on a graph, we plot $\left(\int_0^1 \! \pri_{11}\,\mathrm{d}z + 12X^{-}\right) \Darcy$ for varying $X^{-}$ in figure~\ref{fig: pressure drop inflow}. We find thereby that $\presfny_1 = \inpdrop/\Darcy$, where $\inpdrop = -0.117$ to three significant figures.

\subsubsection{Inner-inner \region{IIa}}
\label{sec: Smaller boundary layer}

We now investigate the inner-inner boundary layer closer to the interface, \region{IIa} in figure~\ref{fig: inflow inner BL}. This boundary layer is required to satisfy the no-tangential-slip condition on the interface, and captures a balance between the relevant inertial terms and the viscous terms induced by the presence of the porous interface.

The relevant scalings are $X^{-} = \epsb \Ri$ and $\wi_0 = \epsb \wii_0$, and we introduce the new variables $\ui_0 = \uii_0$ and $\pri_1= \prii_1$ to signify that we are working in \region{IIa}. Under these scalings, in \region{IIa} (where $-\infty < \Ri < 0$ and $0 < z < 1$) the exterior flow equations in figure~\ref{fig: inflow leading order} become
\begin{subequations}
\label{eq: inner inner region ref}
\begin{align}
\label{eq: inner inner region x-mom ref}
\epsb^{-1} \term{\uii}{0} \term{\uii}{0\Ri} + \epsb \term{\wii}{0} \term{\uii}{0z} &= - \term{\prii}{1\Ri} +  \epsb^{-1} \term{\uii}{0\Ri\Ri} + \epsb \term{\uii}{0zz}, \\
\label{eq: inner inner region y-mom ref}
 \term{\uii}{0} \term{\wii}{0\Ri} + \epsb^2 \term{\wii}{0} \term{\wii}{0z} &= - \epsb \term{\prii}{1z} + \term{\wii}{0\Ri\Ri} +  \epsb^2 \term{\wii}{0zz}, \\
\label{eq: inner inner region cont ref}
0 &= \term{\uii}{0\Ri} + \epsb^2 \term{\wii}{0z},
\end{align}
\end{subequations}
and the interfacial boundary conditions become
\refstepcounter{equation}
\[
\uii_0 = \Ui_0, \quad \prii_1 = \Pri_1, \quad \term{\wii}{0} = 0 \quad \text{on } \Ri = 0 \text{ for } 0 < z < 1.
\eqno{(\theequation{\mathit{a}-\mathit{c}})}\label{eq: Interfacial BC in region IIb}
\]

The purpose of this boundary layer is to determine the tangential velocity $\wii_0$, and to ensure that it is possible to satisfy the no-tangential-slip boundary condition on $X^{-} = 0$. Therefore, we form asymptotic expansions as follows:
\begin{subequations}
\label{eq: Region IIa asymptotic series}
\begin{align}
\term{\uii}{0}  &= 6z(1-z) + \order{\epsb}, \\
\term{\wii}{0} &= \term{\wii}{01} + \order{\epsb^{4/3}}, \\
\prii_1 &= \term{\pri}{11}(0,z) + \order{\epsb^{1/3}}
\end{align}
\end{subequations}
as $\epsb \ttz$, where $\term{\pri}{11}$ is known from \S\ref{sec: Regions II and III}. Substituting the asymptotic expansions \eqref{eq: Region IIa asymptotic series} into the equations \eqref{eq: inner inner region ref} and equating powers of $\epsb$, we find that~\eqref{eq: inner inner region x-mom ref} and~\eqref{eq: inner inner region cont ref} are automatically satisfied at leading order, whilst the leading-order terms in~\eqref{eq: inner inner region y-mom ref} are
\begin{align} 
\label{eq: inner inner region y-mom 1O}
6z(1-z) \term{\wii}{01\Ri} = \term{\wii}{01\Ri\Ri}.
\end{align}
Using~\eqref{eq: Solution to Pri_11}, the leading-order interfacial boundary conditions~\eqref{eq: Interfacial BC in region IIb} become
\refstepcounter{equation}
\[
6z(1-z) = -\Darcy \Pri_{11\Ri},  \quad \term{\pri}{11} = \Pri_{11}, \quad \term{\wii}{01} = 0  \quad \text{on } \Ri = 0 \text{ for } 0 < z < 1.
\eqno{(\theequation{\mathit{a}-\mathit{c}})}\label{eq: Interfacial BC in region IIb 1O}
\] 
The relevant far-field condition arises via matching with \region{II}, giving
\begin{align}
\label{eq: Matching BC in region IIb 1O}
\wii_{01} \to \term{\wi}{01}(0,z) \quad \text{as } \Ri \ttni \text{ for } 0 < z < 1,
\end{align}
where $\term{\wi}{01}$ is determined in \S\ref{sec: Regions II and III}.

The interfacial conditions (\ref{eq: Interfacial BC in region IIb 1O}a,b) are used to couple regions $\mathrm{II}$ and $\mathrm{III}$ in \S\ref{sec: Regions II and III}. The remaining terms in the system \eqref{eq: inner inner region y-mom 1O}--\eqref{eq: Matching BC in region IIb 1O} yield $\wii_{01}$, and are solved by
\begin{align}
\label{eq: region IIb soln 1O}
\wii_{01} = \term{\wi}{01}(0,z) \left(1 - \exp(6 z(1-z) \Ri) \right).
\end{align}
We see that~\eqref{eq: region IIb soln 1O} can only be a solution for inflow, as we require $6 z(1-z) \Ri < 0$ as we move back into \region{II}. For outflow, this inner-inner region is not a valid boundary layer, and we must tackle the problem in a different manner. 

We note that this boundary layer persists for alternate tangential slip conditions. We can showcase the robust nature of the boundary layer via two examples. The first example uses the general tangential-slip condition derived in \citet{carraro2015effective}. That is, instead of~(\ref{eq: Interfacial BC in region IIb 1O}c), we consider a condition of the form
\begin{align}
\wi = \BJ \ui \quad \text{on } X^{-} = 0, \text{ for } 0 < z < 1,
\end{align}
which reduces to~(\ref{eq: Interfacial BC in region IIb 1O}c) when $\BJ = 0$. Here, $\BJ$ is a slip coefficient, obtained by solving the cell problem given in \citet{carraro2015effective}. Using this condition, the transverse velocity is
\begin{align}
\wi \sim 6 \BJ z(1-z) \exp(6 z(1-z) \Ri) + \epsb \term{\wi}{01}(0,z) \left(1 - \exp(6 z(1-z) \Ri) \right) + \order{\epsb^{4/3}},
\end{align}
in \region{IIa}. In the second example, we use a Beavers and Joseph slip condition~\citep{porousBC} of the form
\begin{align}
\wi_{\Ri} = \dfrac{\hat{\BJ}}{\sqrt{\Darcy}} \left(\Wi - \wi \right) \quad \text{on } X^{-} = 0, \text{ for } 0 < z < 1,
\end{align}
where $\hat{\BJ}$ is the experimentally determined slip coefficient. In this case, the transverse velocity would be
\begin{align}
\label{eq: NO slip soln}
\wi \sim \epsb \left(\term{\wi}{01}(0,z) + \dfrac{\hat{\BJ} \Wi_0}{6 \sqrt{\Darcy} z(1-z)}  \right) + \order{\epsb^{4/3}}.
\end{align}
Hence the boundary layer structure persists if the no-slip condition is modified in either of these two ways. Physically, we can conclude that the thickness of each boundary layer involved in inflow is unchanged for any of the three boundary conditions applied at the interface. Moreover, the flow induced by each of the three boundary conditions is only different in \region{IIa}, the boundary layer closest to the interface.

\begin{figure}
\centering
\includegraphics[width=\textwidth]{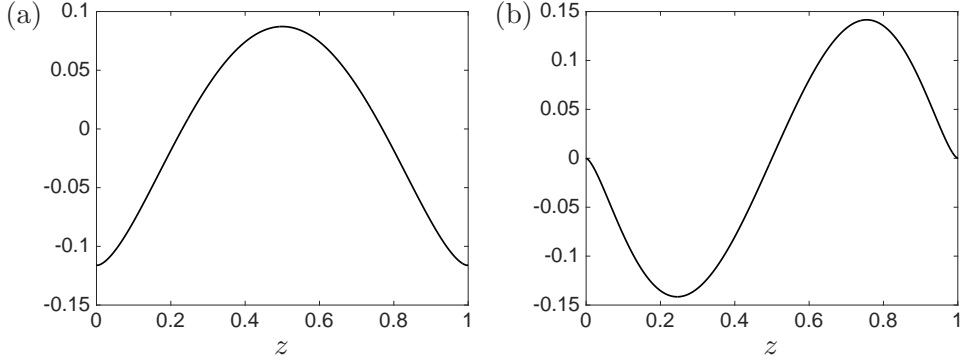}
\caption{The interfacial stress predicted by the inner flows (from \eqref{eq: Stresses for inflow}) at $\order{\eps}$. (a) The normal stress (we plot $\Darcy \pri_{11}(0,z)$) and (b) the shear stress (we plot $6z(1-z) \Darcy \term{\wi}{01}(0,z)$). As both the normal and shear stress scale with $1/\Darcy$ at this order, both plotted terms are independent of $\Darcy$.}
\label{fig: stress_high_Rey_inflow}
\end{figure}

Returning to the original no-slip boundary condition (\ref{eq: Interfacial BC in region IIb}c), it is straightforward to calculate that the components of stress on the interface at $x = -1$ are given by
\begin{subequations}
\label{eq: Stresses for inflow}
\begin{align}
\label{eq: Normal stress for inflow}
\stress_{xx} &=  -\left(12 + \eps \term{\pri}{11}(0,z) \right) + \order{\epsb^{1/3} \eps}, \\
\label{eq: Shear stress for inflow}
\stress_{xz} &= -\eps \left(6z(1-z) \term{\wi}{01}(0,z) \right) + \order{\epsb^{1/3} \eps}
\end{align}
\end{subequations}
as $\eps, \epsb \ttz$, with $\eps \ll \epsb \ll 1$. Whilst the normal stress~\eqref{eq: Normal stress for inflow} is nominally $\order{1}$, the leading-order term comes from the decision to impose $\Pres(0,0.5) = 0$ to uniquely define the pressure. Thus, the more interesting result occurs at $\order{\eps}$, which is the lowest order at which the normal stress varies in the $z$-direction. For the shear stress~\eqref{eq: Shear stress for inflow}, the leading-order terms are $\order{\eps}$, and this varies in $z$. Those terms which vary in $z$ cannot be determined solely from the outer solution: they are obtained from our boundary layer analysis and are each plotted in figure~\ref{fig: stress_high_Rey_inflow}. We show in the previous section that $\pri_{11}$ and $\wi_{01}$ both scale with $1/\Darcy$, thus the stress also scales with $1/\Darcy$ at $\order{\eps}$.

Additionally, since we have imposed $\Pres(0,z) = 0$, we have no $\order{\eps/\epsb}$ constant term from the pressure in the normal stress~\eqref{eq: Normal stress for inflow}. In three-dimensions, the tangential (that is, in the $(\bs{e}_x \times \bs{e}_z)$-direction) variation in pressure ensures that there will be an $\order{\eps/\epsb}$ term in the corresponding normal stress component no matter how we define the pressure (as we show in \S\ref{sec: Inflow 3D}).

%/Dropbox/MATLAB code for project/Rotating Darcy/Flow in channel\inflow_stress_high_Rey

We have now solved for the flow variables within the inflow boundary layers for the regions which affect the stress and the pressure drop. As shown in figure~\ref{fig: inflow inner BL}, there are further boundary layers which are important for inflow, and we discuss these in Appendix~\ref{sec: Inflow app}.

\subsubsection{Summary}

The main part of the leading-order inflow may be summarized as follows. The leading-order axial flow and pressure are unchanged until they reach the interface, after which they transition to plug flow inside the obstacle in \region{III}, whose length is comparable to the channel height. This has an effect on the correction to normal flow and pressure in \region{II}. Meanwhile, the tangential velocity satisfies the no-tangential-slip condition on the obstacle surface in \region{IIa} - a boundary layer of even smaller width, which is comparable to the length of the channel height multiplied by the reciprocal of the Reynolds number.

Combining the original asymptotic expansion~\eqref{eq: inner expansions} with the asymptotic expansion in \region{II}, namely \eqref{eq: regions II and III pressure ae}, we deduce that the pressure jump between outer inflow regions $\mathrm{I}$ and $\mathrm{IV}$ is $\order{\eps}$. Specifically, we have determined that
\begin{align}
\label{eq: Inflow pressure drop}
\presfny \sim \dfrac{\inpdrop}{\Darcy} \quad \text{as } \eps, \epsb \ttz \text{ with } \eps \ll \epsb \ll 1,
\end{align}
where $\inpdrop \approx -0.117$.

The argument presented in this section will not hold if the fluid is leaving the obstacle rather than entering it. This is because the boundary layer \region{IIa} is only valid for inflow (as can be seen from (\ref{eq: region IIb soln 1O}b)). We reconsider our approach for outflow in the next section.

\subsubsection{Outflow}
\label{sec: Outflow}

The outflow inner problem (within regions $\mathrm{V}$ and $\mathrm{VI}$) is shown in figure~\ref{fig: outflow leading order}. We again seek a solution as $\epsb \ttz$ with $\eps \ll \epsb \ll 1$ using the method of matched asymptotic expansions.

\begin{figure}
\centering
\includegraphics[width=\textwidth]{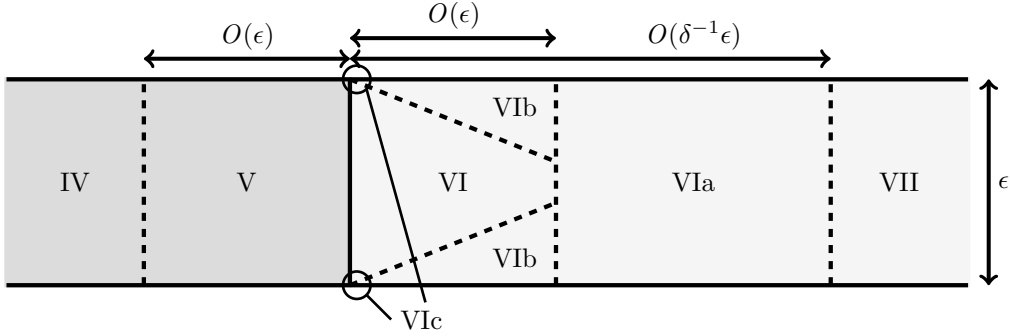}
\caption{Schematic diagram of the inner regions for outflow - $\mathrm{VIa}$, $\mathrm{VIb}$, and $\mathrm{VIc}$. The size of the boundary \region{VIb} has been exaggerated for illustrative purposes, is of height $\order{\eps \epsb^{1/2}}$, and grows proportional to the square root of the distance from the porous obstacle. Region $\mathrm{VIc}$ lies within \region{VIb} and has height and length $\order{\eps \epsb}$.}
\label{fig: outflow BL}
\end{figure}

The asymptotic structure for outflow is shown schematically in figure~\ref{fig: outflow BL}. The flow in \region{V} is unchanged at leading order and reaches the interface as uniform plug flow with magnitude $1$. There is an inviscid core in the centre of \region{VI}, flanked by Prandtl boundary layers near $z=0, 1$ whose thicknesses are of $\order{\left(\epsb X^{+}\right)^{1/2}}$. These boundary layers (\region{VIb}) grow downstream, and the transition to Poiseuille flow occurs when their thicknesses become of $\order{1}$. This occurs on a horizontal lengthscale of $\order{1/\epsb}$ in a transition region we denote \region{VIa}. There is a third boundary layer, denoted \region{VIc}, which occurs in the corner between the channel wall and the porous interface. This region deals with a singularity at the corner that arises in \region{VIb}, as in the classic Prandtl problem.

In this section, we investigate the regions $\mathrm{VI}$ and $\mathrm{VIa}$ to understand the transition from plug to Poiseuille flow and to calculate the pressure jump between the outer regions $\mathrm{IV}$ and $\mathrm{VII}$. Regions $\mathrm{VIb}$ and $\mathrm{VIc}$ are discussed in Appendix~\ref{sec: regions VIb and VIc} since they are not important to the transition. Recall that, from~\eqref{eq: Inflow pressure drop}, we calculated an $\order{\eps}$ pressure jump between outer regions~$\mathrm{I}$ and $\mathrm{IV}$ for inflow. We show that the pressure jump between outer regions $\mathrm{IV}$ and $\mathrm{VII}$ is of $\order{\eps / \epsb}$ for outflow.

We start by investigating how the fluid moves through \region{VI}. The leading-order behaviour is plug flow, and we determine the effect of the interfacial no-tangential-slip condition on the correction to the plug flow. In particular, we determine the far-field behaviour of the flow, which allows us to impose the correct matching conditions for the problem of transition to Poiseuille flow in \region{VIa}.

\subsubsection{Inner regions $\mathrm{V}$ and $\mathrm{VI}$}
\label{sec: Region VI}

To determine the outflow pressure jump~(\ref{eq: asymptotic series for pres jump}b), we must consider first-order flow terms. Therefore, we pose the asymptotic expansions
\refstepcounter{equation} 
\[
\term{\bs{\ui}}{0} \sim \bs{e}_x + \epsb^{1/2} \term{\bs{\ui}}{0a}, \quad
\term{\bs{\Qi}}{0} \sim \bs{e}_x + \epsb^{1/2} \term{\bs{\Qi}}{0a}, \quad
\term{\pri}{1} \sim \epsb^{-1/2} \term{\pri}{1a}, \quad
\term{\Pri}{1} \sim -\dfrac{X^{+}}{\Darcy} + \epsb^{1/2} \term{\Pri}{1b},
\eqno{(\theequation{\mathit{a}-\mathit{d}})}\label{eq: region VI asymptotic expansions}
\]
as $\epsb \ttz$, where the non-integer powers of $\epsb$ arise due to the correction terms from the boundary layers in \region{VIb} near the channel walls. We determine the leading-order behaviour within the coupled regions $\mathrm{V}$ and $\mathrm{VI}$, but do not fully solve for the higher-order terms due to their complexity. Instead, we analyse the higher-order problems to obtain the information required for the matching condition with the transition \region{VIa}.

Substituting the asymptotic expansions \eqref{eq: region VI asymptotic expansions} into the system outlined in figure~\ref{fig: outflow leading order}, the leading-order equations are trivially satisfied. The $\order{\epsb^{1/2}}$ exterior flow equations are
\refstepcounter{equation}
\[
\term{\ui}{0aX^{+}} = -\term{\pri}{1 a X^{+}}, \quad \term{\wi}{0aX^{+}} = -\term{\pri}{1 a z}, \quad 0 = \term{\ui}{0aX^{+}} + \term{\wi}{0az} \quad \text{in region $\mathrm{VI}$},
\eqno{(\theequation{\mathit{a}-\mathit{c}})}\label{eq: Outflow inner regions governing equations 2O}
\]
and the $\order{\epsb^{1/2}}$ interior flow equations are
\refstepcounter{equation}
\[
\term{\Qi}{0a} = -\Darcy \nabla \term{\Pri}{1b}, \quad 0 =  \nabla \bcdot \term{\Qi}{0a} \quad \text{in region $\mathrm{V}$}.
\eqno{(\theequation{\mathit{a},\mathit{b}})}\label{eq:  porous flow inner outflow 2O}
\]
In \region{V}, the first-correction to the boundary condition on the channel walls is given by
\begin{align}
\label{eq: no flux or slip BC Darcy inner outflow 2O}
\Wi_{0a} &= 0 \quad \text{on } z = 0, 1 \, \text{ for } X^{+} < 0,
\end{align}
while the far-field conditions are obtained by matching with \region{IV}, to yield
\refstepcounter{equation}
\[
\term{\Pri}{1b} \ttz, \quad \term{\Qi}{0a} \to \bs{0} \quad \text{ as } X^{+} \ttni \, \text{ for } 0 < z < 1.
\eqno{(\theequation{\mathit{a},\mathit{b}})}\label{eq:  region IV to V matching conditions 2O}
\]
In \region{VI}, the boundary conditions on the channel walls are obtained by matching with \region{VIb}. We show in Appendix~\ref{sec: regions VIb and VIc} that, at this order, the flow problem within \region{VIb} is equivalent to the standard Prandtl problem of uniform flow past a flat plate~\citep{prandtl1904uber}. The matching conditions are therefore given by
\begin{subequations}
\label{eq: no flux or slip BC inner outflow 2O}
\begin{alignat}{6}
\label{eq: no flux or slip BC NS inner outflow 2O z=0}
\term{\wi}{0a} &= \leak \left(4 X^{+}\right)^{-1/2} &\quad &\text{on }  z = 0 \, \text{ for } X^{+}> 0, \\
\label{eq: no flux or slip BC NS inner outflow 2O z=1}
\term{\wi}{0a} &= -\leak \left(4 X^{+}\right)^{-1/2} &\quad &\text{on }  z = 1 \, \text{ for } X^{+}> 0,
\end{alignat}
\end{subequations}
where $\leak \, (\approx 1.721)$ is a constant that is determined numerically from a nonlinear ordinary differential equation (discussed further in Appendix~\ref{sec: regions VIb and VIc}); the parameter $\beta$ used in~\citet{van1970entry} and~\citet{wilson1971entry} is related to $\leak$ via $\leak = 2^{1/2} \beta$. Finally, the interfacial conditions on $X^{+} = 0$ are
\refstepcounter{equation}
\[
\ui_{0a} = \Ui_{0a}, \quad \pri_{1a} = 0, \quad \wi_{0a} = 0 \quad \text{on } X^{+} = 0 \text{ for } 0 < z < 1.
\eqno{(\theequation{\mathit{a}-\mathit{c}})}\label{eq: BC Cont of flux outflow inner 2O}
\]

We do not solve the coupled system of equations~\eqref{eq:  porous flow inner outflow 2O}--\eqref{eq: BC Cont of flux outflow inner 2O} here, though we note that it is possible to obtain an implicit representation for $\wi_{0a}$ via a Fourier transformation. Instead, we acquire the necessary information for the matching with \region{VIa}, which allows us to determine $\presfnz_0$ in the next section. We note that~\eqref{eq: Outflow inner regions governing equations 2O} can be rearranged to deduce the following three results: firstly, that
\begin{align}
\label{eq: the correction to w in the outflow inner region}
\nabla^2 \term{\wi}{0a} = 0 \quad \text{in region $\mathrm{VI}$};
\end{align}
secondly, that $\pri_{1a}$ is the harmonic conjugate of $\term{\wi}{0a}$; and, thirdly, that $\pri_{1a} + \ui_{1a}$ is independent of $X^{+}$. We deduce that the leading-order far-field behaviour of $\wi_{0a}$ by considering the governing equation~\eqref{eq: the correction to w in the outflow inner region} together with the channel wall boundary conditions~\eqref{eq: no flux or slip BC NS inner outflow 2O z=0} and~\eqref{eq: no flux or slip BC NS inner outflow 2O z=1} as $X^{+} \tti$, to obtain the expansion
\begin{align}
\label{eq: w outflow inner region matching condition}
\term{\wi}{0a} \sim \dfrac{\leak(1 - 2z)}{\left(4 X^{+}\right)^{1/2}} \quad \text{as } X^{+} \tti \text{ for } 0 < z < 1.
\end{align}
Using the far-field condition~\eqref{eq: w outflow inner region matching condition} in the governing equations~(\ref{eq: Outflow inner regions governing equations 2O}a,c), we further deduce that
\refstepcounter{equation}
\[
\ui_{0a} \sim \leak \left(4 X^{+} \right)^{1/2}, \quad \term{\pri}{1a} \sim -\leak \left(4 X^{+} \right)^{1/2} \quad \text{as } X^{+} \tti \text{ for } 0 < z < 1.
\eqno{(\theequation{\mathit{a},\mathit{b}})}\label{eq:  region VI to VII matching conditions 2O}
\]

We note that the far-field expansions~\eqref{eq: w outflow inner region matching condition}--\eqref{eq:  region VI to VII matching conditions 2O} have no dependence on the permeability $\Darcy$. As these far-field expansions are used to form matching conditions, which transfer information into the transition \region{VIa} (which is where we calculate the pressure drop $\presfnz_0$), we note that $\presfnz_0$ will be independent of $\Darcy$. We carry out this matching in Appendix~\ref{sec: Small X analysis in regionVIa}.

The components of stress on the interface at $x = 1$ are given by
\begin{subequations}
\begin{align}
\label{eq: Normal stress outflow 2d}
\stress_{xx} &= -\left(12 - \dfrac{2}{\Darcy} + \eps \epsb^{-1/2} \term{\pri}{1 a} \right) + \mathit{o}(\eps \epsb^{-1/2}), \\
\label{eq: Shear stress outflow 2d}
\stress_{xz} &= \eps \epsb^{1/2} \left(\term{\ui}{0az} + \term{\wi}{0aX^{+}} \right) + \mathit{o}(\eps \epsb^{1/2})
\end{align}
\end{subequations}
as $\eps, \epsb \ttz$ with $\eps \ll \epsb \ll 1$, where we have not explicitly calculated $\term{\pri}{1 a}$, $\term{\ui}{0a}$, or $\term{\wi}{0a}$. In terms of the first order at which the stress varies in $z$, we find that the normal stress~\eqref{eq: Normal stress outflow 2d} for outflow is a factor of $\order{\epsb^{-1/2}}$ larger than for inflow~\eqref{eq: Normal stress for inflow}, while the shear stress~\eqref{eq: Shear stress outflow 2d} for outflow is a factor of $\order{\epsb^{1/2}}$ smaller than for inflow~\eqref{eq: Shear stress for inflow}.

\subsubsection{Transition \region{VIa}}
\label{sec: Region VIa}
We now consider the region in which the flow transitions from plug to Poiseuille flow. This occurs when the growing Prandtl boundary layers with thickness of $\order{(\epsb X^{+})^{1/2}}$ become of $\order{1}$, that is, when $X^{+} = \order{1/\epsb}$, bringing about a balance in the exterior flow equations between fluid momentum and viscosity across the full channel width. Recall that we have called this \region{VIa}, as illustrated in figure~\ref{fig: outflow BL}.

The relevant scalings are $X^{+} = \epsb^{-1} \Rii$ and $\wi_0 = \epsb \wiii_0$, and we introduce the new variables $\ui_0 = \uiii_0$ and $\pri_1= \priii_1$ to signify that we are working in \region{VIa}. Then, in \region{VIa} (where $0 < \Rii < \infty$ and $0 < z < 1$) the exterior flow equations given in figure~\ref{fig: outflow leading order} become
\begin{subequations}
\label{eq: inner inner inner large Reya}
\begin{align}
		\label{eq: inner inner inner x-mom in large Reya}
		\term{\uiii}{0} \term{\uiii}{0\Rii} + \term{\wiii}{0} \term{\uiii}{0z} &= - \epsb \term{\priii}{1\Rii} + \epsb^2 \term{\uiii}{0\Rii \Rii} + \term{\uiii}{0zz}, \\
\label{eq: inner inner  inner z-mom in large Reya}
		\epsb^2\left(\term{\uiii}{0} \term{\uiii}{0\Rii} + \term{\wiii}{0} \term{\wiii}{0z}\right) &= - \epsb \priii_{1z} + \epsb^4 \term{\wiii}{0\Rii \Rii} + \epsb^2\term{\wiii}{0zz}, \\
		\label{eq: inner inner inner cont in large Reya}
\term{\uiii}{0\Rii} + \term{\wiii}{0z} &= 0.
\end{align}
\end{subequations}

We pose the asymptotic expansions
\refstepcounter{equation}
\[
\uiii_0 = \term{\uiii}{00} + \order{\epsb^{1/2}}, \quad \wiii_0 = \term{\wiii}{01} + \order{\epsb^{1/2}}, \quad \priii_1 = \epsb^{-1} \term{\priii}{10} + \order{\epsb^{-1/2}},
\eqno{(\theequation{\mathit{a}-\mathit{c}})}\label{eq: region VIa asymptotic expansions}
\]
as $\epsb \ttz$, and substitute them into the governing equations \eqref{eq: inner inner inner large Reya} to yield the following leading-order governing equations in \region{VIa}:
\refstepcounter{equation}
\[
\term{\uiii}{00} \term{\uiii}{00\Rii} + \term{\wiii}{01} \term{\uiii}{00z} = - \term{\priii}{10\Rii} + \term{\uiii}{00 zz}, \quad 0 = - \term{\priii}{10z}, \quad \term{\uiii}{00 \Rii} + \term{\wiii}{01 z} = 0.
\eqno{(\theequation{\mathit{a}-\mathit{c}})}\label{eq: inner inner inner x-mom in large Reya leading order}
\]
The boundary conditions on the channel walls become
\begin{align}
\label{eq: no slip for region VIb}
\bs{\ui}_{00} = \bs{0} \quad \text{on } z = 0, 1 \text{ for } \Rii> 0.
\end{align}

The leading-order matching conditions are derived in Appendix~\ref{sec: Small X analysis in regionVIa}. They are obtained by forming composite expansions between regions $\mathrm{VI}$ and $\mathrm{VIb}$, then using Van Dyke's matching rule~\citep{van1975perturbation}. We use the multiplicative composite expansions~\citep{van1975perturbation} to ensure that the matching condition for the velocity satisfies the no-slip condition on the channel walls. This allows a greater accuracy when we solve numerically the resulting system. The matching conditions are given by
\begin{subequations}
\label{eq: Composite exp for vel near region VIa entrance text}
\begin{alignat}{5}
\label{eq: Composite exp for u text}
\term{\uiii}{00} &\sim \left(1 + 2 \leak \Rii^{1/2}\right)f'\left(\etaa\right)f'\left(\etab\right), \\
\label{eq: Composite exp for w text}
\term{\wiii}{01} &\sim \dfrac{\left(1 - 2z \right)}{2 \leak \Rii^{1/2}}\left(\etaa f'\left(\etaa\right)- f\left(\etaa\right)\right)\left(f\left(\etab\right) - \etab f'\left(\etab\right)\right), \\
\label{eq: Pressure for region VIa entrance text}
\priii_{10} &\sim -2 \leak \Rii^{1/2},
\end{alignat}
\end{subequations}
as $\Rii \downarrow 0$, where $f$ is the standard Blasius similarity solution (defined in~\eqref{eq: Blasius equation}), $\etaa = z/\Rii^{1/2}$ and $\etab = (1-z)/\Rii^{1/2}$. The matching conditions~\eqref{eq: Composite exp for vel near region VIa entrance text} are consistent with a small $\Rii$ expansion (for fixed $z$) of the governing equations~\eqref{eq: inner inner inner x-mom in large Reya leading order}. The matching conditions between regions $\mathrm{VIa}$ and $\mathrm{VII}$ are given by
\refstepcounter{equation}
\[
\term{\uiii}{00} \sim 6 z(1-z), \quad \term{\wiii}{01} \ttz, \quad \term{\priii}{10} \sim -12 \Rii + \presfnz_0 \quad \text{as } \Rii \tti \text{ for } 0 < z < 1,
\eqno{(\theequation{\mathit{a}-\mathit{c}})}\label{eq: Matching condition for pressure from region VII to I}
\]
where $ \presfnz_0$ is determined from the solution in \region{VIa}. The full system for the flow in \region{VIa} is given by~\eqref{eq: inner inner inner x-mom in large Reya leading order}--\eqref{eq: Matching condition for pressure from region VII to I}.

In~\citet{bodoia1961finite}, this same transition region was considered, though was not derived formally through asymptotic methods. The authors used the boundary conditions
\refstepcounter{equation}
\[
\uiii_{00} = 1, \quad \wiii_{01} =  0, \quad \priii_{10} = 0 \quad \text{on } \Rii = 0 \text{ for } 0 < z < 1,
\eqno{(\theequation{\mathit{a}-\mathit{c}})}\label{eq: uniform flow boundary condition for region VII before scaling}
\]
instead of the matching conditions~\eqref{eq: Composite exp for vel near region VIa entrance text}. We note that, unlike~\eqref{eq: Composite exp for w text}, the condition (\ref{eq: uniform flow boundary condition for region VII before scaling}b) is not consistent with the small $\Rii$ approximation to the governing equations~\eqref{eq: inner inner inner x-mom in large Reya leading order}, from which it can be deduced that $\wiii_{01}$ is singular at the corners at $\Rii = 0$, $z = 0, 1$. With the boundary conditions~\eqref{eq: uniform flow boundary condition for region VII before scaling}, it was found in~\citet{bodoia1961finite} that the pressure jump $\presfnz_0 \approx -0.338$.

\begin{figure}
\centering
\includegraphics[width=0.8\textwidth]{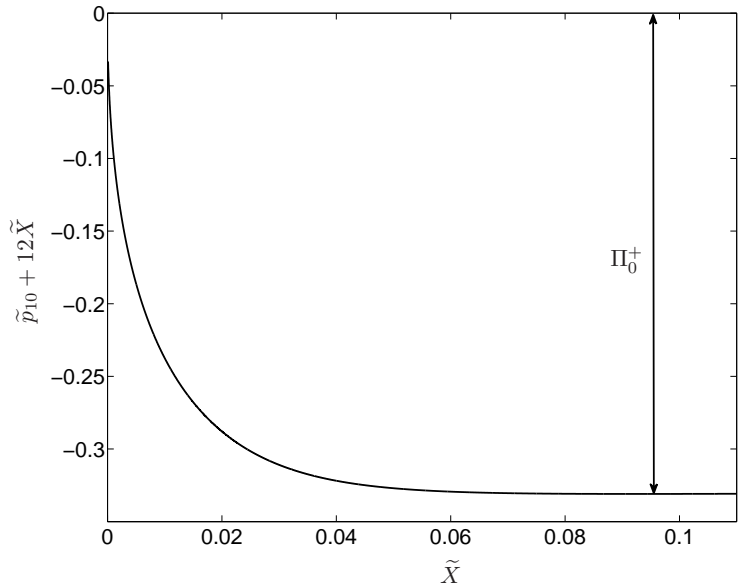}
\caption{Calculating the outflow pressure drop along the channel by plotting $\priii_{10} + 12 \Rii$. See text for further details.}
\label{fig: Pressure difference}
\end{figure}

We now calculate $\presfnz_0$ numerically using the formal matching conditions~\eqref{eq: Composite exp for vel near region VIa entrance text}. We start the numerical simulation at $\Rii = 10^{-4}$ to avoid the inverse square-root singularities in $\wiii_{01}$ and in $\priii_{10\Rii}$, and use the finite-difference scheme outlined in~\citet{bodoia1961finite}. In figure~\ref{fig: Pressure difference}, we show the numerical solutions for $\priii_{10} + 12\Rii$ as a function of $\Rii$, and we emphasise that $\priii_{10}$ is independent of $z$. From the matching condition~(\ref{eq: Matching condition for pressure from region VII to I}c), we determine that the value of the leading-order outflow pressure jump between outer regions $\mathrm{IV}$ and $\mathrm{VII}$ is given by $\presfnz_0 = -0.331$ to 3 significant figures. We note that the value of $\presfnz$ found by~\citet{bodoia1961finite} using their inconsistent boundary conditions~\eqref{eq: uniform flow boundary condition for region VII before scaling} only differs by about $2\%$ from the value of $\presfnz$ obtained using formal matching conditions.

Combining the original asymptotic expansion~\eqref{eq: inner expansions} with the asymptotic expansion in \region{VIa}~(\ref{eq: region VIa asymptotic expansions}c), we deduce that the pressure jump between outer outflow regions $\mathrm{IV}$ and $\mathrm{VII}$ is $\order{\eps/ \epsb }$. Specifically, we have determined that
\begin{align}
\label{eq: Outflow pressure drop}
\presfnz \sim \dfrac{\presfnz_0}{\epsb} \quad \text{as } \eps, \epsb \ttz \text{ with } \eps \ll \epsb \ll 1,
\end{align}
where $\presfnz_0 \approx -0.331$.

\subsection{Summary}

We have thus far determined the boundary-layer structure for two-dimensional flow through a channel with stationary walls containing a porous obstacle. We formulated the full problem for an $\order{1}$ Reynolds number, then considered the asymptotic sub-limits in which $\eps \ll \Rey \ll 1$, and $1 \ll \Rey \ll 1/\eps$. In the first sub-limit, the inflow and outflow boundary-layer structure was found to be the same, whereas in the latter sub-limit, a rich boundary-layer structure was unveiled, which differed for inflow and outflow. We determined how flow transitions from Poiseuille to plug for inflow and from plug back to Poiseuille for outflow, which allowed us to determine the leading-order asymptotic values of the pressure jumps $\presfny$ and $\presfnz$ across regions $\mathrm{II}$/$\mathrm{III}$ and regions $\mathrm{V}$/$\mathrm{VI}$, respectively. For inflow, where the flow problems in regions $\mathrm{II}$ and $\mathrm{III}$ decouple, we calculated the stress acting on the interface up to the order at which there is a variation in $z$. For outflow, where the relevant flow problems in regions $\mathrm{V}$ and $\mathrm{VI}$ are coupled, we determined the scalings of the stress acting on the interface up to the order at which there is a variation in $z$.

In the next section, we generalize the problem and consider unsteady three-dimensional flow in a Hele-Shaw cell past a tight-fitting, highly permeable cylindrical porous obstacle, with a smooth boundary. As the fluid can now travel \emph{around} the porous obstacle, the high-Reynolds-number limit has an entrainment effect for outflow. We show that this entrainment can lead to a net force acting on the obstacle, even when the unidirectional far-field forcing is periodic with a zero mean. 

\section{Unsteady three-dimensional flow}
\label{sec: 3D flow}

\begin{figure}
\centering
\includegraphics[width=0.8\textwidth]{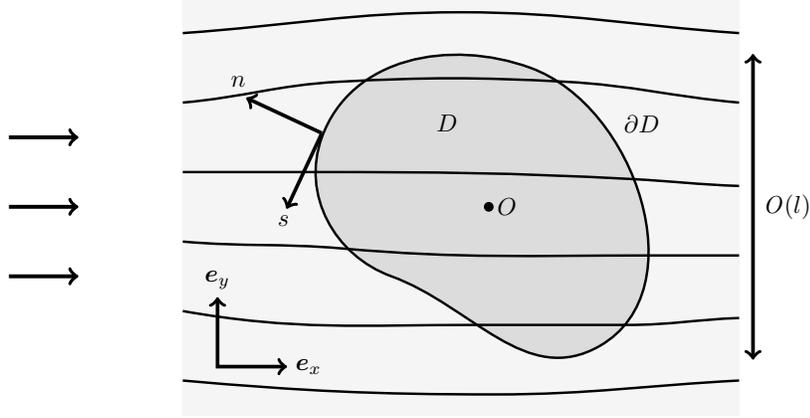}
\caption{Plan view of streamlines through and around a general porous obstacle model with a uniform far-field flow. The $z$-axis points out of the page. The Hele-Shaw cell has a depth of length $\hgt$ in the $z$-direction (not shown). We show the unit vectors $\bs{e}_x$ and $\bs{e}_y$ in the $x$- and $y$-directions respectively (for the Cartesian coordinate system), and the curvilinear $(\nrmouter,\tng)$-coordinate system for a given point on the boundary of the obstacle.}
\label{fig: bio general 3D}
\end{figure}

A schematic of the three-dimensional problem is illustrated in figure~\ref{fig: schematic set up model}b, and a two-dimensional plan view is illustrated in figure~\ref{fig: bio general 3D}. We initially work in a Cartesian coordinate system $(x,y,z)$ with origin at the obstacle centre of mass projected onto the planar plate of the Hele-Shaw cell in the plane $z=0$, the other plate lying in the plane $z = \hgt$. The obstacle axis is normal to the two parallel flat sides of the Hele-Shaw cell, and is parallel to the $z$-axis. The typical cross-sectional extent of the obstacle is $\leng$ and the Hele-Shaw cell height is $\hgt$, and we shall also assume that $\eps = \hgt/\leng \ll 1$.

The exterior and interior fluid velocities are denoted by $\uut = \ut \bs{e}_x + \vt \bs{e}_y + \wt \bs{e}_z$ and $\QQt = \Ut \bs{e}_x + \Vt \bs{e}_y + \Wt \bs{e}_z$, respectively. The exterior and interior fluid pressures are denoted by $\prest$ and $\Prest$, respectively. We use the superscript $\tng$ to denote variables in the three-dimensional problem, and we later reduce the three-dimensional problem by relating these variables to their two-dimensional counterparts in the three-dimensional analogue of the inner regions considered in  \S\ref{sec: General asymptotic structure}--\ref{sec: Rey gg 1}. The typical magnitude of the unidirectional far-field flow is $\Uinf$. We nondimensionalize by scaling the variables as follows: $(x, y) \sim \leng$, $z \sim \eps \leng$, $t \sim \leng/\Uinf$ (corresponding to an $\order{1}$ Strouhal number), $(\prest,\Prest) \sim \viscosity \Uinf/(\eps\hgt)$, $(\ut, \vt, \Ut, \Vt) \sim \Uinf$, and $(\wt, \Wt) \sim \eps \Uinf$. With these scalings, the dimensionless Navier--Stokes equations, which hold inside the Hele-Shaw cell but outside the porous obstacle, are given by
\begin{subequations}
\label{eq: 3D bulk flow BL}
\begin{align}
\eps \Reyo \left(\ut_t + \uut \bcdot \bnabla  \ut \right) &= -\prest_x + \eps^2 \nabla^2_{\perp} \ut + \ut_{zz}, \\
\eps \Reyo  \left(\vt_t + \uut \bcdot \bnabla  \vt \right) &= -\prest_y + \eps^2 \nabla^2_{\perp} \vt + \vt_{zz}, \\
\eps^3 \Reyo  \left(\wt_t + \uut \bcdot \bnabla  \wt \right) &= -\prest_z + \eps^4 \nabla^2_{\perp} \wt + \eps^2 \wt_{zz}, \\
\label{eq: 3D bulk flow cont BL}
0 &= \nabla \bcdot \uut,
\end{align}
\end{subequations}
where $\Reyo = \fluiddensity \hgt \Uinf /\viscosity$ is the global Reynolds number, $\bnabla$ is the three-dimensional gradient operator, and $\nabla^2_{\perp} = \p_{xx} + \p_{yy}$ is the two-dimensional Laplacian operator. The dimensionless Darcy equations, which hold inside the porous obstacle, are given by
\refstepcounter{equation}
\[
\Ut = -\Darcy \Prest_x, \quad \Vt = -\Darcy \Prest_y, \quad \eps^2 \Wt = -\Darcy \Prest_z, \quad 0 = \nabla \bcdot \QQt,
\eqno{(\theequation{\mathit{a}-\mathit{d}})}\label{eq: 3D porous flow BL}
\]
where $\Darcy = \permeability/\hgt^2 \ll 1$, as before.

\subsection{Boundary conditions}
The boundary conditions on the plates of the Hele-Shaw cell at $z = 0$, $1$ are those of no slip and no flux for the exterior flow and of no flux for the interior flow, so that
\begin{subequations}
\label{eq: no flux or slip BC 3D BL}
\begin{alignat}{5}
\label{eq: no flux or slip BC NS 3D BL}
\uut &= \bs{0} &\quad &\text{on } z = 0, 1 &\quad &\text{outside the obstacle}, \\
\label{eq: no flux or slip BC Darcy 3D BL}
\Wt &= 0 &\quad &\text{on } z = 0, 1 &\quad &\text{inside the obstacle}.
\end{alignat}
\end{subequations}

On the interface, we impose three-dimensional equivalents of the two-dimensional interfacial conditions given by~\eqref{eq: BC Cont of flux inflow}. These are continuity of normal flux, continuity of pressure, and a no-tangential-slip condition, as follows
\begin{subequations}
\label{eq: BC Interface 3D BL}
\begin{alignat}{5}
\label{eq: BC Interface 3D flux}
\uut \bcdot \bs{n} &= \QQt \bcdot \bs{n} &\quad &\text{on the interface}, \\
\prest &= \Prest &\quad &\text{on the interface}, \\
\uut - \left(\uut \bcdot \bs{n}\right)\bs{n} &= \bs{0} &\quad &\text{on the interface}.
\end{alignat}
\end{subequations}
where $\bs{n}$ is the unit normal pointing out of the porous obstacle.

The far-field condition is
\begin{align}
\label{eq: Far-field}
\prest \sim -12x G(t) \quad \text{as } x^2 + y^2 \to \infty,
\end{align}
where $G(t)$ is a dimensionless function of time. We take $|G(t)|$ and $|G'(t)|$ to be of $\order{1}$, so that the far-field forcing does not have a large velocity or acceleration.

\subsection{Asymptotic structure}

The dimensionless problem~\eqref{eq: 3D bulk flow BL}--\eqref{eq: Far-field} is characterised by two lengthscales: the typical cross-sectional extent of the obstacle, $1$, and the Hele-Shaw cell height, $\eps \ll 1$. We again seek a solution using the method of matched asymptotic expansions in terms of the small parameter $\eps$. We note that, in contrast to \S\ref{sec: 2D Flow}, not all of the fluid has to pass \emph{through} the porous obstacle as it is now able to flow \emph{around} the porous obstacle. In this section, we show that this difference results in a Darcy velocity of $\order{\Darcy}$ with an $\order{1}$ pressure drop, in comparison to the two-dimensional case where the Darcy velocity was of $\order{1}$ with an $\order{1/\Darcy}$ pressure drop. As the continuity of flux condition \eqref{eq: BC Interface 3D flux} ensures that $\uut \bcdot \bs{n} = \order{\Darcy}$ near the interface, we are able to deduce that $\Darcy = \order{\eps}$ corresponds to an obstacle which is impermeable at leading order in $\eps$. As discussed previously, we wish to investigate the case where the obstacle is as permeable as possible, and we therefore consider the limit where $\eps \ll \Darcy \ll 1$. The local Reynolds number (defined later) characterises the flow close to the interface. We initially consider the distinguished limit whereby the local Reynolds number is of $\order{1}$, then consider the asymptotic sub-limits of a large and small local Reynolds number in more detail, as before. 

The outer problems are characterised by an $\order{1}$ lengthscale in the $(x,y)$-plane (with the Hele-Shaw cell height, $\eps$, being the lengthscale in the $z$-direction), where we have at leading order Poiseuille and plug flow in the exterior and interior regions, respectively. We assume that the obstacle is a cylinder whose cross-section has a curvature of $\order{1}$, so that the surface near a point on the obstacle boundary is well-approximated by the tangent plane at that point. The inner regions are then characterised by an $\order{\eps}$ extension of each local normal plane to the interface and occur on either side of the interface, so that the inner problems are similar to those considered in \S\ref{sec: 2D Flow}, and the boundary layer structure is its three-dimensional equivalent.

\subsection{Outer regions}

In the outer regions, we take asymptotic expansions in the limit as $\eps \ttz$, of the form
\begin{align}
f = f_0 + \eps f_1 + \order{\eps^2},
\end{align}
for $f \in \{\ut, \vt, \wt, \Ut, \Vt, \Wt, \prest, \Prest \}$. At leading-order we find that, exterior to the obstacle,
\refstepcounter{equation}
\[
\ut_0 = g_1(z) \prest_{0x},\quad \vt_0 = g_1(z) \prest_{0y}, \quad \wt_0 = 0, \quad \bnabla \bcdot \uut_0 = 0,
\eqno{(\theequation{\mathit{a}-\mathit{d}})}\label{eq: u_0 in outer 3d variables soln}
\]
where $g_1(z) = z(z-1)/2$, while interior to the obstacle,
\refstepcounter{equation}
\[ 
\Ut_0 = -\Darcy \Prest_{0x}, \quad \Vt_0 = -\Darcy \Prest_{0y}, \quad \Wt_0 = 0, \quad \bnabla \bcdot \QQt_0 = 0,
\eqno{(\theequation{\mathit{a}-\mathit{d}})}\label{eq: U_0 in outer 3d variables soln}
\]
so that both the interior and exterior pressure satisfy Laplace's equation in two dimensions. We define the two-dimensional cross-section of the obstacle as $\obs$ and the one-dimensional boundary of this as $\bdy$. Thus, the governing equations are
\refstepcounter{equation}
\[
\nabla_\perp^2 \prest_{0} = 0 \quad \text{outside } \obs, \qquad \nabla_\perp^2 \Prest_{0} = 0 \quad \text{in } \obs.
\eqno{(\theequation{\mathit{a},\mathit{b}})}\label{eq: pressure in outer 3d variables soln}
\]
The far-field condition is
\begin{align}
\prest_0 \sim -12x G(t) \quad \text{as } x^2 + y^2 \to \infty.
\end{align}
In \S\ref{sec: 3D inner regions}, we find that the leading-order coupling conditions on the interface are
\refstepcounter{equation}
\[
\prest_0 = \Prest_0, \quad \pbyp{\prest_0}{\nrmouter} = 12 \Darcy \pbyp{\Prest_{0}}{\nrmouter} \quad \text{on } \bdy,
\eqno{(\theequation{\mathit{a},\mathit{b}})}\label{eq: pressure BC in outer 3d variables}
\]
where (\ref{eq: pressure BC in outer 3d variables}b) is equivalent to the continuity of total flux condition given in~(\ref{eq: BC outer interfacial}a). The leading-order problem is defined by \eqref{eq: pressure in outer 3d variables soln}--\eqref{eq: pressure BC in outer 3d variables}, and can be solved for a given obstacle boundary $\bdy$. For future reference, we deduce from \eqref{eq: U_0 in outer 3d variables soln} that the segments of the boundary where we have inflow/outflow at leading order occur when $\partial \Prest_0/ \partial n$ evaluated on the boundary is positive/negative.

For the $\order{\eps}$ correction terms we find that, outside $\obs$,
\begin{subequations}
\label{eq: solution for first corr outer 3D}
\begin{align}
\label{eq: solution for u_1 and v_1}
\ut_1 &= \pbyp{}{x} \left(\prest_{1} g_1(z) + \Reyo \left( \prest_{0 t} \, g_2(z) + \left|\nabla \prest_0\right|^2 g_3(z)\right)\right), \\
\vt_1 &= \pbyp{}{y} \left(\prest_{1} g_1(z) + \Reyo \left( \prest_{0 t} \, g_2(z) + \left|\nabla \prest_0\right|^2 g_3(z) \right) \right), \\
\label{eq: solution for w_1}
\wt_1 &= -\nabla_\perp^2\left(\prest_1 \int_0^z \! g_1(\zz) \, \mathrm{d}\zz + \Reyo \left|\nabla \prest_0\right|^2 \int_0^z \! g_3(\zz) \, \mathrm{d}\zz \right), \\
\bnabla \bcdot \uut_1 &= 0,
\end{align}
where
\begin{align}
g_2(z) &= \dfrac{g_1(z)}{12}\left(z^2 - z + 1 \right), \\
\label{eq: definition of g(z)}
g_3(z) &= \dfrac{g_1(z)}{240}\left(2z^4 - 4z^3 + z^2 + z + 1 \right),
\end{align}
\end{subequations}
while in the obstacle,
\refstepcounter{equation}
\[
\Ut_1 = -\Darcy \Prest_{1x}, \quad \Vt_1 = -\Darcy \Prest_{1y}, \quad \Wt_1 = 0, \quad \bnabla \bcdot \QQt_1 = 0 \quad \text{in } \obs,
\eqno{(\theequation{\mathit{a}-\mathit{d}})}\label{eq: U_1 in outer 3d variables soln}
\]
so that the pressures satisfy
\refstepcounter{equation}
\[
\nabla_\perp^2 \left(\prest_1 + \dfrac{3\Reyo}{560} \left|\nabla \prest_0\right|^2\right) = 0 \quad \text{outside } \obs, \qquad \nabla_\perp^2 \Prest_{1} = 0 \quad \text{in } \obs.
\eqno{(\theequation{\mathit{a},\mathit{b}})}\label{eq: Governing equation for p_1}
\]
The time derivative of $\nabla^2 \prest_0$ does not appear in (\ref{eq: Governing equation for p_1}a) due to the governing equation (\ref{eq: pressure in outer 3d variables soln}a). The far-field condition is
\begin{align}
\prest_1 \to 0 \quad \text{as } x^2 + y^2 \to \infty.
\end{align}
The first-order coupling conditions on the interface are determined in \S\ref{sec: 3D inner regions} and Appendix~\ref{sec: Higher-order average flux}, and are given by
\begin{subequations}
\label{eq: first order BC in 3d variables}
\begin{align}
\label{eq: pressure BC in outer 3d variables eps}
\prest_1 - \Prest_1 &= \presfnyt  H(-\uu_0 \bcdot \bs{n})+ \presfnzt H(\uu_0 \bcdot \bs{n}), \\
\label{eq: flux BC in outer 3d variables eps}
\pbyp{}{\nrmouter}\left(\prest_1 - \dfrac{\Reyo}{10} \left( \prest_{0 t} - \dfrac{3}{56} \left|\nabla \prest_0\right|^2\right)\right) &- 12 \Darcy \pbyp{\Prest_{1}}{\nrmouter} \notag \\
&= \gamfy  H(-\uu_0 \bcdot \bs{n})+ \gamfz H(\uu_0 \bcdot \bs{n}),
\end{align}
\end{subequations}
for $\uu_0 \bcdot \bs{n} \neq 0$, where a superscript ending in $-$/$+$ refers to an inflow/outflow quantity, and $H(x)$ is the Heaviside step function. These inflow/outflow segments of $\bdy$ are determined from \eqref{eq: u_0 in outer 3d variables soln}--\eqref{eq: pressure BC in outer 3d variables} as previously discussed.  The presence of the Heaviside functions in~\eqref{eq: first order BC in 3d variables} introduces the possibility of pressure singularities appearing at points on the boundary at which $\uu_0 \bcdot \bs{n} = 0$. We discuss this in more detail once we determine the asymptotic approximations of the coefficients pre-multiplying these functions.

The coupling condition~\eqref{eq: flux BC in outer 3d variables eps} represents conservation of mass across the interface at this order and, in contrast to the two-dimensional versions (\ref{eq: BC outer interfacial HO inflow}a), (\ref{eq: BC outer interfacial HO outflow}a), has additional terms which correspond to an entrainment effect at this order. That is, the jump in the $\order{\eps}$ average normal velocity across the interface (between outer regions) has a contribution from the variation of the leading-order tangential velocity in the inner regions.

\subsection{Inner regions}
\label{sec: 3D inner regions}

As the inner regions are close to the obstacle boundary, it will be convenient to work in general curvilinear coordinates $(\nrmouter,\tng,z)$ such that $\nrmouter = 0$ on the boundary, with $\nrmouter >0$  corresponding to the exterior of $\obs$, and $\tng$ being the arc length along the boundary measured anticlockwise, as illustrated in figure~\ref{fig: bio general 3D}.

The relevant inner scalings are $\nrmouter = \eps \nrm$ and $(\wt, \Wt) = (\uizt,\Uizt)/\eps$, and we define the components of fluid velocity in curvilinear coordinates by $\uuit = \uirt \bs{n} + \uitht \bs{t} + \uizt \bs{e}_z$ and $\Qit = \Uirt \bs{n} + \Uitht \bs{t} + \Uizt \bs{e}_z$ in the exterior and interior regions, respectively. Under these scalings, the exterior-flow equations~\eqref{eq: 3D bulk flow BL} become (see, for example, \citet{schlichting2000boundary})
\begin{subequations}
\label{eq: BL eq general coord}
\begin{align}
\label{eq: BL eq general coord nrm}
		\Reyo \left(\uirt \uirt_\nrm + \uizt \uirt_z \right) + \order{\eps \Reyo} &= -\eps^{-1}\prit_{\nrm} + \uirt_{\nrm\nrm} + \uirt_{zz} +\order{\eps}, \\
\label{eq: BL eq general coord tng}
		\Reyo\left(\uirt \uitht_\nrm + \uizt \uitht_z \right) + \order{\eps \Reyo} &= -\prit_{\tng} + \uitht_{\nrm\nrm} + \uitht_{zz} + \order{\eps}, \\
\label{eq: BL eq general coord z}
		\Reyo\left(\uirt \uizt_\nrm + \uizt \uizt_z \right) + \order{\eps \Reyo} &= -\eps^{-1} \prit_{z} + \uizt_{\nrm\nrm} + \uizt_{zz} + \order{\eps}, \\
\label{eq: BL eq general coord cont}
		0 &= \uirt_{\nrm} + \uizt_z + \eps \left( \curv(\tng) \uirt + \uitht_{\tng} \right) + \order{\eps^2},
\end{align}
\end{subequations}
where $\curv(\tng)$ denotes the curvature of the obstacle boundary (positive if the centre of the osculating circle lies in the region in which $\nrmouter < 0$ for a given $\tng$), and we assume that $|\curv| = \order{1}$. Similarly, the interior-flow equations~\eqref{eq: 3D porous flow BL} become
\begin{subequations}
 \label{eq: 3D BL Darcy inner}
\begin{align}
 \label{eq: 3D BL Darcy r inner}
		\eps \Uirt &= - \Darcy \Prit_\nrm, \\
		\label{eq: 3D BL Darcy theta inner}
		\Uitht  &= - \Darcy \Prit_\tng + \order{\eps}, \\
		\label{eq: 3D BL Darcy z inner}
		\eps \Uizt &=  - \Darcy \Prit_z, \\
		\label{eq: 3D BL Darcy continuity inner}
		0 &= \Uirt_{\nrm} + \Uizt_z + \eps  \left(\curv(\tng) \Uirt + \Uitht_{\tng} \right) + \order{\eps^2}.
\end{align}
\end{subequations}
The boundary conditions on the Hele-Shaw cell walls~\eqref{eq: no flux or slip BC 3D BL} become
\begin{subequations}
\label{eq: no flux or slip BC 3D BL Inner}
\begin{alignat}{5}
\label{eq: no flux or slip BC NS 3D BL Inner}
\uuit &= \bs{0} &\quad &\text{on } z = 0, 1& &\text{ for } \nrm > 0, \\
\label{eq: no flux or slip BC Darcy 3D BL Inner}
\Uizt &= 0 &\quad &\text{on } z = 0, 1& &\text{ for } \nrm < 0,
\end{alignat}
\end{subequations}
The interfacial conditions~\eqref{eq: BC Interface 3D BL} on $\nrm = 0$ become
\refstepcounter{equation}
\[
\uirt = \Uirt, \quad \prit = \Prit, \quad \uitht = 0, \quad \uizt = 0.
\eqno{(\theequation{\mathit{a}-\mathit{d}})}\label{eq: BC Interface 3D BL inner}
\]

We form inner expansions in powers of $\eps$ as follows:
\begin{align}
\label{eq: inner expansions 3D BL}
(\uuit,\Qit,\prit,\Prit) &= (\uuit_{0},\Qit_{0},\prit_{0},\Prit_{0}) + \eps (\uuit_{1},\Qit_{1},\prit_{1},\Prit_{1}) + \order{\eps^2} \quad \text{as } \eps \ttz;
\end{align}
substituting these expansions into the inner exterior-flow equations~(\ref{eq: BL eq general coord}a,c) and the interior-flow equations~(\ref{eq: 3D BL Darcy inner}a,c), then equating powers of $\eps$, yields the leading-order pressure equations
\refstepcounter{equation}
\[
0 =-\prit_{0\nrm}, \quad 0 = -\prit_{0z}, \quad 0 =-\Prit_{0\nrm}, \quad 0 = -\Prit_{0z}.
\eqno{(\theequation{\mathit{a}-\mathit{d}})}\label{eq: 3D leading-order inner pressure}
\]
This leading-order system is the same as in the two-dimensional case, and we proceed in exactly the same way. Matching with the outer pressures, and using the leading-order version of the continuity of pressure condition~(\ref{eq: BC Interface 3D BL inner}b), we deduce that the leading-order pressures are unchanged through the inner regions and we are justified in writing the coupling condition~(\ref{eq: pressure BC in outer 3d variables}a). In a similar manner, integrating the leading-order version of the continuity equations~\eqref{eq: BL eq general coord cont} and~\eqref{eq: 3D BL Darcy continuity inner} over the cell height, using the cell-wall boundary conditions~\eqref{eq: no flux or slip BC 3D BL Inner}, and then matching with the outer regions, justifies the coupling condition~(\ref{eq: pressure BC in outer 3d variables}b). The entrainment effect does not occur at this order.

The first-order flow equations in the $(\nrm,z)$-plane decouple from the tangential velocities $\uitht$ and $\Uitht$, and the tangential coordinate $\tng$ can be treated as a parameter in the problem. We can further reduce these problems to the two-dimensional cases considered previously using the scalings
\begin{subequations}
\begin{align}
\left(\uirt_0(\nrm,\tng,z,t), \Uirt(\nrm,\tng,z,t)\right) &= -\scalu(\tng,t) \left(\ui_0(\nrm,z), \Ui_0(\nrm,z)\right), \\
\left(\uizt_0(\nrm,\tng,z,t), \Uizt_0(\nrm,\tng,z,t)\right) &= |\scalu(\tng,t)| \left(\wi_0(\nrm,z), \Wi_0(\nrm,z)\right), \\
\left(\prit_1(\nrm,\tng,z,t), \Prit_1(\nrm,\tng,z,t)\right) &= \left(\Pres_1(0,\tng,t),\Pres_1(0,\tng,t)\right) + |\scalu(\tng,t)| \left(\pri_1(\nrm,z), \Pri_1(\nrm,z)\right),
\end{align}
\end{subequations}
where the terms on the right-hand side without a superscript $\tng$ are variables from the two-dimensional problem, and $\scalu(\tng,t) = -\int_0^1 \! \uu_0 \bcdot \bs{n} \, \mathrm{d}z = \prest_{0 \nrmouter}(0,\tng,t)/12 = \Darcy \Prest_{0 \nrmouter}(0,\tng,t)$ is positive for inflow and negative for outflow. With $\nrm = - X^{-}$ for inflow and $\nrm = X^{+}$ for outflow we obtain the systems shown in figures~\ref{fig: inflow leading order} and~\ref{fig: outflow leading order}, but with $\Rey$ replaced by the appropriate local Reynolds number, namely $\Rey(\tng,t) = |\scalu| \Reyo$. The three-dimensional pressure jump functions in~\eqref{eq: pressure BC in outer 3d variables eps} are related to their two-dimensional equivalents via the expression
\begin{align}
\label{eq: 3D pres jump to 2D}
\left(\presfnyt(\tng,t), \presfnzt(\tng,t) \right) = |\scalu(\tng,t)| \left(\presfny, \presfnz \right).
\end{align}
We have, therefore, already calculated the asymptotic behaviour of $\presfnyt$ and $\presfnzt$ for the pressure coupling condition~\eqref{eq: pressure BC in outer 3d variables eps} in \S\ref{sec: Inflow} and \S\ref{sec: Outflow}. We see that $\presfnyt, \presfnzt \ttz$ as $\scalu \ttz$ and thus \eqref{eq: pressure BC in outer 3d variables eps} does not impose a pressure singularity at points where $\scalu = 0$. However, the possibility still remains that~\eqref{eq: flux BC in outer 3d variables eps} could still impose a pressure singularity when $\scalu = 0$, in which case an additional inner region would be present. We proceed by assuming that $|\scalu| = \order{1}$, and discuss any pressure singularities when they occur. We now calculate the behaviour of the tangential velocities $\uitht_0$ and $\Uitht_0$ within the inner regions (which are new to the three-dimensional problem). This will allow us to determine local properties near the interface, for example, the stress, and to couple the outer regions by determining the functions $\gamfy$ and $\gamfz$ in the final coupling condition~\eqref{eq: flux BC in outer 3d variables eps}.

Within the obstacle, the tangential velocity is given at leading order by
\begin{align}
\label{eq: uitht_0 3D}
\Uitht_0 \equiv -\Darcy \Prest_{0 \tng}(0,\tng,t).
\end{align}
The scaling
\begin{align}
\label{eq: tang scal}
\uitht_0(\nrm,\tng,z,t) =  -(\prest_{0 \tng}(0,\tng,t)/12) \uith_0(\nrm,z;\tng,t),
\end{align}
where we reiterate that $\tng$ and $t$ are now parameters in the inner three-dimensional problem, results in the exterior-flow equation
\begin{align}
\label{eq: scaled out uith}
\Rey(\tng,t)\left(\uir_0 \uith_{0X} + \uiz_0 \uith_{0z} \right) &= 12 + \uith_{0XX} + \uith_{0zz}.
\end{align}
Here, we introduce $X$ such that $X  = X^{-} < 0$ for inflow, and $X  = X^{+} > 0$ for outflow, thus remaining consistent with the two-dimensional analysis. We emphasise that, as with the two-dimensional case, we expect the behaviour of the tangential velocity to vary depending on whether we have inflow or outflow. The cell-wall boundary conditions for $\uith_0$ are given by
\begin{align}
\label{eq: no slip 3D azi}
\uith_0 = 0 \quad \text{on } z = 0, 1 \text{ for } X^{-} <0 \text{ and } X^{+} >  0,
\end{align}
the interfacial conditions are
%\begin{subequations}
\begin{align}
\label{eq: BJ slip 3D azi}
\uith_{0} &= 0 \quad \text{on } X = 0 \text{ for } 0 < z < 1,
%\uith_{0X^{+}} &= \dfrac{\BJ}{\sqrt{\Darcy}}\left(\uith_0 - 12\Darcy \right) \quad \text{on } X^{+} = 0 \text{ for } 0 < z < 1,
\end{align}
%\end{subequations}
and the matching conditions are
\begin{align}
\label{eq: matching 3D azi}
\uith_0 \to 6z(1-z) \quad \text{as } X^{-} \ttni \text{ and } X^{+} \tti \text{ for } 0 < z < 1.
\end{align}
%We note that the cell-wall condition~\eqref{eq: no slip 3D azi} and the Beavers and Joseph slip condition~\eqref{eq: BJ slip 3D azi} are incompatible with each other, and lead to a discontinuity at the corners at $X = 0$, $z = 0, 1$.

In Appendix~\ref{sec: Higher-order average flux}, we integrate the $\order{\eps}$ continuity conditions~\eqref{eq: BL eq general coord cont}, \eqref{eq: 3D BL Darcy continuity inner} over the channel height to obtain the expression
\begin{align}
\label{eq: int cont 2O final}
\pbyp{}{\nrmouter}&\left(\prest_1 - \dfrac{\Reyo}{10} \left(\prest_{0 t} - \dfrac{3}{56} \left|\nabla \prest_0\right|^2\right)\right) - 12 \Darcy \pbyp{\Prest_{1}}{\nrmouter} \notag \\
&= 12 \int_0^\infty \! \int_0^1 \! \left(\uitht_{0\tng}(\nrm,\tng,z,t) - \uut_{0 \tng}(0,\tng,z,t) \bcdot \bs{\tng}\right) \, \mathrm{d}z \, \mathrm{d} \nrm ,
\end{align}
encompassing the entrainment effect which occurs at this order. Using the outer solutions~\eqref{eq: u_0 in outer 3d variables soln} and the tangential velocity scaling~\eqref{eq: tang scal} in~\eqref{eq: int cont 2O final}, we deduce that the inflow and outflow flux jumps $\gamfy$ and $\gamfz$ in~\eqref{eq: flux BC in outer 3d variables eps} are given by
\begin{subequations}
\label{eq: gamfy and gamfz in terms of v}
\begin{align}
\label{eq: gamfy in terms of v}
\gamfy &= -\int\displaylimits_{-\infty}^0 \! \int\displaylimits_0^1 \! \pbyp{}{\tng} \left( \prest_{0 \tng}(0,\tng,t) \left(\uith_{0}(X^{-},z;\tng,t) - 1\right)\right) \, \mathrm{d}z \, \mathrm{d} X^{-}, \\
\label{eq: gamfz in terms of v}
\gamfz &= -\int\displaylimits_{0}^\infty \! \int\displaylimits_0^1 \! \pbyp{}{\tng} \left( \prest_{0 \tng}(0,\tng,t)\left(\uith_{0}(X^{+},z;\tng,t) - 1\right)\right) \, \mathrm{d}z \, \mathrm{d} X^{+},
\end{align}
\end{subequations}
where $\uith_0$ satisfies the system of equations~\eqref{eq: scaled out uith}--\eqref{eq: matching 3D azi}. We proceed by determining the leading-order asymptotic behaviour of $\gamfy$ and $\gamfz$ in the sub-limits in which $\Rey \ll 1$ and $1 \ll 1/\Darcy \ll \Rey \ll 1/\eps$.

\subsection{Small local Reynolds number: $\Rey \ll 1$}
\label{sec: Small Rey 3D}

For the small local Reynolds number case, $\Rey \ll 1$, the inertial terms in the governing equation~\eqref{eq: scaled out uith} are not present at leading-order in $\Rey$. Therefore, $\uith_0$ has no dependence on $\tng$ or $t$ (which come from $\Rey(\tng,t)$), nor on whether we have inflow or outflow. The local Reynolds number uses the local velocity component normal to the interface averaged over the channel height. Thus, the problem we present in this section governs the case when the normal local velocity is small, as well as the case when the global Reynolds number is small. The former case can occur, for example, in the transition between inflow and outflow for a large global Reynolds number, or when the obstacle is close to impermeable. For the latter reason, this problem also appears in the impermeable obstacle case considered in \citet{thompson1968secondary}.

We consider outflow without loss of generality, the $\order{1}$ terms in~\eqref{eq: scaled out uith} yielding the governing equation
\begin{align}
\label{eq: low rey azi vel}
-12 = \term{\uith}{0XX}  + \term{\uith}{0zz}, \quad \text{with } X^{+} = X > 0,
\end{align}
It follows from~\eqref{eq: low rey azi vel} and the outflow versions of the boundary conditions given in~\eqref{eq: no slip 3D azi}--\eqref{eq: matching 3D azi} that
\begin{align}
\label{eq: v sol in 3D small Rey}
\uith_0 = 6z(1-z) - 24 \displaystyle\sum_{k=0}^\infty \alpk^{-3} \exp(-\alpk X) \sin \alpk z,
\end{align}
where $\alpk = (2k + 1)\pi$ for non-negative integers $k$. Using this flow result, we can deduce that the shear stress in the tangential direction is
\begin{align}
\label{eq: 3D low Rey stress}
\term{\stress}{\nrmouter \tng} \sim \eps \uitht_{0 X}(0,z;\tng,t) = \eps \left(-2 \prest_{0 \tng}(0,\tng,t)  \displaystyle\sum_{k=0}^\infty \alpk^{-2} \sin \alpk z \right),
\end{align}
as $\eps, \Reyo \ttz$.

%are given by
%\[
%A_k = -\dfrac{24}{\alpk^3}, \quad
%\alpk = (2k + 1)\pi.
%\eqno{(\theequation{\mathit{b},\mathit{c}})}\label{eq: v in 3D small Rey coeff}
%\]

Substituting the tangential flow solution (\ref{eq: v sol in 3D small Rey}a) into the outer flux jump conditions \eqref{eq: gamfy and gamfz in terms of v}, we obtain at leading order an analytic expression for the coupling condition \eqref{eq: flux BC in outer 3d variables eps}, namely
\begin{align}
\label{eq: flux BC in outer 3d variables eps small Rey}
\pbyp{\prest_1}{\nrmouter} - 12 \Darcy \pbyp{\Prest_{1}}{\nrmouter} = \dfrac{93 \zeta(5)}{2 \pi^5} \prest_{0 \tng \tng} \quad \text{on } \bdy,
\end{align}
where $\zeta$ is the Riemann zeta function.
As inflow and outflow are reversible in Stokes equations, the Heaviside functions in \eqref{eq: flux BC in outer 3d variables eps} disappear and we can deduce that, in the limit in which $\eps \ll \Rey \ll 1$, no pressure singularities exist up to $\order{\eps}$.

\subsection{Large local Reynolds number: $1 \ll 1/\Darcy \ll \Rey \ll 1/\eps$}
 
Now we consider the local asymptotic sub-limit $1 \ll 1/\Darcy \ll \Rey \ll 1/\eps$ and, as in \S\ref{sec: Rey gg 1}, we find it useful to work with the inverse local Reynolds number, $\epsbt = \Rey^{-1} \ll 1$, for ease of notation. The asymptotic structure here is the same as in \S\ref{sec: Rey gg 1}; our goal is to determine the behaviour of the tangential velocity $\uith_0$ from the system~\eqref{eq: scaled out uith}--\eqref{eq: matching 3D azi}, and use this to determine the asymptotic behaviour of the flux jump functions from~\eqref{eq: gamfy and gamfz in terms of v}.

We scale the first correction terms in the outer regions with the global Reynolds number as follows
\begin{align}
\label{eq: asymptotic series for large Rey 3D}
(\uut_1, \QQt_1, \prest_1, \Prest_1) = \Reyo(\uut_{10}, \QQt_{10}, \prest_{10}, \Prest_{10}) + (\uut_{11}, \QQt_{11}, \prest_{11}, \Prest_{11}) + \order{\Reyo^{-1}},
\end{align}
but the inner regions will be asymptotic expansions in the local Reynolds number. We split the analysis into inflow and outflow, starting with the former.

\subsubsection{Inflow}
\label{sec: Inflow 3D}
The inflow behaviour is similar to the flow considered in \S\ref{sec: Inflow}, for which the asymptotic structure is given in figure~\ref{fig: inflow inner BL}. We show that the leading-order tangential velocity is independent of $X^{-}$ through \region{II} and \region{III}, but the next-order flow depends on $X^{-}$ in \region{II}. In this section we also show that the no-tangential-slip condition~\eqref{eq: BJ slip 3D azi} is satisfied in \region{IIa}, and we calculate the asymptotic behaviour of $\gamfy$ as $\epsbt \ttz$.

In \region{II}, the tangential flow is unchanged at leading order and satisfies the asymptotic expansion
\begin{align}
\label{eq: ae v in inflow inner}
\uith_0(X^{-},z;\tng,t) = 6z(1-z) + \epsbt \uith_{01} + \order{\epsbt^{4/3}} \quad \text{as } \epsbt \ttz.
\end{align}
Substituting~\eqref{eq: ae v in inflow inner} into the governing equation~\eqref{eq: scaled out uith} and equating terms of $\order{1}$, gives
\begin{align}
\label{eq: uith_01 eq}
z(1-z) \uith_{01 X^{-}} + (1-2z) \uiz_{01} = 0,
\end{align}
with the matching condition
\begin{align}
\label{eq: uith_01 matching}
\uith_{01} \ttz \quad \text{as } X^{-} \ttni.
\end{align}
Equation~\eqref{eq: uith_01 eq}, with far-field condition~\eqref{eq: uith_01 matching}, can be integrated numerically with respect to $X^{-}$ to determine $\uith_{01}$. The resulting correction to the tangential velocity induces a slip on the cell walls, which is remedied by a boundary layer in \region{IIb} (as illustrated in figure~\ref{fig: inflow inner BL}), the details of which are omitted from this paper for the sake of brevity.

The coordinate scaling in \region{IIa} is $X^{-} = \epsbt \Ri$, and we use hats to denote all variables in this region. We use the flow variables calculated in \S\ref{sec: Smaller boundary layer}, and pose an asymptotic expansion for the tangential velocity of the form
\begin{align}
\label{eq: uiith 3d ae}
\uiith_0(\Ri,z;\tng,t) = \uiith_{00}(\Ri,z;\tng,t) + \order{\epsbt}, % \uiith_{01}(\Ri,z;\tng,t) + \order{\epsbt^{4/3}},
\end{align}
where the leading-order term satisfies the governing equation
\begin{align}
\label{eq: uiith_01 eq}
6z(1-z) \uiith_{00\Ri} = \uiith_{00\Ri \Ri} \quad \text{for } \Ri < 0, \, 0 < z < 1,
\end{align}
with no-tangential-slip condition~\eqref{eq: BJ slip 3D azi} yielding
\begin{align}
\label{eq: BJ slip uiith 3D}
\uiith_{00} = 0 \quad \text{on } \Ri = 0, \, 0 < z < 1,
\end{align}
and a matching condition with \region{II} given by
\begin{align}
\label{eq: uiith_01 matching}
\uiith_{00}(\Ri,z;\tng,t) \sim 6z(1-z) \quad \text{as } \Ri \ttni, \text{ for } 0 < z < 1.
\end{align}
The system~\eqref{eq: uiith_01 eq}--\eqref{eq: uiith_01 matching} is solved by
\begin{align}
\label{eq: tng vel 3D inflow region IIa}
\uiith_{00} = 6z(1-z) \left(1 - \exp(6z(1-z)\Ri)\right).
\end{align}

The leading-order contribution to $\gamfy$ in~\eqref{eq: gamfy in terms of v} is of $\order{\epsbt}$, and comes from both the $\order{\epsbt}$ term in \region{II} and the $\order{1}$ term in \region{IIa}, as \region{IIa} is $\order{\epsbt}$ smaller than \region{II}. Therefore, substituting the asymptotic expansion~\eqref{eq: ae v in inflow inner} into~\eqref{eq: gamfy in terms of v}, and using the governing equation~\eqref{eq: uith_01 eq} to relate the velocity $\uith_{01}$ to $\uiz_{01}$ (which has already been calculated numerically in \S\ref{sec: Regions II and III}), along with the leading-order solution for the tangential velocity in \region{IIa} \eqref{eq: tng vel 3D inflow region IIa}, the leading-order contribution to \eqref{eq: gamfy in terms of v} becomes
\begin{align}
\label{eq: gamfz inflow large Rey}
\gamfy \sim \dfrac{1}{\Reyo} \pbyp{}{\tng}\left(\dfrac{\prest_{0 \tng}(0,\tng,t)}{\scalu(\tng,t)}\left(1 - \int\displaylimits_{-\infty}^0 \! \int\displaylimits_0^1 \!
\left(\dfrac{2z - 1}{z(1-z)}\int\displaylimits_{-\infty}^{X^{-}} \! \uiz_{01}(\xi,z;\tng,t)  \mathrm{d}\xi \right)
\, \mathrm{d}z \, \mathrm{d} X^{-} \right)\right),
\end{align}
%\begin{align}
%\label{eq: gamfz inflow large Rey}
%\gamfy \sim -\epsb \int\displaylimits_{-\infty}^0 \! \int\displaylimits_0^1 \! \pbyp{}{\tng} \left( \dfrac{\prest_{0 \tng}(0,\tng,t)}{\scalu(\tng,t)}\left(\dfrac{2z - 1}{z(1-z)}\int\displaylimits_{-\infty}^{X^{-}} \! \uiz_{01}(\xi,z;\tng,t)  \mathrm{d}\xi \right)\right) \, \mathrm{d}z \, \mathrm{d} X^{-} \quad \text{as } \epsbt \ttz,
%\end{align}
as $\eps, 1/\Reyo \ttz$ with $\eps \ll 1/\Reyo \ll \Darcy \ll 1$. We note that the integral is finite because $\uiz_{01} \ttz$ as $z \ttz, 1$.

In contrast to the two-dimensional case, the pressure now varies in the $\tng$-direction. Therefore, the component of the interfacial stress in the normal direction is different from the two-dimensional case~\eqref{eq: Normal stress for inflow}, and in three-dimensions is given by
\begin{align}
\label{eq: Normal stress for inflow 3D}
\stress_{\nrmouter \nrmouter} = -\left(\Prest_0(0,\tng,t) + \eps\left(\Reyo \Prest_{10}(0,\tng,t) + \prit_{11}(0,\tng,z,t)\right) \right) + \order{ \eps / \Reyo^{1/3}}
\end{align}
as $\eps, 1/\Reyo \ttz$ with $\eps \ll 1/\Reyo \ll \Darcy \ll 1$. We note that $\Prest_0$ and $\Prest_{10}$ have no $z$-dependence. Therefore, as in two dimensions, the first order at which the normal stress varies in $z$ is at $\order{\eps}$. Finally, we determine the leading-order term in the component of the shear stress on the obstacle boundary in the $\tng$-direction for inflow, $\term{\stress}{\nrmouter \tng}$. From the tangential flow solution in \region{IIa} given by~\eqref{eq: tng vel 3D inflow region IIa}, and recalling the scaling~\eqref{eq: tang scal}, we deduce that
\begin{align}
\label{eq: 3D inflow stress}
\term{\stress}{\nrmouter \tng} \sim \dfrac{\eps}{\epsbt} \dfrac{\prest_{0 \tng}(0,\tng,t)}{12}\uiith_{0 \Ri}(0,z;\tng,t) = - \eps \Reyo \left(3 \scalu(\tng,t) \prest_{0\tng}(0,\tng,t) \left(z(1-z) \right)^2 \right),
\end{align}
as $\eps, 1/\Reyo \ttz$ with $\eps \ll 1/\Reyo \ll \Darcy \ll 1$.

\subsubsection{Outflow}
\label{sec: Outflow 3D}

Outflow occurs in a similar manner to \S\ref{sec: Outflow}, for which the asymptotic structure is illustrated in figure~\ref{fig: outflow BL}. The tangential velocity within \region{V} is plug flow (as given by~\eqref{eq: uitht_0 3D}) and the tangential flow in \region{VI} is of $\order{\epsbt}$. The tangential flow then transitions to Poiseuille flow in \region{VIa}, where it becomes of $\order{1}$. In this section we solve for the leading-order tangential flow for outflow and calculate the asymptotic behaviour of $\gamfz$ as $\epsbt \ttz$.

In \region{VI}, we use the flow variables calculated in \S\ref{sec: Region VI} as well as the asymptotic expansion
\begin{align}
\label{eq: ae v in outflow inner}
\uith_0(X^{+},z;\tng,t) = \epsbt \left(12 X^{+}\right)  + \order{\epsbt^{3/2}} \quad \text{as } \epsbt \ttz,
\end{align}
which satisfies the no-tangential-slip condition~\eqref{eq: BJ slip 3D azi}. The no-slip condition on the cell walls is satisfied in \region{VIb} (as illustrated in figure~\ref{fig: outflow BL}) via a similarity solution, as described in Appendix~\ref{sec: regions VIb and VIc}. As with the three-dimensional inflow case considered in the previous section, the component of the interfacial stress in the normal direction is now
\begin{align}
\label{eq: 3D outflow normal stress}
\stress_{n n} = -\left(\Prest_0(0,\tng,t) + \eps \Reyo \Prest_{10}(0,\tng,t) \right) + \order{\eps /\epsbt^{1/2}},
\end{align}
as $\eps, 1/\Reyo \ttz$ with $\eps \ll 1/\Reyo \ll 1$, and where $\Prest_0$ and $\Prest_{10}$ have no $z$-dependence. Therefore, as in two dimensions, the first order at which the normal stress varies in $z$ is at $\order{\eps/\epsbt^{1/2}}$. From \eqref{eq: ae v in outflow inner} and the scaling~\eqref{eq: tang scal}, we deduce that the stress on the interface in the tangential direction is
\begin{align}
\label{eq: 3D outflow stress}
\term{\stress}{\nrmouter \tng} \sim -\eps \dfrac{\prest_{0\tng}(0,\tng,t)}{12} \uiith_{0 X^{+}}(0,z;\tng,t) = \dfrac{\eps}{\scalu(\tng,t)\Reyo }\left(\prest_{0\tng}(0,\tng,t) \right),
\end{align}
as $\eps, 1/\Reyo \ttz$ with $\eps \ll 1/\Reyo \ll \Darcy \ll 1$.

The transition to Poiseuille flow occurs in \region{VIa}, within which the relevant coordinate scaling is $X^{+} = \epsbt^{-1} \Rii$. We further use $\uith_0 = \uiiith_0$ to indicate we are in this intermediate region, and use the variables introduced in \S\ref{sec: Region VIa}. We substitute the asymptotic expansion
\begin{align}
\label{eq: region VIa ae 3D BL uth}
\uiiith_0 = \uiiith_{00} + \order{\epsbt^{1/2}} \quad \text{as } \epsbt \ttz,
\end{align}
into the governing equation~\eqref{eq: scaled out uith} to obtain the leading-order governing equation
\begin{align}
\label{eq: region VIa 3D gov eq v}
\uiiir_{00} \uiiith_{00\Rii} + \term{\uiiiz}{01} \uiiith_{00z} = 12 +  \uiiith_{00zz}.
\end{align}
The cell wall boundary conditions~\eqref{eq: no slip 3D azi} become
\begin{align}
\label{eq: no flux or slip BC NS 3D BL region VIa LO tng}
\uiiith_{00} = 0 \quad \text{on } z = 0, 1 \text{ for } \Rii > 0,
\end{align}
the matching conditions with \region{VI} are
\begin{align}
\label{eq: v in VIa near 0}
\uiiith_{00} \ttz \quad \text{as } \Rii \downarrow 0 \text{ for } 0 < z < 1,
\end{align}
and the far-field behaviour is determined via the matching condition~\eqref{eq: matching 3D azi}, and is given by
\begin{align}
\label{eq: v in VIa near infty}
\term{\uiiith}{00} &\sim 6z(1-z) \quad \text{as } \Rii \tti \text{ for } 0 < z < 1.
\end{align}

The system~\eqref{eq: region VIa 3D gov eq v}--\eqref{eq: v in VIa near infty} details the transition from plug to Poiseuille tangential flow. To determine the flux jump $\gamfz$ at leading order, the system must be solved numerically. We note that the system~\eqref{eq: region VIa 3D gov eq v}--\eqref{eq: v in VIa near infty} is independent of $\Darcy$, and hence only needs to be solved once to determine $\gamfz$ at leading order.

Accounting for the change of variable $X^{+} = \epsbt^{-1} \Rii$ being dependent on $\tng$, we find
\begin{align}
\pbyp{}{\tng} \mapsto \pbyp{}{\tng} - \dfrac{\Rii}{\scalu}\dbyd{\scalu}{\tng} \pbyp{}{\Rii},
\end{align}
and it follows that the leading-order behaviour of $\gamfz$ is given by
\begin{subequations}
\label{eq: gamfz outflow large Rey both}
\begin{align}
\label{eq: gamfz outflow large Rey}
\gamfz \sim -\Reyo \Lambda_a\pbyp{}{\tng}\left(\scalu \prest_{0 \tng}(0,\tng,t) \right),
\end{align}
where
\begin{align}
\label{eq: Lambda a and Lambda b}
\Lambda_a = \int\displaylimits_{0}^\infty \! \int\displaylimits_0^1 \! \left(1 - \uiiith_{00}(\Rii,z)\right) \, \mathrm{d}z \, \mathrm{d} \Rii.
\end{align}
\end{subequations}
We calculate $\uiiith_{00}$ numerically using a second-order central finite-difference scheme, then determine $\Lambda_a$ using the trapezium rule. We find that $\Lambda_a \approx 0.125$ to three significant figures.

Combining the outflow flux jump~\eqref{eq: gamfz outflow large Rey both} with its inflow equivalent~\eqref{eq: gamfz inflow large Rey}, and recalling that $\scalu = \Darcy\Prest_{0\nrmouter}(0,\tng,t)$, we find that the leading-order (in $\epsb$) coupling condition~\eqref{eq: flux BC in outer 3d variables eps} in the large Reynolds number asymptotic sub-limit is given by
\begin{subequations}
\label{eq: BC in outer 3d variables large Rey full}
\begin{align}
\label{eq: flux BC in outer 3d variables large Rey full}
\pbyp{}{\nrmouter}&\left(\prest_{10} - \dfrac{1}{10} \prest_{0 t} + \dfrac{3}{560} \left|\nabla \prest_0\right|^2\right) - 12 \Darcy \pbyp{\Prest_{10}}{\nrmouter} \notag \\
&= -\Darcy \Lambda_a \pbyp{}{\tng}\left(\Prest_{0\nrmouter}\prest_{0 \tng}\right) H(\uu_0 \bcdot \bs{n}) \quad \text{on } \bdy.
\end{align}
The leading-order pressure jump condition~\eqref{eq: pressure BC in outer 3d variables eps} in the large Reynolds number sub-limit is given by
\begin{align}
\label{eq: pres BC in outer 3d variables large Rey full}
\prest_{10} - \Prest_{10} &= \left(\Darcy\Prest_{0\nrmouter}\right)^2\presfnz_0 H(\uu_0 \bcdot \bs{n}) \quad \text{on } \bdy.
\end{align}
\end{subequations}
We see that the coefficient pre-multiplying the Heaviside function in \eqref{eq: flux BC in outer 3d variables large Rey full} does not necessarily vanish when $\uu_0 \bcdot \bs{n} = 0$, whereas the coefficient pre-multiplying the Heaviside function in \eqref{eq: pres BC in outer 3d variables large Rey full} does vanish. Thus, at these points the pressure is continuous but the pressure derivative along the boundary will have a $\log$ singularity.

\subsection{Summary}
\label{sec: 3D summary}

We now summarise the main results for the three-dimensional case. The general outer problem is governed by the \emph{two-dimensional} linear equations: outside $\obs$
\refstepcounter{equation}
\[
\nabla^2_{\perp} \prest_0 = 0, \quad
\nabla^2_{\perp} \left(\prest_1 + \dfrac{3 \Reyo}{560} |\nabla \prest_0|^2 \right) = 0;
\eqno{(\theequation{\mathit{a},\mathit{b}})}
\]
and inside $\obs$
\refstepcounter{equation}
\[
\nabla^2_{\perp} \Prest_0 = 0, \quad
\nabla^2_{\perp} \Prest_1 = 0.
\eqno{(\theequation{\mathit{a},\mathit{b}})}
\]
The coupling conditions (in outer variables) on the boundary of the obstacle cross-section $\bdy$ are given by
\begin{subequations}
\label{eq: Overview jump conditions}
\begin{align}
\label{eq: pressure BC LO}
\prest_0 - \Prest_0 &= 0, \\
\label{eq: pressure BC FO}
\prest_1 - \Prest_1 &= \Darcy \left| \pbyp{\Prest_{0}}{n}\right| \presfn(s,t;\Reyo,\Darcy), \\
\pbyp{\prest_0}{n} - 12 \Darcy \pbyp{\Prest_0}{n} &= 0, \\
\label{eq: flux BC FO}
\pbyp{}{n} \left(\prest_1 - \dfrac{\Reyo}{10}\left(\prest_{0 t} - \dfrac{3}{56} |\nabla \prest_0|^2\right) \right) - 12 \Darcy \pbyp{\Prest_1}{n} &= \Lambda(s,t;\Reyo,\Darcy).
\end{align}
\end{subequations}
Here, $\p / \p n$ denotes the outward normal derivative to the obstacle, and we define the functions
\begin{subequations}
\begin{align}
\presfn(s,t;\Reyo,\Darcy) &= \presfny(s,t;\Reyo,\Darcy)  H(\Prest_{0n})+ \presfnz(s,t;\Reyo,\Darcy) H(-\Prest_{0n}), \\
\Lambda(s,t;\Reyo,\Darcy) &= \gamfy(s,t;\Reyo,\Darcy)  H(\Prest_{0n})+ \gamfz(s,t;\Reyo,\Darcy) H(-\Prest_{0n}),
\end{align}
\end{subequations}
where $\presfny$ and $\presfnz$ are outputs to the nonlinear systems presented in figures~\ref{fig: inflow leading order} and~\ref{fig: outflow leading order} (using $\Reyo$ instead of $\Rey$), $\gamfy$ and $\gamfz$ are solutions to~\eqref{eq: gamfy in terms of v} and~\eqref{eq: gamfz in terms of v} respectively, and $H(x)$ denotes the Heaviside step function.
 
For $\eps, \Reyo \ttz$, we found in \S\ref{sec: Small Rey 3D} that $\presfny = - \presfnz$ and these functions can be numerically determined by solving a system of linear equations (shown in figure~\ref{fig: All_low_Rey}(a)). Additionally, from~\eqref{eq: flux BC in outer 3d variables eps small Rey}, we found that
\begin{align}
\gamfy, \gamfz \sim \dfrac{93 \zeta(5)}{2 \pi^5} \prest_{0 \tng \tng}(0,\tng,t) \quad \text{as } \eps, \Reyo \ttz,
\end{align}
where $\zeta$ is the Riemann zeta function. %$\tng$ is the tangential arclength in the anticlockwise direction, and the constants $A_k$ and $\alpk$ are defined in~(\ref{eq: v sol in 3D small Rey}b,c).

For $\eps, 1/\Reyo \ttz$ with $\eps \ll 1/\Reyo \ll \Darcy \ll 1$, we found in \S\ref{sec: Inflow} and \S\ref{sec: Outflow}, respectively, that
\refstepcounter{equation}
\[
\presfny \sim \inpdrop / \Darcy, \quad
\presfnz \sim \Reyo \Darcy |\Prest_{0\nrmouter}(0,\tng,t)| \presfnz_0,
\eqno{(\theequation{\mathit{a},\mathit{b}})}
\]
recalling that $\Rey(\tng,t) = \Darcy|\Prest_{0 \nrmouter}|\Reyo$, where $\inpdrop \approx -0.117$ and $\presfnz_0 \approx -0.331$.  In the same limit, we also found that $\gamfy = \order{1/( \Reyo \Darcy)}$ (the prefactor at this order of magnitude is given in~\eqref{eq: gamfz inflow large Rey}), and that
\begin{align}
\gamfz \sim - \Reyo \Darcy \Lambda_a \pbyp{}{\tng} \left(\Prest_{0 \nrmouter}(0,\tng,t) \prest_{0 \tng}(0,\tng,t)\right)
\end{align}
as $\eps, 1/\Reyo \ttz$ with $\eps \ll 1/\Reyo \ll \Darcy \ll 1$, where $\Lambda_a \approx 0.125$ (defined in \eqref{eq: Lambda a and Lambda b}).

In the limit in which $\eps, \Reyo \ttz$, we found that the interfacial stress $\stresstensor(\bs{n})|_{\bdy} = -\prest_0 \bs{n}|_{\bdy} + \order{\eps}$, where $\bs{n}$ is the unit normal vector on the interface directed towards the exterior fluid. The prefactors in the error term can be determined by solving a linear system of equations. For these stress components in the $\nrmouter$- and $z$-directions, we must solve the problem of Stokes flow coupled to Darcy flow, and present the system to be solved numerically in \S\ref{sec: Small Rey}. The error term due to the stress component in the $\tng$-direction is given in~\eqref{eq: 3D low Rey stress}. For the stress components in the $\nrmouter$- and $z$-directions, we showed in figures~\ref{fig: All_low_Rey}(b) and (c), respectively, the leading-order difference between the stress predicted by the boundary layer analysis and the stress predicted by a naive application of the outer solutions.

In the limit in which $\eps, 1/\Reyo \ttz$ with $\eps \ll 1/\Reyo \ll \Darcy \ll 1$, the interfacial stress is given by
\begin{align}
\label{eq: stress for large Rey}
\stresstensor(\bs{n})|_{\bdy} &\sim - \left(\Prest_0(0,\tng,t) + \eps \Reyo \Prest_{10}(0,\tng,t) \right)\bs{n} \notag \\
& \qquad - \eps \Reyo \Darcy \left(3 \Prest_{0 \nrmouter}(0,\tng,t) \prest_{0\tng}(0,\tng,t) \left(z(1-z) \right)^2 \right) H(\Prest_{0 \nrmouter}) \bs{t},
\end{align}
where $\bs{t} = \bs{e}_z \times \bs{n}$ is the unit tangent vector on the interface in the anticlockwise direction. The given stress components in the normal direction in \eqref{eq: stress for large Rey} do not vary in $z$, and the order of magnitude of the error is different for inflow and for outflow. For inflow, the error is of $\order{\eps}$ and the correction terms in the $\nrmouter$- and $z$-directions are given in \eqref{eq: Normal stress for inflow 3D} and \eqref{eq: Shear stress for inflow}, respectively. In figure~\ref{fig: stress_high_Rey_inflow}, we show the difference between the $\order{\eps}$ stress (in the $\nrmouter$- and $z$-directions) from the inner flow and that predicted by a naive application of the outer solution. For outflow, the errors in the interfacial stress are of $\order{\eps (\Reyo \Darcy)^{1/2}}$ in the $\nrmouter$-direction, $\order{\eps / (\Reyo \Darcy)}$ in the $\tng$-direction, and $\order{\eps /(\Reyo \Darcy)^{1/2}}$ in the $z$-direction, as deduced in \eqref{eq: 3D outflow normal stress}, \eqref{eq: 3D outflow stress}, and \eqref{eq: Shear stress outflow 2d}, respectively.

\section{Application of results to a cylinder with a circular cross-section within a Hele-Shaw cell}
\label{sec: Example}
We now apply our results to the problem of a forced time-dependent far-field flow in a Hele-Shaw domain containing a porous cylinder whose cross-section is a circle with dimensionless radius $1$. Such a set-up can arise in tissue engineering, where the interior of the porous obstacle is seeded with cells. Another application is in modelling the erosion of a porous biofilm, for example, the attempted removal of dental plaque from between teeth.

In these applications, quantities of physical interest can be determined from our analysis. Firstly, the net force acting on the porous obstacle determines the movement of the porous obstacle were it free to move. Although a periodic forcing in the low Reynolds number regime would lead to no net movement over one oscillation \citep{purcell1977life}, the same is not true for a large Reynolds number regime, and we investigate the latter case here. Secondly, in tissue engineering applications, cell growth is often coupled to the shear stress that cells experience, which is one part of a process known as mechanotransduction. Thus, to help with cell placement within a porous scaffold, it is important to know how the internal shear stress varies within an obstacle. We use results from \citet{modelperfusionbio} to estimate the spatial variation of shear stress within the porous obstacle, and thus the spatial variation of shear stress experienced by cells placed within such an obstacle. Finally, biofilm erosion is often taken to be proportional to the square root of the shear stress acting on the interface \citep{duddu2009two} and, to this end, we determine the interfacial shear stress. 

We use cylindrical polar coordinates $(r,\theta ,z)$, where $r=0, z = 1/2$ corresponds to the centre of the porous obstacle, and define the components of the exterior and interior fluid velocities as $\uut = \uer \bs{e}_r + \ueth \bs{e}_\theta + \uez \bs{e}_z$ and $\QQt = \Uer \bs{e}_r + \Ueth \bs{e}_\theta + \Uez \bs{e}_z$, respectively.

We consider the (more interesting) case where $\eps, 1/\Reyo \ttz$ with $\eps \ll 1/\Reyo \ll \Darcy \ll 1$, and use the corresponding outer system as outlined in \S\ref{sec: 3D summary}. That is, we consider asymptotic expansions of the form
\begin{align}
\uut \sim \uut_0 + \eps \Reyo \uut_{10} \quad \text{as } \eps, 1/\Reyo \ttz \text{ with } \eps \ll 1/\Reyo \ll \Darcy \ll 1,
\end{align}
with similar expressions for $\QQt$, $\prest$, and $\Prest$. The governing equations for the exterior fluid are
\refstepcounter{equation}
\[
\nabla^2_{\perp} \prest_0 = 0, \quad
\nabla^2_{\perp} \left(\prest_{10} + \dfrac{3}{560} |\nabla \prest_0|^2 \right) = 0 \quad \, \text{for } r>1, \, 0 < \theta \leq 2 \pi,\,  0 < z < 1;
\eqno{(\theequation{\mathit{a},\mathit{b}})}
\]
and, for the interior flow,
\refstepcounter{equation}
\[
\nabla^2_{\perp} \Prest_0 = 0, \quad
\nabla^2_{\perp} \Prest_{10} = 0 \quad \, \text{for } r<1, \, 0 < \theta \leq 2 \pi,\,  0 < z < 1.
\eqno{(\theequation{\mathit{a},\mathit{b}})}
\]
The coupling conditions on $r=1$ are given by
\begin{subequations}
\label{eq: Example jump conditions}
\begin{align}
\label{eq: ex pressure BC LO}
\prest_0 - \Prest_0 &= 0, \\
\label{eq: ex pressure BC FO}
\prest_{10} - \Prest_{10} &= \presfnz_0 \left(\Darcy  \pbyp{\Prest_{0}}{r}\right)^2  H(-\Prest_{0r}), \\
\pbyp{\prest_0}{r} - 12 \Darcy \pbyp{\Prest_0}{r} &= 0, \\
\label{eq: ex flux BC FO}
\pbyp{}{r} \left(\prest_{10} - \dfrac{1}{10}\left(\prest_{0 t} - \dfrac{3}{56} |\nabla \prest_0|^2\right) \right) - 12 \Darcy \pbyp{\Prest_{10}}{r} &= -\Darcy \Lambda_a \pbyp{}{\theta} \left(\Prest_{0 r} \prest_{0 \theta}\right) H(-\Prest_{0r}),
\end{align}
\end{subequations}
where $\Darcy$ is the dimensionless permeability, $\presfnz_0$, $\Lambda_a$, and $\Lambda_b$ are constants defined in \S\ref{sec: 3D summary}, and $H(x)$ is the Heaviside step function. The far-field conditions are
\begin{align}
\prest_0 \sim -12x G(t), \quad \prest_{10} \to 0 \quad \text{as } r \tti.
\end{align}

The leading-order problem is solved by
\refstepcounter{equation}
\[
\prest_0 = -12\left(r + \dfrac{A}{r} \right) G(t) \cos \theta , \quad
\Prest_0 = -12(1 + A) r G(t) \cos \theta ,
\eqno{(\theequation{\mathit{a},\mathit{b}})}\label{eq: LO flow sol}
\]
where $A = (1 - 12\Darcy)/(1 + 12 \Darcy)$. The points on the obstacle boundary at which the flow transitions between inflow and outflow are when $\Prest_{0r} = 0$ which, for this specific problem, is when $\theta = \pi/2, 3\pi/2$, and the region $\theta \in (\pi/2, 3\pi/2)$ admits inflow when $G>0$, and outflow when $G<0$. The first-order problem is solved by 
\begin{subequations}
\label{eq: First order pressure ex}
\begin{align}
\prest_{10} &= -\dfrac{27 A^2}{35} \dfrac{G^2(t)}{r^4} + \sumjinf{n} a_{n} (t) r^{-n} \cos n \theta, \\
\label{eq: First order pressure Darcy}
\Prest_{10} &= \displaystyle\sum_{n=0}^\infty b_{n} (t) r^{n} \cos n \theta,
\end{align}
\end{subequations}
where the coefficients $a_{n}$, $b_{n}$ can be determined via a standard application of Fourier series, and we given $b_n$ (which are used in the following analysis) in Appendix \ref{sec: Flow coefficients}. We note that the even modes (apart from $n=0, 2$) vanish, and $(a_{2n+1}, b_{2n+1}) \sim (-1)^n/(2n+1)^2 (\alpha,\beta)$ as $n \tti$ (where $\alpha$ and $\beta$ are constant). Thus, at the points $\theta = \pi/2, 3\pi/2$ (where there is a transition between inflow and outflow) the $\order{\eps \Reyo}$ pressures are continuous but their derivatives with respect to $\theta$ are singular, as expected. Using~\eqref{eq: stress for large Rey}, the total force acting on the obstacle, $\bs{F}$, is given by
\begin{align}
\label{eq: Force ex}
\bs{F} &\sim -\int_0^{2\pi} \! \left(\Prest_0(1,\theta,t) + \eps \Reyo \Prest_{10}(1,\theta,t) \right) \left(\cos \theta \bs{e}_x + \sin \theta \bs{e}_y\right)\, \mathrm{d}\theta \notag \\
&-  \eps \Reyo \Darcy \int_{0}^{2\pi} H(\Prest_{0r}) \int_0^1 \! 3  \left(z(1-z) \right)^2 \Prest_{0r}(1,\theta,t) \prest_{0\theta}(1,\theta,t) \left(-\sin \theta \bs{e}_x + \cos \theta \bs{e}_y\right)\,\mathrm{d}z \,\mathrm{d}\theta ,
\end{align}
as $\eps, 1/\Reyo \ttz$ with $\eps \ll 1/\Reyo \ll \Darcy \ll 1$. Thus the only contribution to the total force acting on the obstacle from $\Prest_{10}$ is from the coefficient $b_{1}(t)$, given by
\begin{align}
b_1(t) = \dfrac{6 (1-A)}{5(1 + 12 \Darcy) }G'(t) - \dfrac{96 \Darcy (1+A)^2}{\pi(1 + 12 \Darcy)} \left(2\Darcy \presfnz_0 + \Lambda_a \right)  G(t)|G(t)|.
\end{align}
Thus, the force~\eqref{eq: Force ex} is given by
\begin{align}
\label{eq: Actual force on cylinder}
\bs{F} \sim  \left(12 \pi (1+A) G(t) + \eps \Reyo \left(\dfrac{48 \Darcy}{5}\left(1+A\right)^2 G(t)|G(t)| - \pi b_1(t)\right) \right) \bs{e}_x,
\end{align}
as $\eps, 1/\Reyo \ttz$ with $\eps \ll 1/\Reyo \ll \Darcy \ll 1$.

If the far-field forcing is periodic, with zero mean, the force acting on the obstacle does not necessarily vanish. That is, if there exists $T>0$ such that $G(t + T) \equiv G(t)$ for all $t$, and where $\int_0^T \! G(t) \, \mathrm{d}t = 0$, the net force experienced by the obstacle over one oscillation is given by
%\begin{align}
%\label{eq: net force}
%\int_0^T \! \bs{F}(t) \, \mathrm{d}t \sim \eps \Reyo \Darcy \dfrac{96 (1+A)^2}{1 + 12 \Darcy} \left(2\Darcy \presfnz_0 + \Lambda_a \right) \left( \int_0^T \! G(t)|G(t)| \, \mathrm{d}t \right) \bs{e}_x,
%\end{align}
\begin{align}
\label{eq: net force}
\int_0^T \! \bs{F}(t) \, \mathrm{d}t \sim 96 (1+A)^2 \left(\eps \Reyo \Darcy \right) \left(\dfrac{1}{10} + \dfrac{2\Darcy \presfnz_0 + \Lambda_a}{1 + 12 \Darcy} \right) \left(\int_0^T \! G(t)|G(t)| \, \mathrm{d}t \right) \bs{e}_x
\end{align}
as $\eps, 1/\Reyo \ttz$ with $\eps \ll 1/\Reyo \ll \Darcy \ll 1$. The integral on the right-hand side of \eqref{eq: net force} vanishes for single mode oscillations (for example, $G(t) = \cos t$) but, in general, not for more complicated periodic functions (for example, $G(t) = \cos t - \cos 2 t$). Even though the forcing is periodic, the effect of fluid inertia is to impart a net force over one oscillation. This would cause the obstacle to drift were it free to move. We note that the right-hand side of~\eqref{eq: net force} vanishes as $\Darcy \ttz$, and hence the net force would not appear at this order for impermeable obstacles. We can attribute the appearance of the net force to the entrainment effect that occurs for outflow, which we were able to capture via a formal boundary layer analysis.

In tissue engineering applications, the shear stress $S$ experienced \emph{within} the porous obstacle is of interest, as cells are placed within the porous obstacle and their growth may be dependent on the shear stress. As different cells will require different levels of shear stress to grow optimally, it is important to understand how the shear stress experienced by cells will vary across the porous obstacle. In \citet{modelperfusionbio}, it was shown that the shear stress experienced by cells within the individual porous scaffold pores can be estimated from the Darcy velocity. In particular, the shear stress is proportional to the magnitude of the Darcy velocity, which can be obtained by the pressure solutions (\ref{eq: LO flow sol}b) and \eqref{eq: First order pressure Darcy}, to obtain
\begin{align}
\label{eq: Shear stress interior}
S \propto |\Darcy \nabla \Prest| \sim 12 \Darcy (1 + A) |G| - \eps \Reyo \Darcy \dfrac{G}{|G|} \displaystyle\sum_{n=0}^\infty (n+1) b_{n + 1} (t) r^{n} \cos n \theta.
\end{align}

We can see from \eqref{eq: Shear stress interior} that the spatial variation of shear stress through the obstacle occurs at $\order{\eps \Reyo \Darcy}$. We show that there can be significant spatial variation of shear stress within the obstacle (figure \ref{fig: internal stress 3D}). From the coefficients $b_n(t)$, given in Appendix \ref{sec: Flow coefficients}, we can deduce that the spatial variation is symmetric across $x = 0$ when $\int_0^T \! G|G| \, \mathrm{d}t = 0$. Thus, all single-mode oscillations will cause an internal shear stress which is symmetric across $x = 0$ at $\order{\eps \Reyo}$, for example $G(t) = (\cos t)/\sqrt{\pi}$ (figure \ref{fig: internal stress 3D}a), but this will not be the case for more complicated forcing functions, such as $G(t) = (\cos t - \cos 2 t)/\sqrt{2\pi}$ (figure \ref{fig: internal stress 3D}b). The prefactors of these functions are chosen such that $\int_0^{2\pi} \! G^2(t) \, \mathrm{d}t = 1$. We see that there are apparent singularities in the internal shear stress at $(x,y) = (0, \pm 1)$, these are due to the logarithmic singularities that we have previously discussed. Although the shear stress will increase near these points, their singular nature is unphysical and the shear stress will be bounded; we expect that a formal realization of this could be achieved by considering boundary regions near these points if required.
  
\begin{figure}
\centering
\includegraphics[width=\textwidth]{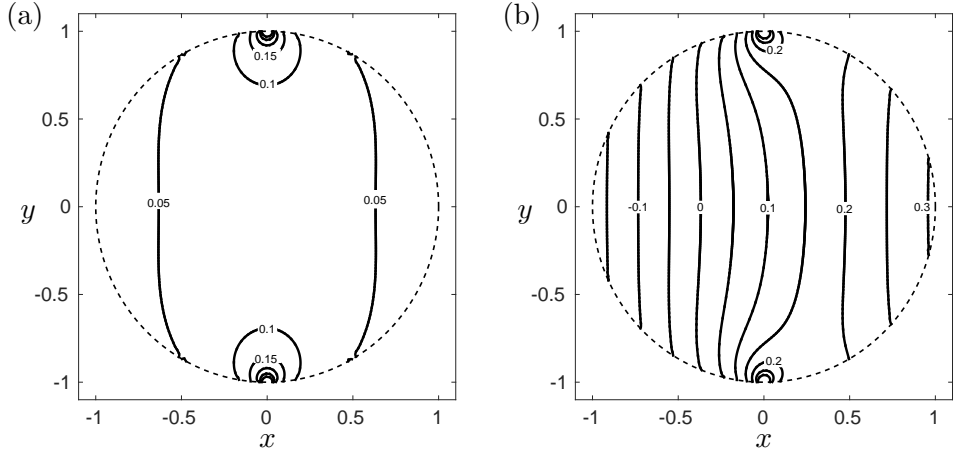}
\caption{The spatial variation of the Darcy velocity averaged over one temporal oscillation of the far-field, given by the $\order{\eps \Reyo \Darcy}$ term in \eqref{eq: Shear stress interior}. This term is proportional to the shear stress experienced within the porous obstacle. We use far-field forcing functions of (a) $G(t) = \cos(t)/\sqrt{\pi}$, (b) $G(t) = (\cos t - \cos 2t)/\sqrt{2 \pi}$. The functions are chosen such that $\int_0^{2\pi} \! G^2(t) \, \mathrm{d}t = 1$. In both figures, we take $\Darcy = 0.01$, we use contour spacings of $0.05$ and, as there are logarithmic singularities at $(x,y) = (0, \pm 1)$, we do not plot contours of values larger than $0.3$. We see that the symmetry exhibited in (a) is not present in (b). This is due to the effect of fluid inertia.}
\label{fig: internal stress 3D}
\end{figure}

\begin{figure}
\centering
\includegraphics[width=0.8\textwidth]{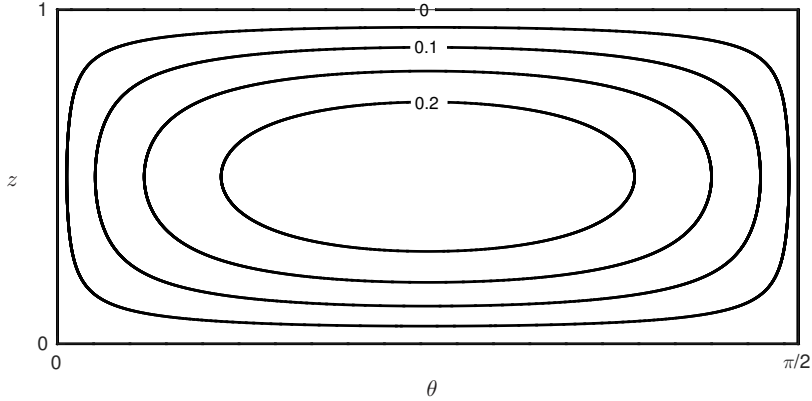}
\caption{A contour plot of the function $z(1-z)  |\sin 2 \theta|^{1/2}$, showing the shape of erosion in the region $(\theta,z) \in [0, \pi/2] \times [0, 1]$. This shape is repeated in each quadrant of the circular cylinder interface, but the magnitude will vary, as discussed in the text. We use contour spacings of $0.05$.}
\label{fig: interface stress 3D}
\end{figure}

We can also apply the results of our boundary layer analysis to predict the erosion of a porous biofilm, by calculating the shear stress acting on the interface. The square root of this shear stress is proportional to the interfacial erosion due to the flow \citep{duddu2009two}. Using the flow solution \eqref{eq: LO flow sol} in \eqref{eq: stress for large Rey}, we see that the leading-order shear stress, $\bs{\tau}$, is
\begin{align}
\label{eq: Shear stress example}
\bs{\tau} = 216 \left(\eps \Reyo \Darcy\right) H(\Prest_{0r}) \left( (1+A) G(t)z(1-z)\right)^2 \sin 2 \theta \, \bs{t},
\end{align}
where $\bs{t} = \bs{e}_z \times \bs{n}$ is the tangential vector on the interface in the direction of increasing $\theta$. The Heaviside step function in \eqref{eq: Shear stress example} means that $\bs{\tau}$ vanishes for $\theta \in (\pi/2, 3\pi/2)$ when $G<0$, and for $\theta \in (-\pi/2, \pi/2)$ when $G>0$. For a periodic far-field forcing with zero mean, the temporal average over one oscillation of the square root of the magnitude of shear stress is
%\begin{subequations}
%\begin{align}
%\label{eq: erosion}
%\dfrac{1}{T}\int_0^T \! |\bs{\tau}|^{1/2} \, \mathrm{d}t &=  \Gamma z(1-z)  |\sin 2 \theta|^{1/2}, \\
%\Gamma &= \left(\dfrac{3 \eps \Reyo \Darcy}{2}\right)^{1/2} 12 (1+A) \overline{G}, \\
%T \overline{G} &= \begin{cases}
%\int_0^T \! |G| H(G(t)) \, \mathrm{d}t \quad \text{for } \theta \in (\pi/2, 3\pi/2),\\
%\int_0^T \! |G| H(-G(t)) \, \mathrm{d}t \quad \text{for } \theta \in (-\pi/2, \pi/2),
%\end{cases}
%\end{align}
%\end{subequations}
\begin{subequations}
\begin{align}
\label{eq: erosion}
\dfrac{1}{T}\int_0^T \! |\bs{\tau}|^{1/2} \, \mathrm{d}t &=  \Gamma z(1-z)  |\sin 2 \theta|^{1/2}, \\
\label{eq: Gamma erosion}
\Gamma &= \left(216 \left(\eps \Reyo \Darcy\right) \right)^{1/2} \dfrac{1+A}{T} \begin{cases}
\int_0^T \! |G| H(G(t)) \, \mathrm{d}t \quad \text{for } \theta \in (\pi/2, 3\pi/2),\\
\int_0^T \! |G| H(-G(t)) \, \mathrm{d}t \quad \text{for } \theta \in (-\pi/2, \pi/2),
\end{cases}
\end{align}
\end{subequations}
Here, \eqref{eq: erosion} is proportional to the initial biofilm erosion. Thus, we have deduced that this erosion has a shape of $z(1-z)|\sin 2 \theta|^{1/2}$, repeated in each quadrant of the interface. The magnitude of this erosion will be the product of $\Gamma$ and the constant of proportionality linking erosion to the square root of shear stress. From the form of $\Gamma$, given in \eqref{eq: Gamma erosion}, we see that the magnitude of erosion is not necessarily symmetric across $x = 0$ and the asymmetry will depend on $G(t)$. The interfacial erosion is locally maximal at $(\theta,z) = (\pi(1 +  2n)/4,0.5)$ for $n \in \mathbb{Z}$ and, in each quadrant, decreases from these points of maxima with two lines of symmetry, one with constant $z$ and the other with constant $\theta$ (figure \ref{fig: interface stress 3D}).

%We note that \eqref{eq: net force} has a more aesthetic form at leading order in $\Darcy \ll 1$, given by
%\begin{align}
%\label{eq: net force K 0}
%\int_0^T \! \bs{F}(t) \, \mathrm{d}t \sim \dfrac{192}{5} \left(\eps \Reyo \Darcy \right) \left(1  + 10 \Lambda_a\right) \left(\int_0^T \! G(t)|G(t)| \, \mathrm{d}t \right) \bs{e}_x.
%\end{align}

%Thus, for example, if $G(t) = \cos 2 \pi t$ with $T = 1$, the net force is given by
%\begin{align}
%\int_0^1 \! \bs{F}(t) \, \mathrm{d}t \sim \dfrac{96 \eps \Reyo \Darcy}{5} \bs{e}_x.
%\end{align}

\section{Discussion}
\label{sec: Discussion}

We make extensive use of the method of matched asymptotic expansions to investigate the laminar flow around and through a porous obstacle in both a channel and a Hele-Shaw cell. In particular, we determine how the flow behaves close to an interface (in a region whose width is of the order of the small aspect ratio) between single-phase and porous flows (governed by the Navier--Stokes and Darcy equations, respectively) for both small and large Reynolds numbers.

Our analysis allows us to resolve everywhere the leading-order fluid velocity, a necessary condition to start investigating the nutrient transport in the system, with the eventual goal to optimize nutrient delivery to cells within the porous obstacle. Furthermore, we obtain important characteristics of the inner flow (in a sense made formal within the main text), such as the interfacial stress, and to determine suitable conditions to couple the (simpler) outer flows. The analytical approach reveals the dependence of the flow on the Reynolds number and the dimensionless permeability, without resorting to numerically expensive parameter sweeps for an inherently nonlinear problem. Importantly, we note that any variation of the interfacial stress in the direction transverse to the channel/cell cannot be determined solely from the outer solutions, and requires an in-depth boundary layer analysis. In particular, we must continue past the leading-order terms in the asymptotic expansions to obtain the relevant contributions to the stress. 

In the first part of this paper we consider two-dimensional steady flow through a channel fully blocked by a finite length porous obstacle and, in particular, examine the regions near the interface between single-phase and porous flow. These inner regions satisfy the full two-dimensional Navier--Stokes equations, and we calculate the pressure and flux jumps between outer regions in the asymptotic limits of small and large Reynolds number.

Whilst we include results for the small-Reynolds-number limit for comparison and completeness, the case which is more physically relevant to tissue engineering applications is the large-Reynolds-number limit. In this paper, we show that the latter case also exhibits more interesting flow behaviour, and we unveil a rich boundary-layer structure. The leading-order (in terms of the large Reynolds number) transition between Poiseuille and plug flow occurs within the porous medium for inflow, and outside the porous medium for outflow. Therefore, the boundary layer structure is different for inflow and outflow, reflecting the irreversible nature of the large Reynolds number flow. We determined that the pressure jump was an order of the Reynolds number larger in magnitude for outflow than for inflow. There is no flux jump between outer regions, due to the restriction of fluid movement to a two-dimensional plane.

In the second part of this paper we extend the two-dimensional work to consider unsteady three-dimensional flow in a Hele-Shaw cell containing a cylindrical porous obstacle (which fits tightly between the parallel planar plates of the cell) whose cross-sectional boundary is smooth. Due to the smooth boundary, the flow behaviour and boundary layer structure close to the interface are similar to the two-dimensional channel flow considered previously and we can reduce much of the three-dimensional problem to the two-dimensional problem considered in the first part of the paper. However, in contrast to the two-dimensional channel flow case, a small flux jump between the outer flows is now present, as the tangential flow allows a small amount of fluid to move around, instead of through, the porous obstacle. We summarise the main results of the three-dimensional problem in~\S\ref{sec: 3D summary}, including the results for the interfacial stress and the derived conditions required to couple the outer flows.

In \S\ref{sec: Example} we apply the general results to a three-dimensional cylinder with a circular cross-section in a flow with a far-field forcing. From this, we are able to make several physical predictions. We deduce that, due to the fluid inertia, a periodic forcing with zero mean could cause a drift on the cylinder were it free to move. This is relevant to the tissue engineering application we discussed in \S\ref{sec: intro}, where it is important to control the movement of the porous obstacle, which has implications for the delivery of nutrients to cells within the construct. Moreover, knowledge of this long-term drift may be helpful in flushing blockages from pipes. We also calculate the spatial variation in internal shear stress through the obstacle. As cell growth can be coupled to shear stress, this has important implications for cell placement in tissue engineering applications. We show that the spatial variation in shear stress is symmetric through the obstacle for a single-mode far-field oscillatory forcing, but can break symmetry for multiple-mode far-field oscillatory forcings. We are also able to determine the interfacial stress and use this result to obtain the initial interfacial erosion of a porous biofilm, assuming that interfacial erosion is proportional to the square root of the modulus of shear stress with a given constant of proportionality.

In the large-Reynolds-number limit, we apply three different tangential slip conditions on the interface to see the effect of varying boundary condition. We find that any of the three tangential slip conditions we apply only produce the same leading-order coupling conditions for inflow. Thus, knowledge of the tangential velocity condition is important for inflow if local information about, say, the interfacial stress were required, and it is important for outflow if global information about the outer problem is required. Additionally, we note that although we imposed continuity of pressure on the full problem, the coupling condition~\eqref{eq: pressure BC LO} would remain the same if we had imposed continuity of stress instead. This is because the viscous term in the normal stress does not contribute to the stress at leading order. The higher order pressure coupling condition \eqref{eq: pressure BC FO} would remain unchanged in the large-Reynolds-number limit for the same reason, but would be different in the small-Reynolds-number limit.

In this work, we have considered a porous medium within which the flow is governed by Darcy's equations. We restrict ourselves to these governing equations because the tissue engineering application we are interested in requires a porous scaffold which maintains its structural integrity whilst moving within the flow. Thus, at the very least, the underlying solid matrix of which the porous medium consists must be connected, and Brinkman's equations will not apply \citep{auriault2009domain}. However, there may be applications which are not as restrictive, where Brinkman's equations may apply. In such cases, the coupling conditions for the outer problem would be different to those derived in this paper, and a new analysis would be necessary to determine these. Whilst the issue of coupling the plain and porous flow regions is numerically simpler (as Brinkman's equations are able to be applied to both plain and porous fluid domains), the presence of the Laplacian viscous term means that it is more difficult to make analytic progress with the resulting partial differential equations. For example, the boundary layer structure may be significantly more complex within the porous medium.

Although we have considered a porous obstacle whose interfacial boundary is normal to the channel or bioreactor walls, this may not be the case. We show in \S\ref{sec: Example} that erosion of an initially straight wall is not uniform in any tangent direction to the interface. More generally, manufacturing constraints could cause the porous insert to have non-straight walls.  Such imperfections will only change the flow problem if the gradient of the walls is significant in one of the boundary layers. If this occurs, the change in problem geometry yields a complicated extension to the problem we have presented. However, this extension is vital if the full dynamics of erosion are to be considered in, for example, the shape evolution of a biofilm.

The tissue engineering experiment discussed in \S\ref{sec: intro} involves a moving porous obstacle, which is possible as there are small gaps between the flat sides of the porous obstacle and the flat sides of the bioreactor (figure~\ref{fig: HARV_schematic}). Whilst our work has considered unsteady flow, we have restricted ourselves to a pinned obstacle in this paper, and neglected the small gaps. To consider the dynamic problem fully, we must investigate the role of these gaps. The presence of gaps would have no effect on the flow up to the asymptotic orders considered if these gaps were smaller than the size of the boundary layers presented~\citep{Dalwadi2014Thesis}. However, a gap height of the same order as the width of one of these boundary layers would significantly complicate the boundary layer problem, as a result of the significant change in the geometry of one or more domains in which we solve various submodels arising in the boundary layer analysis. Knowledge of the work presented here allows the porous obstacle trajectory to be calculated once the effect of the small gaps has been calculated. As the obstacle movement would affect the flow, this would allow a deeper understanding of the dynamics of nutrient transport and, ultimately, tissue growth.

The analytic nature of our results can significantly reduce the numerical expense of solving such dynamical problems. This is because the coupling conditions and stress components that we have determined are all in terms of the outer variables. Hence, we have reduced a nonlinear three-dimensional problem to a linear two-dimensional problem. More generally, this work highlights how the exploitation of an underlying separation of scales can significantly reduce the computational complexity of a physical problem.

%Once we understand the effect of the gaps on the flow, we are able to consider the coupled problem where the porous obstacle moves, and we are able to extend this work to consider Hele-Shaw cell walls moving in rigid-body translation, rotation, or a combination of both~\citep{Dalwadi2014Thesis}. In \S4, we determine the stress acting on the interface of the porous obstacle. Were the obstacle not pinned, this information could be used in conjunction with the moving cell wall results to determine an equation of motion for the obstacle. As the obstacle movement would affect the flow, this would allow a deeper understanding into the dynamics of nutrient transport and, ultimately, tissue growth. In particular, the analytic nature of these results can reduce the numerical expense involved in implementing the relevant dynamic mechanics and, in general, this work highlights how the exploitation of an underlying separation of scales can significantly reduce the computational complexity of a physical problem.

\section*{Acknowledgements}
This work was carried out whilst M.P.D. held a Doctoral Training Grant funded by the Engineering and Physical Sciences Research Council at the Mathematical Institute, University of Oxford.

\appendix

\section{Additional inflow regions}
\label{sec: Inflow app}
%\subsection{Region $\mathrm{IIb}$}
%\label{sec: Smaller wall boundary layer}

In this Appendix we resolve the additional boundary layers for inflow. We first introduce a boundary layer near $z=0$ to resolve the issue of the $\order{\epsb}$ slip velocity induced on the channel wall in \region{II} in \S\ref{sec: Regions II and III}. We call this boundary layer \region{IIb}. We use $X^{-} = X$ for convenience, and note that there will be a similar boundary layer near $z=1$.

%The classical Prandtl result for high Reynolds number flow past a flat plate is that a boundary layer of height $\order{\epsb^{1/2}}$ is formed near the plate, which grows proportionally to the square root of the distance from the plate tip. This may seem like the canonical example from which to proceed (and, indeed, it plays a role in \region{VIb} for outflow as we discuss in \S\ref{sec: Outflow}, and in Appendix~\ref{sec: regions VIb and VIc}), but there are complications in our problem which mean that a different scaling and solution near the boundary are required. These complications stem from the fact that the induced slip velocity $\ui_0 = \order{\epsb}$, and that the leading-order Poiseuille flow is linear in $z$ near the wall. Therefore, the correct scaling comes from looking for distinguished limits when $\order{\ui_0} = \order{z}$ close to the wall, and then scaling $\wi_0$ such that we have a non-trivial incompressibility condition. The problem of induced slip is then resolved at a higher order.
 
Before we proceed, it will be useful to analyse the fluid velocities $\ui_{01}$ and $\wi_{01}$ in \region{II} in more detail. In the separable solutions to the governing equations~\eqref{eq: High Rey Inner inflow bulk 2O} and \eqref{eq: wi_01 governing equation}, the fluid velocities $\ui_{01}$ and $\wi_{01}$ are given by
\refstepcounter{equation}
\[
\term{\ui}{01} = \displaystyle\sum_{k=1}^{\infty} C_k \exp(\eigvalII X^{-}) g'_k(z), \quad \term{\wi}{01} = -\displaystyle\sum_{k=1}^{\infty} \eigvalII C_k \exp(\eigvalII X^{-}) g_k(z),
\eqno{(\theequation{\mathit{a},\mathit{b}})}\label{eq: Region II velocities at epsa}
\]
where the eigenfunctions $g_k(z)$ satisfy the eigenvalue problem
\begin{align}
\label{eq: SL eig val problem}
g''_k(z) + \left(\dfrac{2}{z(1-z)} + \eigvalII^2\right)g_k(z) = 0 \quad \text{for } z \in (0,1)
\end{align}
together with boundary conditions $g_k(0) = g_k(1) = 0$ (obtained from the no-slip channel wall boundary conditions given in figure~\ref{fig: inflow leading order}). The coefficients $C_k$ would have to be determined numerically using the orthogonality of the eigenfunctions.
%, and we note that $\eigvalII$ are all real as~\eqref{eq: SL eig val problem} is in Sturm-Liouville form.

%We note that the eigenvalue problem~\eqref{eq: SL eig val problem} is in Sturm-Liouville form. However, as there are singularities in the coefficients within~\eqref{eq: SL eig val problem} at the end-points of the domain, we cannot blindly apply the standard results of Sturm-Liouville theory. In Appendix~\ref{sec: modified SL theory}, we modify certain Sturm-Liouville results to determine that $\eigvalII^2 \geqslant 0$ in general, and that $\eigvalII^2 >  0$ for the expansion used in~\eqref{eq: Region II velocities at epsa}. We are then able to use the complete spectrum of positive roots in~\eqref{eq: Region II velocities at epsa}. We now continue with the solution in \region{IIb}.

The relevant scalings are $\ui_{0} = \epsb^{1/3} \uiwall$, $\wi_0 = \epsb^{4/3} \wiwall$, and $z = \epsb^{1/3} Z$, and we introduce the new variable $\pri_1= \priwall$ to signify that we are working in \region{IIb}. Under these scalings, the bulk flow governing equations in Figure~\ref{fig: inflow leading order} become
\begin{subequations}
\label{eq: Inflow channel BL near wall equations}
\begin{alignat}{5}
\label{eq: Inflow channel x-dir BL near wall}
\uiwall \uiwall_X + \epsb^{2/3} \wiwall \uiwall_Z &= - \epsb^{1/3} \priwall_{X} + \uiwall_{ZZ} + \epsb^{2/3} \uiwall_{XX}, \\
\label{eq: Inflow channel z-dir BL near wall}
\epsb \uiwall \wiwall_X + \epsb^{5/3}\wiwall \wiwall_Z &= - \priwall_{Z} + \epsb \wiwall_{ZZ} + \epsb^{5/3} \wiwall_{XX}, \\
\label{eq: Inflow channel com BL near wall}
0 &= \uiwall_{X} + \epsb^{2/3} \wiwall_{Z}.
\end{alignat}
\end{subequations}
The boundary conditions on the channel wall are
\begin{align}
\label{eq: wall BC region IIb}
\bs{\uiwall} = 0 \quad \text{on } Z = 0 \text{ for } X < 0.
\end{align}

We use asymptotic expansions
\begin{subequations}
\label{eq: asymptotic exp for region IIb}
\begin{align}
\uiwall &\sim 6Z - \epsb^{1/3} 6 Z^2 + \epsb^{2/3} \term{\uiwall}{01} + \order{\epsb}, \\
\wiwall &\sim \term{\wiwall}{02} + \order{\epsb}, \\
\priwall &\sim -12X + \epsb^{1/3} \term{\priwall}{12} + \order{\epsb^{2/3}}
\end{align}
\end{subequations}
as $\epsb \ttz$; substituting into the bulk flow equations~\eqref{eq: Inflow channel BL near wall equations} yields
\begin{subequations}
\label{eq: Inflow channel BL near wall 3}
\begin{alignat}{5}
\label{eq: Inflow channel x-dir BL near wall 3}
6 \left(Z \term{\uiwall}{01X} + \term{\wiwall}{02}\right) &= -\term{\priwall}{12X} +  \term{\uiwall}{01ZZ}, \\
\label{eq: Inflow channel z-dir BL near wall 3}
0 &= -\term{\priwall}{12Z}, \\
\label{eq: Inflow channel com BL near wall 3}
0 &= \term{\uiwall}{01X} + \term{\wiwall}{02Z}.
\end{alignat}
\end{subequations}
The channel wall boundary condition~\eqref{eq: wall BC region IIb} is given by
\begin{align}
\label{eq: wall BC region IIb 3}
\uiwall_{01} =  \wiwall_{02} = 0 \quad \text{on } Z = 0 \text{ for } X < 0,
\end{align}
and, from the general form of the flow solution in \region{II}, \eqref{eq: Region II velocities at epsa}, the matching condition with the horizontal velocity in \region{II} is given by
\begin{align}
\label{eq: Matching conditions in region IIb 3}
\term{\uiwall}{01} \sim \displaystyle\sum_{k=1}^{\infty} C_k \exp(\eigvalII X) g'_k(0), \quad \text{as } Z \to \infty.
\end{align}
%The solution to this system imposes the form of $\wiwall_{02}$ and $\priwall_{12}$ as $z \tti$ and gives the next-order matching condition with \region{II} (though these are not required further in this paper).

We seek solutions of the system~\eqref{eq: Inflow channel BL near wall 3}--\eqref{eq: Matching conditions in region IIb 3} in the form
\begin{align}
\label{eq: SFwall general form}
(\uiwall_{01}, \wiwall_{02}, \priwall_{12X}) = \displaystyle\sum_{k=1}^{\infty} \exp(\eigvalII X) \left(f'_k(Z), -\eigvalII f_k(Z), b_k\right),
\end{align}
%\refstepcounter{equation}
%\[
%\uiwall_{01} = \displaystyle\sum_{k=1}^{\infty} \exp(\eigvalII X) f'_k(Z), \quad \wiwall_{02} = -\displaystyle\sum_{k=1}^{\infty} \eigvalII \exp(\eigvalII X) f_k(Z), \quad \priwall_{12X} = \displaystyle\sum_{k=1}^{\infty} b_k \exp(\eigvalII X),
%\eqno{(\theequation{\mathit{a}-\mathit{c}})}\label{eq: SFwall general form}
%\]
which automatically satisfies~\eqref{eq: Inflow channel z-dir BL near wall 3}, \eqref{eq: Inflow channel com BL near wall 3}, and where $b_k$ and $f_k(Z)$ are to be determined. Substituting~\eqref{eq: SFwall general form} into~\eqref{eq: Inflow channel x-dir BL near wall 3}, we obtain the differential equation
\begin{align}
\label{eq: Inflow channel SF BL near wall}
f'''_k(Z) + a_k^3 \left(Z f'_k(Z) - f_k(Z)  \right)= b_k,
\end{align}
for $f_k(Z)$, where $a_k^3 = 6 \eigvalII > 0$, and with boundary conditions
\refstepcounter{equation}
\[
f_k(0) = 0, \quad f'_k(0) = 0, \quad f'_k(\infty) = C_k g'_k(0).
\eqno{(\theequation{\mathit{a}-\mathit{c}})}\label{eq: BC for reduced problem in region II**}
\]
The system~\eqref{eq: Inflow channel SF BL near wall}--\eqref{eq: BC for reduced problem in region II**} is solved by
\begin{align}
\label{eq: Airy solution}
f_k(Z) = C_k g'_k(0) \dfrac{\int_0^Z \! \int_0^u \! \Ai\left(a_k v\right) \, \mathrm{d}v \, \mathrm{d}u}{\int_0^\infty \Ai\left(a_k s\right) \, \mathrm{d}s}.
\end{align}
where $\Ai$ is the Airy function of the first kind. To determine $b_k$, note that the general solution for~\eqref{eq: Inflow channel SF BL near wall} is
\begin{align}
\label{eq: sol to ODE to region II** with pressure involved}
f_k(Z) = \dfrac{b_k}{a_k^3} + B_{k1} Z + B_{k2} h_1(Z) + B_{k3} h_2(Z),
\end{align}
where the far-field behaviour of $h_1$ and $h_2$ as $Z \to \infty$ is
\refstepcounter{equation}
\[
h_1 \sim Z^{-5/4} \exp\left(\dfrac{2}{3}\left(a_k Z\right)^{3/2}\right), \quad h_2 \sim Z^{-5/4} \exp\left(-\dfrac{2}{3}\left(a_k Z\right)^{3/2}\right) \quad \text{as } Z \tti.
\eqno{(\theequation{\mathit{a},\mathit{b}})}\label{eq: large Z behaviour for h_1 and h_2}
\]
Applying the boundary condition~(\ref{eq: BC for reduced problem in region II**}c) imposes $B_{k1} = C_k g_k'(0)$ and $B_{k2} = 0$. The remaining boundary conditions~(\ref{eq: BC for reduced problem in region II**}a--b) yield the solution for $b_k$ and $B_{k3}$, but the simplest route to determine $b_k$ is to equate~\eqref{eq: sol to ODE to region II** with pressure involved} with~\eqref{eq: Airy solution} as $Z \tti$ to deduce that
\begin{align}
b_k = - 3 a_k^4 C_k g'_k(0) \int\displaylimits_0^\infty \! \int\displaylimits_u^\infty \! \Ai\left(a_k v\right) \, \mathrm{d}v \, \mathrm{d}u,
\end{align}
where we have used the identity $\int_0^\infty \! \Ai(a_k v) \, \mathrm{d} v = (3 a_k)^{-1}$.

We are unable to impose any conditions on the interfacial boundary $X = 0$ within \region{IIb}. This issue is resolved by another boundary layer near $X=0$, which we call \region{IIc}. The scalings from \region{II} to \region{IIc} are $\ui_{0} = \order{\epsb^{1/3}}$, $\wi_0 = \order{\epsb^{4/3}}$, $\pri_1 = \order{\epsb^{1/3}}$, $X = \order{\epsb^{2/3}}$, and $z = \order{\epsb^{1/3}}$. The leading-order equations yield a solution of $\ui_0 \sim 6z \tanh (3z(-X)/\epsb)$, from which $\wi_0$ and $\pri_1$ can be determined.

\section{Additional outflow regions}
\subsection{Regions $\mathrm{VIb}$ and $\mathrm{VIc}$}
\label{sec: regions VIb and VIc}

In this section we solve for the flow near the outflow channel wall (to match with the asymptotic regions considered in two (\S\ref{sec: Region VI}) and three-dimensions (\S\ref{sec: Outflow 3D})), in a region we denote \region{VIb}. For the two-dimensional channel flow problem, this boundary layer is equivalent to the classic uniform high Reynolds number flow past a flat plate considered by~\citet{prandtl1904uber}. We therefore consider the three-dimensional problem, which uses the two-dimensional results. We use $X^{+} = X$ for convenience.

% and \region{VIc} (as shown in Figure~\ref{fig: outflow BL}). The work in this section is a classic result in boundary layer theory, first outlined in~\citet{prandtl1904uber}, and is an excellent candidate for the birth of singular perturbation theory. We emphasise that the work in this section is not original, it is provided for reference and for matching purposes. As a result, we are fairly brief with the details. We use $X^{+} = X$ for convenience.

The relevant scalings from \region{VI} to \region{VIb} are: $z = \epsbt^{1/2} \zsixb$, $\uith_0 = \epsbt \vsixb$, and $\wi_0 = \epsbt^{1/2} \wsixb$. We also use $\uir_0 = \usixb$, and $\pri_1 = \psixb$ to signify that we are working in \region{VIb}.

With these scalings, the bulk flow governing equations given in Figure~\ref{fig: outflow leading order} and~\eqref{eq: scaled out uith} become
\begin{subequations}
\label{eq: region VIb gov eq}
\begin{alignat}{5}
\usixb \usixb_{X} + \wsixb \usixb_{\zsixb} &= - \epsbt \psixb_{X} + \usixb_{\zsixb \zsixb} + \epsbt \usixb_{XX}, \\
\usixb \vsixb_{X} + \wsixb \vsixb_{\zsixb} &= 12 + \vsixb_{\zsixb \zsixb} + \epsbt \vsixb_{XX}, \\
\usixb \wsixb_{X} + \wsixb \wsixb_{\zsixb} &= - \psixb_{\zsixb} + \wsixb_{\zsixb \zsixb} + \epsbt \wsixb_{XX}, \\
0 &= \usixb_{X} + \wsixb_{\zsixb}.
\end{alignat}
\end{subequations}
Substituting the asymptotic expansions
\refstepcounter{equation}
\[
\usixb = \usixb_{0} + \order{\epsbt^{1/2}}, \quad
\vsixb = \vsixb_{0} + \order{\epsbt^{1/2}}, \quad
\wsixb = \wsixb_{0} + \order{\epsbt^{1/2}}, \quad
\psixb = \psixb_{10} + \order{\epsbt^{1/2}}
\eqno{(\theequation{\mathit{a}-\mathit{d}})}\label{eq: asymptotic exp for region VIb}
\]
as $\epsbt \ttz$, into the bulk-flow equations~\eqref{eq: region VIb gov eq}, we obtain the following leading-order equations
\begin{subequations}
\label{eq: region VIb gov eq LO}
\begin{align}
\label{eq: region VIb gov eq x-mom LO}
\usixb_0 \usixb_{0X} + \wsixb_0 \usixb_{0\zsixb} &= -\psixb_{10X} + \usixb_{0\zsixb \zsixb}, \\
\usixb_0 \vsixb_{0X} + \wsixb_0 \vsixb_{0\zsixb} &= 12 + \vsixb_{0\zsixb \zsixb}, \\
\label{eq: region VIb gov eq y-mom LO}
0 &= - \psixb_{10 \zsixb}, \\
\label{eq: region VIb gov eq cont LO}
0 &= \usixb_{0X} + \wsixb_{0\zsixb}.
\end{align}
\end{subequations}
The boundary conditions on the channel wall $\zsixb = 0$ are given by
\begin{align}
\label{eq: region VIb channel wall}
\bs{\usixb}_0 = \bs{0} \quad  \text{on } \zsixb = 0 \text{ for } X > 0,
\end{align}
%\refstepcounter{equation}
%\[
%\usixb_0 = 0, \quad \vsixb_0 = 0, \quad \wsixb_0 = 0, \quad  \text{on } \zsixb = 0 \text{ for } X > 0,
%\eqno{(\theequation{\mathit{a}-\mathit{c}})}\label{eq: region VIb channel wall}
%\]
and the matching conditions into \region{VI} are given by
\refstepcounter{equation}
\[
\usixb_0 \sim 1, \quad \vsixb_0 \sim 12 X, \quad \pri_{10} \to 0, \quad  \text{as } \zsixb \to \infty \text{ for } X > 0.
\eqno{(\theequation{\mathit{a}-\mathit{c}})}\label{eq: region VIb matching}
\]
The system~\eqref{eq: region VIb gov eq LO}--\eqref{eq: region VIb matching} is solved by
\refstepcounter{equation}
\[
\usixb_0 = f'(\eta), \quad \vsixb_0 = 12 X h(\eta), \quad \wsixb_{0} = \dfrac{\eta f'(\eta) - f(\eta)}{2X^{1/2}}, \quad \pri_{10} = 0,
\eqno{(\theequation{\mathit{a}-\mathit{d}})}\label{eq: region VIb solution}
\]
where $\eta = \zsixb X^{-1/2}$ is the similarity variable. The function $f(\eta)$ satisfies the following Blasius ODE:
\begin{align}
\label{eq: Blasius equation}
f'''(\eta) + \dfrac{1}{2}f(\eta) f''(\eta) = 0, \quad f(0) = f'(0) = 0, \quad f'(\infty) = 1;
\end{align}
the function $h(\eta)$ satisfies the following ODE
\begin{align}
h''(\eta) + \dfrac{1}{2} f(\eta)h'(\eta) - f'(\eta) h(\eta)  =  -1, \quad h(0) = 0, \quad h(\infty) = 1.
\end{align}
%\begin{align}
%h''(\eta) + \dfrac{1}{2}f(\eta)h'(\eta) = 0, \quad h(0) = 0, \quad h(\infty) = 1,
%\end{align}
%and is solved by
%\begin{align}
%\label{eq: g(eta) the sol for v in VIb}
%h(\eta) = \dfrac{\int_0^\eta \exp\left(-\frac{1}{2}\int_0^\tau f(s) \, \mathrm{d}s\right) \,\mathrm{d}\tau}{\int_0^\infty \exp\left(-\frac{1}{2}\int_0^\tau f(s) \, \mathrm{d}s\right) \,\mathrm{d}\tau},
%\end{align}

The function $f$ is well studied (see, \eg \citet{boyd2008blasius}), and its behaviour up to exponentially small terms as $\eta \to \infty$ is given by
\begin{align}
\label{eq: Far field behaviour of Blasius function}
f \sim \eta - \leak \quad \text{as } \eta \to \infty,
\end{align}
where $\leak \approx 1.721$. This far-field behaviour, in conjunction with the solution for $\wsixb_0$ in~(\ref{eq: region VIb solution}c), yields the matching conditions~\eqref{eq: no flux or slip BC inner outflow 2O}.

Finally, we note that the scalings from \region{VI} to \region{VIc} are: $z\sim \epsbt$, $X^{+} \sim \epsbt$, $\vi_0 \sim \epsbt^2$ and $\prit_1 \sim \epsbt^{-1}$, thus yielding the full two-dimensional Navier-Stokes equations for $\ui_0$ and $\wi_0$, and the full version of \eqref{eq: scaled out uith} for $\vi_0$. That is, taking $\Rey \equiv 1$ in \eqref{eq: scaled out uith}. We do not consider \region{VIc} further.

\subsection{Matching into \region{VIa}}
\label{sec: Small X analysis in regionVIa}

In this section we form the composite expansion between regions $\mathrm{VI}$ and $\mathrm{VIb}$ in their far-field as $X^{+} \tti$, and thereby obtain the matching conditions~\eqref{eq: Composite exp for vel near region VIa entrance text} and~\eqref{eq: v in VIa near 0} with \region{VIa}. When discussing the abscissa in \region{VI}, we use $X^{+} = X$ for convenience.

Combining the leading-order versions of~(\ref{eq: region VI asymptotic expansions}a) and \eqref{eq: ae v in inflow inner}, together with the far-field first-correction terms \eqref{eq: w outflow inner region matching condition}--\eqref{eq:  region VI to VII matching conditions 2O}, we obtain the \region{V} far-field expansions (as $X \tti$) up to $\order{\epsbt^{1/2}}$ for the flow, and up to $\order{\epsbt^{-1/2}}$ for the pressure as follows
\refstepcounter{equation}
\[
\ui_{0} \sim 1 + 2 \leak \left(\epsbt X\right)^{1/2}, \quad \vi_0 \sim 12 \left(\epsbt X \right), \quad \wi_{0} \sim \dfrac{\leak \left(1-2z \right)}{2 (X/\epsbt)^{1/2}}, \quad  \term{\pri}{1} \sim -2\leak (X/\epsbt)^{1/2}.
\eqno{(\theequation{\mathit{a}-\mathit{d}})}\label{eq: region VI to VII far-field}
\]

The matching conditions are obtained as follows. We first form the multiplicative composite expansion of~\eqref{eq: region VI to VII far-field} with the solution near the channel wall at $z = 0$ (given by~\eqref{eq: region VIb solution}), and the equivalent solution near the channel wall at $z=1$ (obtained by taking $\eta = (1-z) (\epsbt X)^{-1/2}$ and reversing the sign on $\wi_0$). Then, we write these expansions in terms of $\Rii = \epsbt X$, and retain the leading-order terms to yield, as $\Rii \downarrow 0$,
\begin{subequations}
\label{eq: Composite exp for vel near region VIa entrance}
\begin{alignat}{5}
\label{eq: Composite exp for u}
\term{\uiii}{0} &\sim \left(1 + 2 \leak \Rii^{1/2}\right)f'\left(\etaa\right)f'\left(\etab\right), \\
\label{eq: uith composite region VI}
\uiiith_{00} &\sim 12 \Rii h\left(\etaa\right)h\left(\etab\right), \\
\label{eq: Composite exp for w}
\term{\wiii}{0} &\sim \dfrac{\left(1 - 2z \right)}{2 \leak \Rii^{1/2}}\left(\etaa f'\left(\etaa\right)- f\left(\etaa\right)\right)\left(\etab f'\left(\etab\right) - f\left(\etab\right)\right), \\
\label{eq: Pressure for region VIa entrance}
\priii_{10} &\sim -2 \leak \Rii^{1/2},
\end{alignat}
\end{subequations}
where $\etaa = z/\Rii^{1/2}$ and $\etab = (1-z)/\Rii^{1/2}$.

\section{Flux jump condition}
\label{sec: Higher-order average flux}

In this section we determine the flux jump in terms of the tangential velocity, as given in~\eqref{eq: int cont 2O final} by integrating the inner continuity equations~\eqref{eq: BL eq general coord cont} and~\eqref{eq: 3D BL Darcy continuity inner} over the cell height. Using the asymptotic expansions~\eqref{eq: inner expansions 3D BL}, and equating powers of $\eps$, we deduce that
\refstepcounter{equation}
\[
\pbyp{}{\nrm}\int_0^1 \uirt_0(\nrm,\tng,z,t) \, \mathrm{d}z = 0, \quad \pbyp{}{\nrm}\int_0^1 \Uirt_0(\nrm,\tng,z,t) \, \mathrm{d}z = 0,
\eqno{(\theequation{\mathit{a},\mathit{b}})}\label{eq: inner 3D LO cont int}
\]
and
\begin{subequations}
\label{eq: int cont eq 2O}
\begin{align}
\int_0^1 \! \uirt_{1\nrm}(\nrm,\tng,z,t) \, \mathrm{d}z +\curv(\tng) \int_0^1 \! \uirt_0(\nrm,\tng,z,t) \, \mathrm{d}z + \int_0^1 \! \uitht_{0\tng}(\nrm,\tng,z,t) \, \mathrm{d}z &= 0, \\
\int_0^1 \! \Uirt_{1\nrm}(\nrm,\tng,z,t) \, \mathrm{d}z +\curv(\tng) \int_0^1 \! \Uirt_0(\nrm,\tng,z,t) \, \mathrm{d}z + \int_0^1 \! \Uitht_{0\tng}(\nrm,\tng,z,t) \, \mathrm{d}z &= 0,
\end{align}
\end{subequations}
noting from~\eqref{eq: uitht_0 3D}, the scaled solution within the porous obstacle, that $\Uitht_0(\nrm,\tng,z,t) \equiv -\Darcy \Prest_{0 \tng}(0,\tng,t)$.

The matching conditions for $\uitht_0$, $\Uitht_0$, $\uirt_1$ and $\Uirt_1$ are given by
\begin{subequations}
\label{eq: matching 3D 2O}
\begin{alignat}{5}
\label{eq: matching bulk 3D 2O}
\uitht_0 &\sim \bs{t} \bcdot \uut_{0}(0,\tng,z,t) &\quad &\text{as } \nrm \tti,  \\
\Uitht_0 &\sim -\Darcy \Prest_{0 \tng}(0,\tng,t) &\quad &\text{as } \nrm \ttni, \\
\uirt_1 &\sim \bs{n} \bcdot \left(\nrm \uut_{0 \nrmouter}(0,\tng,z,t) + \uut_{1}(0,\tng,z,t)\right) &\quad &\text{as } \nrm \tti, \\
\label{eq: matching porous 3D 2O}
\Uirt_1 &\sim \bs{n} \bcdot \left(\nrm \QQt_{0 \nrmouter}(0,\tng,z,t) + \QQt_{1}(0,\tng,z,t)\right) &\quad &\text{as } \nrm \ttni.
\end{alignat}
\end{subequations}

To formally avoid infinite terms in the limit as $\nrm \tti$ when we integrate with respect to $\nrm$ in \eqref{eq: int cont eq 2O}, it is necessary to subtract the far-field limit of~\eqref{eq: int cont eq 2O} from itself (using~\eqref{eq: inner 3D LO cont int} and \eqref{eq: matching 3D 2O}), yielding
\begin{subequations}
\label{eq: int cont eq 2O before again int}
\begin{align}
\int_0^1 \! \left(\uirt_{1\nrm} - \left. \bs{n} \bcdot \uut_{0 \nrmouter} \right|_{\bdy}\right) \, \mathrm{d}z &= - \int_0^1\! \left(\uitht_{0\tng} - \left. \bs{t} \bcdot \uut_{0\tng} \right|_{\bdy}\right) \, \mathrm{d}z, \\
\int_0^1 \! \left(\Uirt_{1\nrm} - \left. \bs{n} \bcdot \QQt_{0 \nrmouter} \right|_{\bdy}\right) \, \mathrm{d}z &= 0.
\end{align}
\end{subequations}
Integrating~\eqref{eq: int cont eq 2O before again int} with respect to $\nrm$ between $0$ and $\infty$ (using the far-field conditions~\eqref{eq: matching 3D 2O}), we deduce that
\begin{subequations}
\label{eq: int cont eq 2O penult form}
\begin{align}
\label{eq: int cont eq 2O penult bulk}
\int_0^1 \! \left(\left.\bs{n} \bcdot \uut_1\right|_{\bdy} - \left.\uirt_{1}\right|_{\nrm = 0}\right) \, \mathrm{d}z &= - \int_0^\infty \int_0^1 \left(\uitht_{0\tng} - \left.\bs{t} \bcdot \uut_{0\tng}\right|_{\bdy} \right) \, \mathrm{d}z \, \mathrm{d} \nrm, \\
\label{eq: int cont eq 2O penult porous}
\int_0^1 \! \left( \left.\bs{n} \bcdot \QQt_1\right|_{\bdy} - \left.\Uirt_{1}\right|_{\nrm = 0}\right) \, \mathrm{d}z &= 0,
\end{align}
\end{subequations}
where outer/inner variables evaluated on $\bdy$ is meant in the sense of the outer/inner problem. Finally, integrating the leading-order interfacial continuity of flux condition (\ref{eq: BC Interface 3D BL inner}a) across the cell height, we can relate~\eqref{eq: int cont eq 2O penult bulk} and~\eqref{eq: int cont eq 2O penult porous}. Using the outer flow solutions~\eqref{eq: solution for first corr outer 3D} and~\eqref{eq: U_1 in outer 3d variables soln}, we obtain~\eqref{eq: int cont 2O final}.

\section{Flow coefficients}
\label{sec: Flow coefficients}

The flow coefficients $b_n(t)$, given in \eqref{eq: First order pressure Darcy}, are
\begin{subequations}
\begin{align}
b_0(t) &= -G^2(t) \left(\dfrac{27}{35}(1+A^2) + 36 \presfnz_0 \Darcy^2 (1+A)^2 \right), \\
\left(1 + 12 \Darcy \right) b_2(t) &= G^2(t) \left(\dfrac{27 A}{35} - 36 \Darcy (1+A)^2 \left(\Lambda_a + 2 \presfnz_0 \Darcy \right) \right), \\
\left(1 + 12 \Darcy \right) b_{2k + 1}(t) &= \dfrac{6}{5}(1-A) \dbyd{G}{t} \delta_{0k} + \dfrac{288 \Darcy (1+A)^2 (-1)^k}{\pi \left((2k + 1)^2 - 4 \right)}\left(\Lambda_a + \dfrac{2 \presfnz_0 \Darcy}{2k + 1} \right) G(t)|G(t)|, \\
b_{2k} &= 0 \quad \text{for } k \neq 0, 1,
\end{align}
\end{subequations}
for integer values of $k$, and where $\delta_{ij}$ is the Kronecker delta, and all other parameters are defined in \S\ref{sec: Example}.

\bibliographystyle{jfm} 

\bibliography{ref.bib}
 
\end{document}